\documentclass[twocolumn]{revtex4-1}
\usepackage{graphicx,amsmath}
\usepackage[dvipsnames]{xcolor}
\usepackage{subcaption}
\usepackage{enumerate}
\usepackage{float}
\usepackage{color}
\usepackage{calrsfs}

\newcommand{\lettersection}[1]{\emph{#1.---}}

\newcommand{\had}{\hat{a}^+}

\newcommand{\hbd}{\hat{b}^+}

\newcommand{\vac}{|{\rm vac}\rangle}
\newcommand{\ket}[1]{\left| #1 \right\rangle}
\newcommand{\bra}[1]{\left\langle #1 \right|}

\newcommand{\ain}{a_{{\rm in}}}
\newcommand{\bin}{b_{{\rm in}}}
\newcommand{\bout}{b_{{\rm out}}}
\newcommand{\aout}{a_{{\rm out}}}
\newcommand{\adin}{a^\dagger_{{\rm in}}}
\newcommand{\bdin}{b^\dagger_{{\rm in}}}

\newcommand{\eea}{\end{eqnarray}}
\newcommand{\bea}{\begin{eqnarray}}
\newcommand{\ee}{\end{equation}}
\newcommand{\be}{\begin{equation}}

\begin{document}
\title{Quantum Networks for Single Photon Detection}

\author{Tzula B. Propp}
\author{S.J. van Enk}
\affiliation{Department of Physics and
Oregon Center for Optical, Molecular \& Quantum Sciences\\
University of Oregon, Eugene, OR 97403}

\begin{abstract}

Single photon detection generally consists of several stages: the photon has to interact with one or more charged particles, its excitation energy will be converted into other forms of energy,  and amplification to a macroscopic signal must occur, thus leading to a ``click.'' We focus here on the part of the detection process before amplification (which we have studied in a separate publication). We discuss how networks consisting of coupled discrete quantum states and structured continua (e.g. band gaps) provide generic models for that first part of the detection process. The input to the network is a single continuum (a continuum of single-photon states), the output is again a single continuum describing the next irreversible step. The process of a single photon entering the network, its energy propagating through that network and finally exiting into another output continuum of modes can be described by a single dimensionless complex transmission amplitude, $T(\omega)$. We discuss how to obtain from $T(\omega)$ the photo detection efficiency, how to find sets of parameters that maximize this efficiency, as well as expressions for other input-independent quantities such as the frequency-dependent group delay and spectral bandwidth. 
We then study a variety of networks and discuss how to engineer different transmission functions $T(\omega)$ amenable to photo detection.

\end{abstract}

\maketitle
\section{Introduction}

The development of single photon detectors is a state of the art research area \cite{young2018,sunter2018,gemmell2017}. 
Fundamental limits to single photon detector (SPD) performance have yet to be uncovered. 
That is, even though there is a large amount of theory for each type of photo detector,
{\em fundamental} limits, independent of platform and architecture, 
should be derived from a general fully quantum-mechanical model of the whole photo detection process, from the initial physical contact the photon makes with the detector to the final ``click.'' 
Device-specific theories, however, are often at least in part phenomenological in nature.
The underlying basics of photodetection theory was developed in the early 1960s \cite{Glauber, MandelWolf,kelley1964}, with the quantum nature of light being taken into account, and with later additions to the theory also incorporating the backaction of the detector on the detected quantum field \cite{scully1969,yurke1984,ueda1999,schuster2005,clerk2010}. More recent additions to the theory have analyzed more deeply the amplification process by itself \cite{proppamp} and its relation to the absorption and transduction part of the process \cite{young2018,helmer2009}. In particular, it turns out that for an ideal detector one should decouple the two processes [by having an irreversible step in between the two] such that the amplification part does not interfere negatively with the absorption/transduction part \cite{young2018,clerk2010}. This decoupling will be assumed in this paper, too.
In order to develop a useful fully quantum-mechanical theory
we cannot be {\em completely} general; or, rather, if we are completely general, then the only statements on fundamental limits we can make are likely going to be merely examples of Heisenberg's uncertainty relations. So we will make three restrictive but---we think---reasonable assumptions about our quantum theory of photo detection.

  \begin{figure}[h] 
	\includegraphics[width=.8\linewidth]{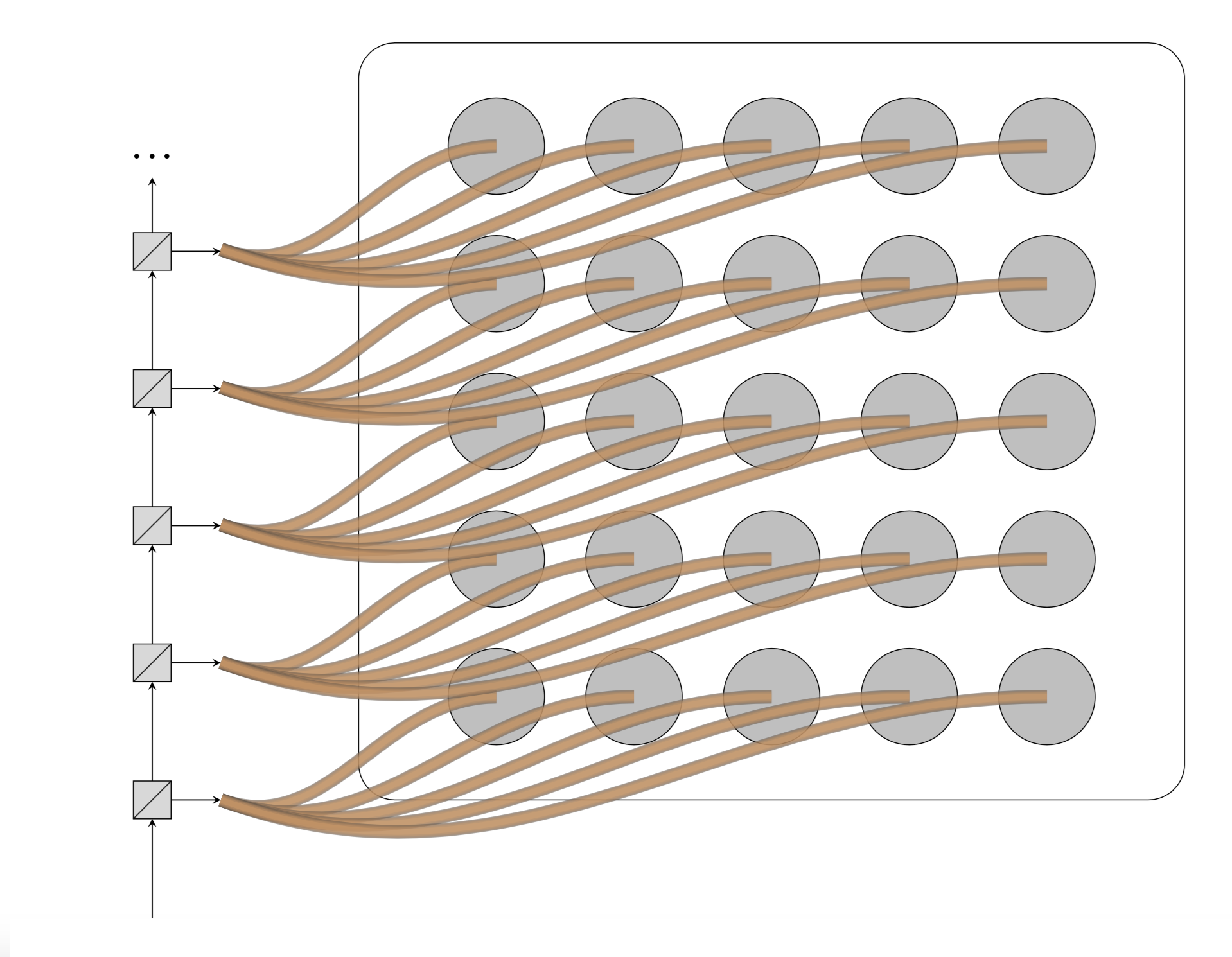} 
	\caption{An array of single photon detecting (SPD) pixels. A series of beamsplitters with low reflectivity ensure that at most one photon is incident on each single photon detector.}
	\label{array}
\end{figure}

First, we focus on {\em single}-photon detection.
The main reason is that number-resolved photo detection is possible using arrays of SPDs where each ``pixel'' receives at most one photon as in Fig. \ref{array} (also see \cite{photonnumber2017}, or \cite{migdall2003} for the time-reversed process of creating a single photon on demand).
So we focus on an individual pixel here. (See Ref.~\cite{combes2017} for a modeling framework for systems with multiple inputs and Refs.~\cite{fan2010,caneva2015,xu2015} for non-linear S-matrix treatments of few-photon transport.)  

Second, although a general state of a single photon is a function of four quantum numbers, one related to the spectral degree of freedom, two related to the two transverse spatial degrees of freedom, and one related to the polarization or helicity degree of freedom, we will restrict ourselves to the spectral (or, equivalently, the temporal) degree of freedom.
That is, the input state can be defined in terms of frequency-dependent creation operators $\had(\omega)$ acting on the vacuum.
The reason is that the other three degrees of freedom can, in principle, if not in practice, be sorted before detection. For example, if one wishes to distinguish between horizontally and vertically polarized photons, one may use a polarizing beam splitter and put two detectors behind each of the two output ports. Similarly, efficient sorting of photons by their orbital angular momentum quantum number \cite{osullivan2012} and spatial mode are also possible \cite{Bouchard2018,Fontaine2019}. It is easier to consider sorting as part of the pre-detection process, rather than a task for the detector itself (see footnote \footnote{To describe sorting as well within this framework, we simply write more transmission functions, e.g. $T_j(\omega)$ for all the different input continua $j$, each leading to their own output continuum. Even more complicatedly, we could consider multiple outputs $i$ for a given input $j$ and write $T_{i|j}(\omega)$. But even in this case we can focus on a particular $i$ and $j$ as we do this in this paper.}). On the other hand, the spectral response of a detector cannot be eliminated; the time/frequency degree of freedom is intrinsic to the resonance-structure of the photo detecting device. 

Third, we are going to assume that each pixel's operation is passive. That is, apart from being turned on at some point, and being turned off at some later point, it operates in a time-independent manner. Thus an incoming photon will interact with a time-independent quantum system. As we will see, active filtering is not needed for perfect detection provided the photo detector has no internal losses (couplings to additional continua).

We can now describe the interaction of a single photon with an arbitrary quantum system as follows. The system may be naturally decomposed into several
subsystems, each of which may have discrete and/or continuous  energy eigenstates. (For example, the photon may be absorbed by a molecule or atom or quantum dot or any structure with a discrete transition that is almost resonant with the incoming photon.)
The continua will in general be structured (for example, containing bands and band gaps in between) \cite{Tellinghuisen1975,Odegard2002,Nygaard2008}, but structured continua can be equivalently described as structureless (flat) continua coupled to (fictitious) discrete states \cite{pseudomodes2,pseudomodes3,pseudomodes4,pseudomodes1}, enabling a Markovian description of the system independent of an input photon's bandwidth. Indeed, it is well known that a non-Markovian open system can always be made Markovian by expanding the Hilbert space (the converse of the usual Stinespring dilation \cite{stinespring}). And so an arbitrary quantum system may be described by a network of discrete states (some physical, some fictitious), coupled to flat continua. The latter coupling makes the time evolution irreversible. Of course, an actual detector is indeed irreversible.
In particular, the amplification process (converting the microscopic input signal into a classical macroscopic output signal) is intrinsically irreversible. 

What we do here is give a general description of the photon entering some network of discrete states [indicated by the black box in Fig.~2] up to and including the first coupling to a flat continuum, that is, up to and including the first irreversible step in the process. 
We analyzed the (irreversible) amplification step in another paper \cite{proppamp} and found the fundamental limits on added noise arising from amplification are so mild that fundamental tradeoffs of a detector are in essence determined by the pre-amplification process, which is the process we analyze here.

 \begin{figure}[h] 
	\includegraphics[width=.6\linewidth]{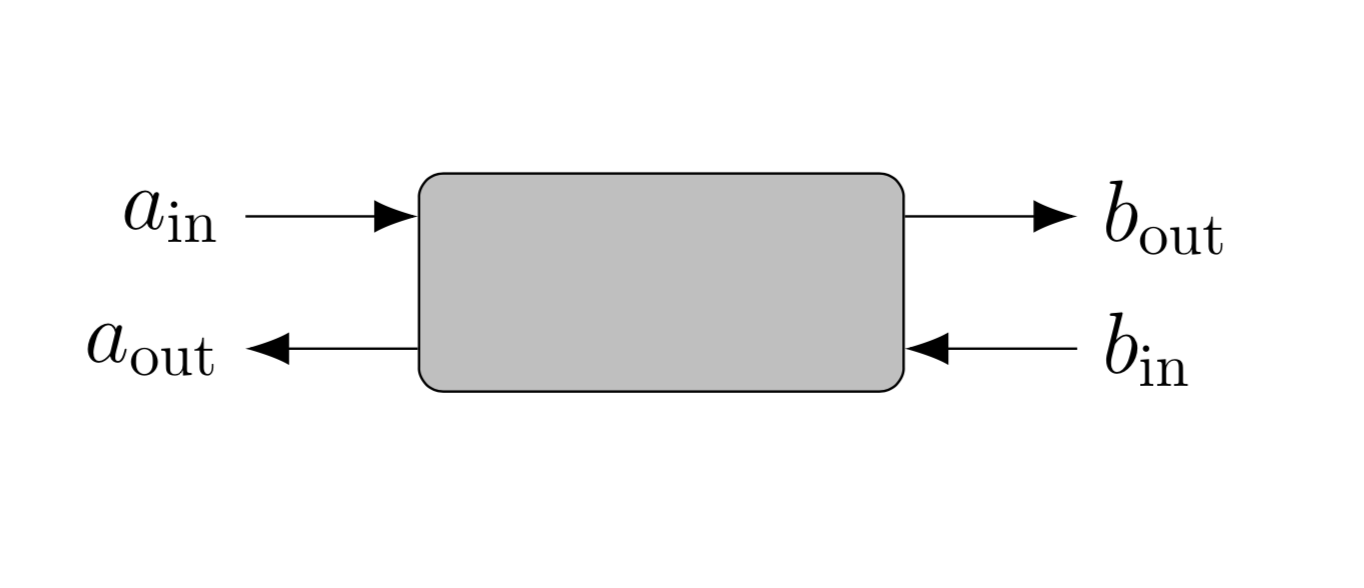} 
	\caption{Input and output fields coupled to a two-port black box quantum network.  The input and output operators are continuous mode (as functions of frequency $\omega$) annihilation operators, satisfying canonical commutation relations \cite{input1985}. Here the mode $\ain(\omega)$ carries the single-photon input state, the reflected mode is described by $\aout(\omega)$, and the output mode $\bout(\omega)$ contains the excitation energy if the input energy successfully traversed the network (and may serve as input to the next part of the photo detection process, e.g., amplification, see Ref.~\cite{proppamp}). $\bin(\omega)$ may contain thermal excitations, but is never occupied by the photon we wish to detect.}
	\label{2ports}
\end{figure}

As we will show below, this first part of the process can then be fully described in terms of a complex transmission amplitude $T(\omega)$, which is simply the probability amplitude for the frequency component of the input signal at frequency $\omega$ to survive the first part of the process and enter the next (amplification) process. (By then the energy $\hbar\omega$ will have been converted into a different type of energy, i.e., a different type of excitation, but that plays no particular role here.) It is straightforward to calculate the transmission coefficient through a network (the equations are linear!). The point here is that from $T(\omega)$ we can then determine three input-independent quantities of interest to the photo detection process. (That one complex function $T(\omega)$ is sufficient is due to our considering only the time/frequency degree of freedom, see again footnote \footnotemark[1]).

First, $|T(\omega)|^2$ gives an upper bound on the probability for the frequency component $\omega$ to be detected. (If there were no losses downstream, it would equal the probability of detection.) 
We are thus particularly interested in identifying quantum systems for which there is at least one frequency $\omega_i$ for which $|T(\omega_i)|=1$.

Second,
an upper bound to the total detectable frequency range is then given by the spectral bandwidth, defined as
\be
\tilde{\Gamma}=\frac{1}{\pi}\int_0^\infty d\omega |T(\omega)|^2.
\ee 
(The $\frac{1}{\pi}$ factor gives agreement with a classical Lorentzian filter; the transmission function is $T(\omega)=\frac{\Gamma}{\Gamma-i(\omega-\omega_0)}$ for a filter with damping factor $\Gamma$ and resonant frequency $\omega_0$ and we find $\tilde{\Gamma}=\Gamma$, see Eq. (\ref{GammaInt}) below.) The inverse of this quantity is also a measure of the time the photon spends in the detector. Indeed, $\tilde{\Gamma}^{-1}$ is a lower bound on the contribution to timing jitter from integrated detection event when quasi-monochromatic photon states are detected with high efficiency (see footnote \footnote{Timing jitter for a photo detection has three contributing components \cite{vanenk2017}. The first comes from the temporal spread of the mode onto which the measurement projects \cite{spectralPOVM} and is intrinsic to the resonance structure of the photo detector. The second comes from integrating a continuously monitored continua to form discrete detection event, which is necessary for an accurate information-theoretic characterization of photo detection and influences photo detection efficiency (see Appendix A). It is this part of the jitter that $\tilde{\Gamma}^{-1}$ is a lower bound for; it sets the timescale for integration where monochromatic states are detected with high efficiency. The third contribution comes from input signals with long temporal wave-packets and, since it is input-dependent, will be ignored in this analysis since here we assume no priors about the single photon input.}).

Third,
we can also define a delay (latency) using the polar decomposition of $T(\omega)$ 
\be
T(\omega)=|T(\omega)|\exp(i\phi(\omega))
\ee
and using the standard definition of group delay
as
\be
\tau_g(\omega)=-\frac{d\phi(\omega)}{d\omega}.
\ee

We can see how this directly relates to an experimentally-measured latency by consider an input single-photon with temporal wave packet $\psi_{\rm in}(t)$ and Fourier transform $\tilde{\psi}_{\rm in}(\omega)$. We can then write the output single-photon state

\bea\label{filtered state}
\ket{\psi'} &=&\int_0^\infty d\omega \tilde{\psi}_{\rm in}(\omega) R(\omega) \had_{\rm out} (\omega) \vac \nonumber \\
&+& \int_0^\infty d\omega \tilde{\psi}_{\rm in}(\omega) T(\omega) \hbd_{\rm out} (\omega) \vac 
\eea where the first and second terms correspond to the reflected and transmitted parts of the single-photon state, respectively  (for details, see Ref.~\cite{spectralPOVM}). From (\ref{filtered state}) we note that, after interacting with the network, the transmitted wave packet will have the form $\psi_{\rm out}(t) = FT^{-1} [\tilde{\psi}_{\rm in}(\omega) T(\omega)]$. For a long input pulse with central frequency $\omega'$, we find $\psi_{\rm out}(t)\approx |T(\omega')| \psi_{\rm in} (t-\tau) e^{i\omega'\tau'}$ with $\tau'$ the difference between the group delay $\tau_g(\omega')$ defined above and the (unmeasurable) phase delay. 

The effect of the group delay on an arbitrary input photon state is to selectively delay and reshape the transmitted wave packet. Of course, if an input photon has a wide spread of frequencies, a differential group delay may increase (or decrease) the temporal spread of the wave packet. This increase (or decrease) in the arrival times of different frequencies is manifestly input-dependent and will not play a role in the input-independent temporal uncertainty or jitter. We can, however, define an additional quantity characterizing the input-independent group delay-induced dispersion 

\be\label{maxdisp}
\mathcal{T}_g =\int_{0}^\infty d\omega \left |\frac{d\tau_g(\omega)}{d\omega}\right | |T(\omega)|^2
\ee whose definition agrees with our physical intuition that a constant group delay over the transmission window will not contribute to dispersion, nor will frequencies that are not transmitted regardless of how large $\frac{d\tau_g(\omega)}{d\omega}$ may be. We find that, for a flat transmission function that is unity over some spectral range $\omega_0\pm\delta\omega$ and zero everywhere else and a monotonic group delay $\tau_g(\omega)$, (\ref{maxdisp}) gives the difference in group delay between the minimum and maximumly transmitted frequencies: $\mathcal{T}_g=|\tau_g(\omega_0-\delta\omega) - \tau_g(\omega_0+\delta\omega)|$. This is clearly the maximum dispersion possible for any input to this system. 

Finding key conditions that change the transmission function $T(\omega)$ and frequency-dependent group delay $\tau_g(\omega)$ are important for the design of coupled-resonator optical waveguide (CROW) networks \cite{yariv1999} for delay-lines \cite{poon2004} and spectral filtering \cite{madsen2000}, where the transmission efficiency, frequency-dependent group delay, and spectral bandwidth will all affect performance. 
These are the three quantities we focus on in the rest of the paper. We'll start with the simplest quantum network to illustrate how we calculate $T(\omega)$ and how our three quantities of interest behave. After that we'll tackle more complicated networks. 

Critically, knowing $T(\omega)$ for a specific photo detector also allows one to construct the positive-operator valued measure (POVM), from which all standard figures of merit can be obtained \cite{vanenk2017}. The POVM element corresponding to a click after the photo detector has been left on for a very long time (in particular, long compared to the bandwidth $\tau\gg \tilde{\Gamma}^{-1}$; see Appendix A for a detailed POVM construction) has the particularly simple form

\bea\label{povmlongtime}
\hat{  \Pi}=\int_{0}^{\infty}\,d\omega\,|T(\omega)|^2 \ket{\omega}\bra{\omega}.
\eea
$\hat{ \Pi}$ is defined such that
the probability of a photon in a state $\hat{ \rho}$ being detected is given by the Born rule $\textnormal{Pr}=\textnormal{Tr}\left( \hat{ \Pi}\hat{ \rho}\right)$. For example,  any photon state $\hat{ \rho}=\sum_i \lambda_i\,\ket{\omega_i '}\bra{\omega_i '}$ where $|T(\omega_i ')|^2 = 1$ and $\sum_i \lambda_i = 1$ will be detected with unit probability. (The states the photo detector can detect perfectly include both pure states [when only one $\lambda_i$ is non-zero] and mixed states comprised entirely of frequencies where $|T(\omega_i')|^2 = 1$. ) Of course, no photon is truly monochromatic (or discretely polychromatic, but it can be effectively so if the wave-packet envelope is long compared to the inverse spectral bandwidth $\tilde{\Gamma}^{-1}$. 

\section{Simple example}

 \begin{figure}[h] 
	\includegraphics[width=.8\linewidth]{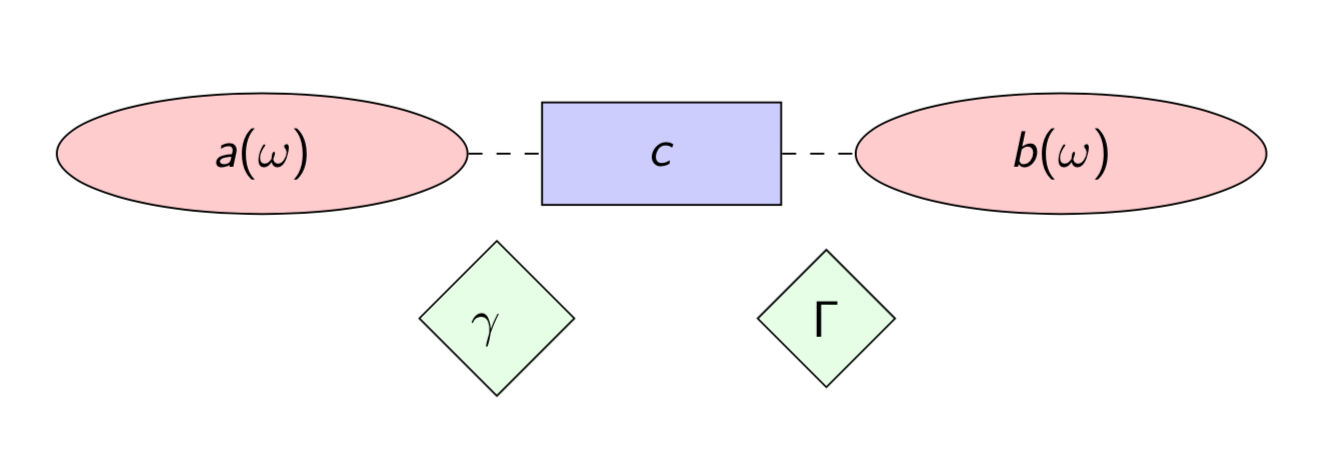} 
	\caption{A simple network comprised of a single discrete state described by an operator $c$ (the bosonic replacement for the two-level system fermionic lowering operator $\sigma^-$) and coupled to left and right continua $a$ and $b$ at rates $\gamma$ and $\Gamma$, respectively. That is, here the black box of Fig. \ref{2ports} contains just one two-level system.} 
	\label{simpleschem}
\end{figure}

The simplest quantum network consists of a single two-level system with a ground state $\ket{g}$ and an excited state $\ket{e}$, with the incoming photon coupling these states (Fig. \ref{simpleschem}). The two-level system is described by fermionic raising and lowering operators $\sigma^+ = \ket{e}\bra{g}$ and $\sigma^- = \ket{g}\bra{e}$ (this is also the simplest model of a photo detector, see \cite{Glauber}). Physically, this excited state could be any discrete state of a single absorber, e.g. the p-state of an atom which then decays to a monitored flat continuum. (Indeed, single absorbers such as atoms \cite{Cirac1997,vanEnk2000,Tey2008,Pinotsi2008,Wang2011}, single molecules \cite{Wrigge2007}, NV-centers \cite{iowaPOVM}, and quantum emitters \cite{Zumofen2008,Chen2011} are known to efficiently couple to and absorb single photons.)

In the Heisenberg picture, the evolution of these raising and lowering operators $\sigma^+$ and $\sigma^-$ will determine whether a photon makes it from one side of the network to the other. By focusing our analysis on cases where at most a single photon is in the network, we can use the equivalence between the two-level system and the simple harmonic oscillator to simplify our problem ab initio: we replace the fermionic raising and lowering operators with bosonic creation and annihilation operators $c^\dagger$ and $c$ \footnote{Since both operators and their expectation values (mode amplitudes) will satisfy the same systems of equations, we omit hats throughout this paper.}. 

We follow standard input-output theory here \cite{input1985}, separating the full evolution of both continua $a$ and $b$ into input and output modes (Fig. \ref{2ports}). By formally solving the Heisenberg evolution equations for the two input continuum mode annihilation operators $\bin$ and $\ain$, we can write the effective system Hamiltonian that governs the evolution of the system operators $c$ and $c^\dagger$

\bea\label{Hamiltonian1}
&\,\\
H&=-\hbar\omega_0 c^\dagger c -  \sqrt{\gamma}(c^\dagger \ain + c\adin) - \sqrt{\Gamma} (c^\dagger \bin + c\bdin)\nonumber
\eea where we have identified $\omega_0$ as the resonance frequency, and $\gamma$ and $\Gamma$ as the left and right side couplings to two continua $a$ and $b$ respectively \footnote{In assuming these couplings to be frequency independent, we are invoking a modified version of the first Markov approximation. Formally, we define $\gamma_i=2\pi\,\kappa_i^2(\omega_i)$ where $\omega_i$ is the resonance frequency of the $i$th discrete state and $\kappa_i(\omega)$ is the coupling between the $i$th discrete state and the left continuum at the discrete state frequency. This implies that, in an experiment, the decays $\gamma_i$ and resonances $\omega_i$ cannot be varied independently, which is well known in the context of the Thomas-Reiche-Kuhn sum rule for electric dipole transitions \cite{kuhn1992}}. In the Heisenberg picture, the time evolution of the discrete state annihilation operator is given by
\bea\label{1state}
\dot{c}=-i\omega_0 c -\frac{\gamma+\Gamma}{2} c - \sqrt{\gamma} \ain - \sqrt{\Gamma}\bin.
\eea
The input mode operators $\ain$ and $\bin$ and output mode operators $\aout$ and $\bout$ are determined by discrete state evolution of the operator $c$ and the two boundary conditions 
\bea\label{1statebound}
\aout-\ain= -\sqrt{\gamma} c\nonumber \\
\bout-\bin= -\sqrt{\Gamma} c.
\eea
It is easiest to solve the equations by taking the Fourier transform. Unitarity implies the existence of a transfer matrix relating in and out fields in the spectral domain

\bea\label{transfer}
\begin{bmatrix}
    \aout (\omega)       \\
    \bout (\omega)      \\
\end{bmatrix} = \begin{bmatrix}
    R(\omega)   & T(\omega)   \\
    T(\omega) & R(\omega)       \\
\end{bmatrix} \,\begin{bmatrix}
    \ain (\omega)       \\
    \bin (\omega)      \\
\end{bmatrix} 
\eea where $|T(\omega)|^2 + |R(\omega)|^2 = 1$ (resulting from our assumption there are no internal losses) \cite{vogelswelsch}. Defining a detuning $\Delta=\omega-\omega_0$, we can easily solve (\ref{1state}) in terms of the Fourier transform of the discrete state annihilation operator 

\bea\label{1statefreq}
c(\omega)= \frac{- \sqrt{\gamma} \ain(\omega) - \sqrt{\Gamma}\bin(\omega)}{\frac{\gamma+\Gamma}{2} -i\Delta}
\eea yielding a transmission function

\bea\label{T1}
T(\omega)= \frac{\sqrt{\gamma\Gamma}}{\frac{\gamma+\Gamma}{2} -i\Delta}.
\eea
We can see from (\ref{T1}) that perfect transmission ($|T(\omega)|^2=1$) occurs only when $\gamma=\Gamma$ and $\Delta=0$. These are the well-known conditions of balanced mirrors and on-resonance required for perfect transmission through a Fabry-Perot cavity \cite{siegman86}. 

We can also calculate the frequency dependent group delay from (\ref{T1})

\bea\label{T1}
\tau_{g} (\omega)=\frac{\frac{\Gamma+\gamma}{2}}{\left(\frac{\Gamma+\gamma}{2}\right)^2 + \Delta^2}.
\eea 
We see that like $T(\omega)$, the group delay is also a Lorentzian with width $\frac{\Gamma+\gamma}{2}$, and that frequencies close to resonance spend the most time in the network with a maximum group delay of $\frac{2}{\Gamma+\gamma}$ on resonance. We similarly find the input-independent group delay-induced dispersion (\ref{maxdisp}) to be $\mathcal{T}_g = \frac{8\gamma\Gamma}{(\gamma+\Gamma)^3}$.

We can also use $T(\omega)$ to calculate a spectral bandwidth (not to be confused with the channel bandwidth discussed in \cite{vanenk2017})

\bea \label{GammaInt}
 \tilde{\Gamma} &= \frac{1}{\pi}\int_{0}^\infty \,d\omega\,|T(\omega)|^2\\
 &= \frac{2\Gamma\gamma}{\Gamma+\gamma}
\eea which is a measure of the number of frequencies that can be efficiently detected. For this simple case, we note that $\tau_g (\omega)=  \tilde{\Gamma}^{-1} |T(\omega)|^2$ and thus
$\int\, d\omega \tau_g(\omega)=\pi$. (This will not be true for a general network, as we shall see shortly.)

\section{Quantum Networks}

We now set up the general problem of an arbitrary network of discrete states connecting two continua. The Hamiltonian is a straightforward generalization of (\ref{Hamiltonian1})

\bea\label{HamiltonianGen}
H=-\sum_i\hbar\omega_i c^\dagger_i c_i - \sum_{ij} g_{ij} (c_ic^\dagger_j + c^\dagger_ic_j)\nonumber\\
- \sum_i \sqrt{\gamma_i}(c^\dagger_i \ain+ c_i\adin) - \sum_i \sqrt{\Gamma_i} (c^\dagger_i \bin + c_i \bdin)\nonumber \\
\eea where we've now defined a real coherent coupling between discrete states $g_{ij}$ (we define $g_{ii}=0$ for each state). Some states may not be coupled to one (or both) continuum, in which case either $\sqrt{\gamma_i}$ or $\sqrt{\Gamma_i}$ (or both) will be zero. 

We can similarly generalize the operator evolution in (\ref{1state}) for an arbitrary network; moving to the spectral domain, we write the spectral dependence of the discrete state operators 

\begin{widetext}\bea\label{quantlangspect}
-i\Delta_i c_i (\omega)= &- \sum\limits_j\,\left(\frac{\sqrt{\gamma_i\,\gamma_j} + \sqrt{\Gamma_i\,\Gamma_j}}{2} + i g_{i\,j}\right) c_j(\omega) -\sqrt{\gamma_i}\,a_{\rm in}(\omega)  -\sqrt{\Gamma_i}\,b_{\rm in}(\omega).
\eea\end{widetext}

Similarly to (\ref{1statebound}), we can write boundary conditions for the two continua with an arbitrary network 

\bea\label{Nstatebound}
\aout(\omega)-\ain(\omega)= -\sum\limits_{i}\sqrt{\gamma_i} c_i(\omega)\nonumber \\
\bout(\omega)-\bin(\omega)= -\sum\limits_i \sqrt{\Gamma_i} c_i(\omega).
\eea

Now going from (\ref{Nstatebound}) to the transfer matrix (\ref{transfer}) involves solving $N$ systems of $N$ coupled first-order differential equations \footnote{Luckily, there is some redundancy. Once one has solved for one $c_i$ in terms of the input and output fields, one can permute the labels to generate the remaining solutions.}. This complication occurs because each discrete state amplitude $c_i(\omega)$ depends on every other amplitude \footnote{One alternative approach for a general network is to take the weak-coupling limit ($2\,g_{ij}\ll\sqrt{\gamma_i\,\gamma_j} + \sqrt{\Gamma_i\,\Gamma_j}\,\forall i,j$) and truncate the solutions after some power in $g_{ij}$, making (\ref{RGen}) the $0$th order approximate solution with higher order corrections. However, this method fails if even a single discrete state decouples from both continua.}.  One can also use numerical techniques to diagonalize the systems of equations and find the transmission function numerically \cite{datta2005}. But this rapidly gets harder with large systems, and masks the analytic conditions for perfect transmission we are interested in identifying. Here, we will instead identify large classes of systems that can be solved exactly with arbitrary couplings, decays, and resonant frequencies.

\underline{Parallel}: each discrete state is directly coupled to both continua ($\gamma_i\neq0$ and $\Gamma_i\neq 0 \,\forall i$) but not to each other. Thus there are multiple parallel paths to the same final state and hence we'll get interference. Physical photo detection platforms described by parallel networks include (i) single atoms with multiple p-states that then decay directly to a continuum (similarly for trapped ions/atoms due to Stark and Zeeman effects, see  Ref.~\cite{Higginbottom2016} for this in generating single photons), (ii) quantum dots \cite{Livache2019}, (iii) structured continua with multiple pseudomodes \cite{Hughes2018}.

\underline{Series}: each of the two continua are coupled to their own discrete state, which are in turn coherently coupled by a chain of intermediate single discrete states. There is just one path from the input continuum to the output continuum. This model describes (i) an atom in a p-state that first decays to a d-state or metastable s-state before decaying to a flat continuum (and analogously for a molecule \cite{young2018}), (ii) coherently coupled frequency filtering in front of a photo detecting platform (for instance, using an antireflective coating \cite{Rosfjord2006} or optical cavities \cite{Dilley2012}), (iii) atomic chains coupled through their p-states (which could be mediated by, for example, a fiber mode \cite{Song2018}).

\underline{Hybrid}: a combination of the above cases. For example, two parallel paths of three steps each or, more generally, layered structured continua (a photon must pass through one before the other). A physical system that can be modeled with a hybrid network is a photosynthetic light-harvesting (i.e. Fenna–Matthews–Olson) complex with multiple paths for coherent transport \cite{Scholes2017,Valleau2017,Chan2018}.

In all cases, we will solve  the systems of equations for $R(\omega)$ directly and make use of the identities $|T(\omega)|^2 = 1-|R(\omega)|^2$ and $T^2(\omega) = -R^2(\omega) \frac{|T(\omega)|^2}{|R(\omega)|^2} $when we calculate the transmission efficiency, spectral bandwidth, and group delay.

\subsection{Parallel Networks}

 \begin{figure}[h] 
	\includegraphics[width=.8\linewidth]{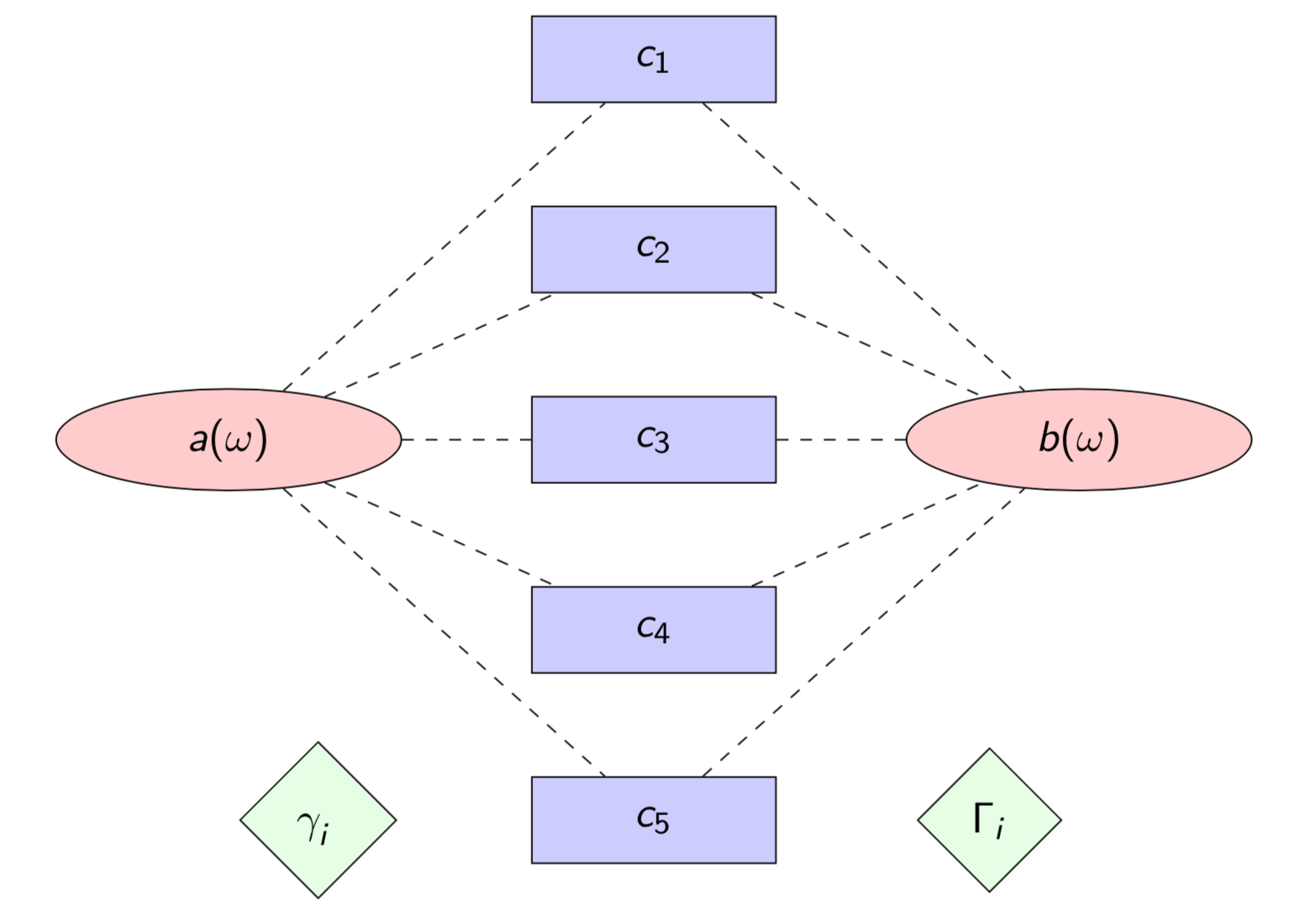} 
	\caption{A parallel network of $N=5$ decoupled discrete states, each described by an operator $c_i $ and coupled to left (input) and right (output) continua $a$ and $b$ at rates $\gamma_i$ and $\Gamma_i$, respectively.}
	\label{pschem}
\end{figure}

We now consider a parallel quantum network where each discrete state is directly coupled to both continua (Fig. \ref{pschem}). If the discrete states are directly coupled to each other, it is helpful to first diagonalize the system in the basis where they are decoupled. We then consider the new modes to be the discrete states a photon can occupy, each with a new resonance frequency and two decay rates to the two continua. (In the strong coupling limit, Rabi splitting comes into effect and these dressed states are the more physical description anyways \cite{steckquoptics}.) This decoupled manifold of discrete states, each decaying at a rate $\gamma_i$ to the input continuum and a rate $\Gamma_i$ to the output continuum (Fig. \ref{pschem}), is the next simplest case to analyze, and provides a simple model of a photo detector with a simple band structure.

We find the frequency dependence of the discrete state operators

\bea\label{quantlangspectpar}
-i\Delta_i c_i (\omega)= &- \sum\limits_j\,\frac{\sqrt{\gamma_i\,\gamma_j} + \sqrt{\Gamma_i\,\Gamma_j}}{2} c_j(\omega) \nonumber\\
&-\sqrt{\gamma_i}\,a_{\rm in}(\omega)  -\sqrt{\Gamma_i}\,b_{\rm in}(\omega).
\eea

A salient feature of Eq. \ref{quantlangspectpar} is the decay terms produce purely virtual coupling between discrete states coupled to the same continua. These \emph{cannot} be thought of as coupling mediated by continua, as the continua are flat (Markovian) and perfectly dissipative, and is instead a purely information-theoretic phenomeona \footnote{We've included the assumption of a single input photon in our theory \emph{ab initio}. Thus, discrete state expectation values must be correlated, manifesting as the purely virtual coupling in Eq. \ref{quantlangspectpar}. If we relax the assumption of a single input photon, we find that these correlations still manifest in the portion of the POVM that projects onto a single photon Hilbert space, that is, the part that is relevant for single photon detection! (See Ref.~\cite{spectralPOVM} for details on constructing POVMs that include general Fock states of photons.)}.

We use (\ref{transfer}) to find $R(\omega)$ by considering an input on only one side of the network (thus setting the expectation value of the other input field operator to zero). This yields an expression $a_{out} (\omega) =R(\omega) a_{in}(\omega)$ (or $b_{out} (\omega) =R(\omega) b_{in}(\omega)$), from which we reconstruct $T(\omega)$. We analytically find the general form of $R(\omega)$ 

\bea\label{RGen}
R(\omega)=\frac{\prod\limits_i \left(\frac{\Gamma_i - \gamma_i}{2} - i\Delta_i\right) - X_-^{(N)}}{\prod\limits_i \left(\frac{\Gamma_i + \gamma_i}{2} - i\Delta_i\right) - X_+^{(N)}}
\eea where $X_\pm^{(N)}$ is a polynomial of order $N-2$ in the detunings $\Delta_i$. (We trivially find $X_\pm^{(1)} = 0$.)  For $N=2$, we can see that $X_\pm^{(2)}$ is symmetric (anti-symmetric) between $\Gamma_i$ and $\gamma_i$

\bea\label{X2}
X_\pm^{(2)} = \left(\frac{\sqrt{\Gamma_1\Gamma_2} \pm \sqrt{\gamma_1\gamma_2}}{2}\right)^2.
\eea
This (anti-)symmetry is also present in higher-$N$ coefficients. From (\ref{RGen}) we can determine a key feature of parallel quantum networks: if some subset of the discrete states have \emph{balanced} decay rates such that $\gamma_i=\Gamma_i$, for large spacings between discrete states compared to the other decay rates $|\omega_i-\omega_j|\gg \gamma_j,\Gamma_j$, we find $R(\omega_i)=0$; input monochromatic photons with frequencies on resonance with those discrete states are transmitted perfectly through the network \footnote{For $N=2$, the requirement for large spacing such that $R(\omega_i)=0$ is less stringent, we only need it much larger than difference in decay rates $|\omega_i-\omega_j|\gg (\sqrt{\Gamma_j}-\sqrt{\gamma_j})^2$.}. 

In general, finding the specific form of $X_\pm^N$ is a numerical task, and we will focus on a simpler case where we can utilize another salient feature of parallel networks: the purely virtual coupling present in (\ref{quantlangspectpar}). Before we consider a network that is uniformly coupled (all decays are the same), we can consider a network with couplings that are inhomogeneous but uniformly unbalanced such that $\Gamma_i = k\,\gamma_i \,\forall i$. We can then write (\ref{quantlangspectpar}) in a simplified form

 \begin{figure}[b] 
	\includegraphics[width=.9\linewidth]{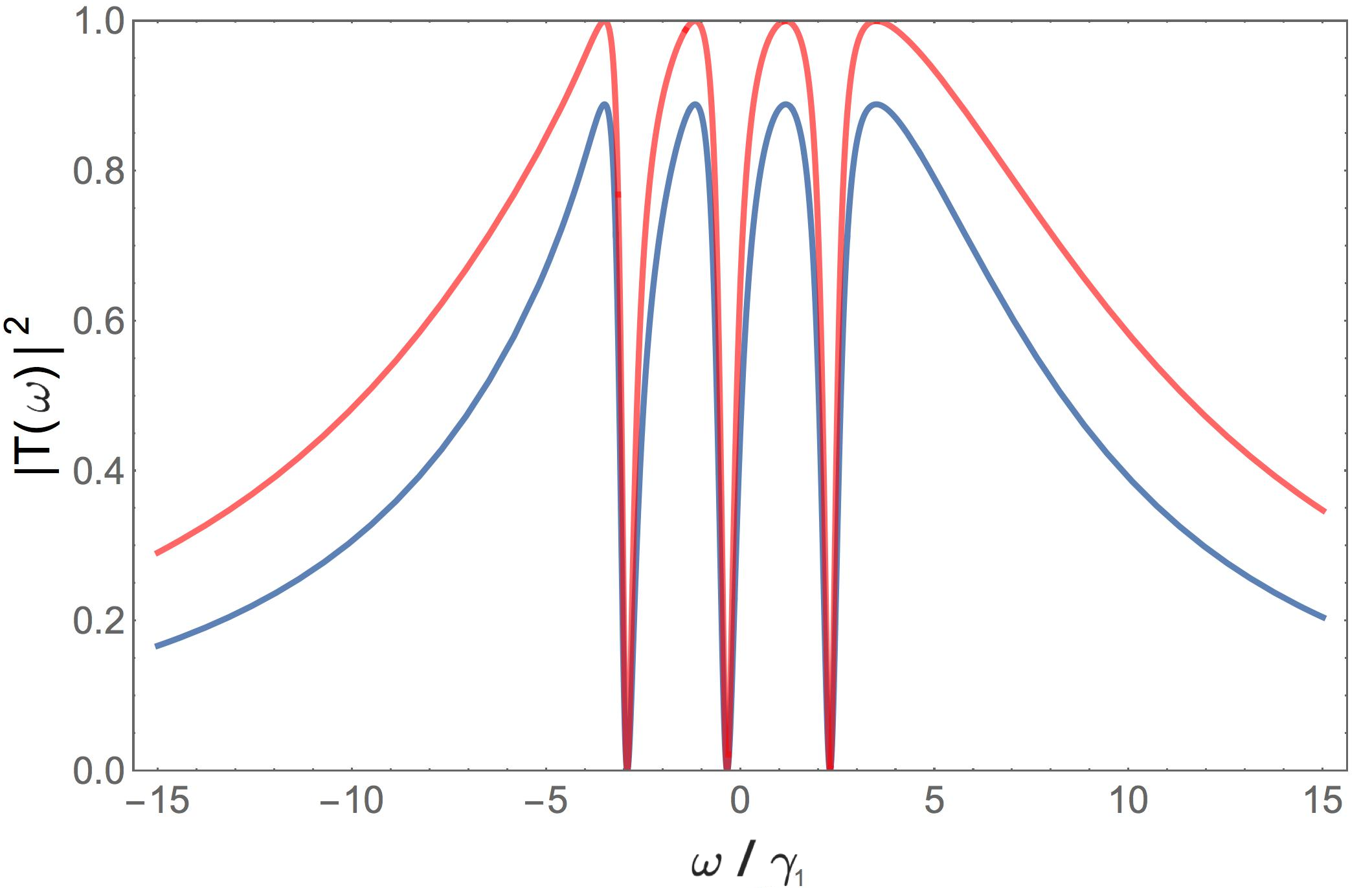} 
	\caption{Transmission probability for parallel networks with $N=4$ equally spaced discrete states, each with a decay rate to the input continuum $\gamma_i=\gamma_1 \,(\frac{7}{5})^{i-1}$. The decays to the monitored output continuum are uniformly unbalanced ($\frac{\Gamma_i}{\gamma_i}=k\,\forall i$), with $k=\frac{1}{2}$ in blue and $k=1$ in red. Frequency is measured w.r.t. the average resonance frequency. We observe the four resonance frequencies $\omega_i$ have maximum transmission probability $|T(\omega_i)|^2=\frac{4k}{(k+1)^2}$. The three frequencies of perfect reflection (i.e, $T(\omega_i)=0$) correspond to solutions of $\sum_i\frac{\gamma_i}{\Delta_i}=0$.}
	\label{parralelk}
\end{figure}

    \begin{figure*}[t]
        \begin{subfigure}[b]{0.32\textwidth}
            \centering
            \includegraphics[width=1.1\textwidth]{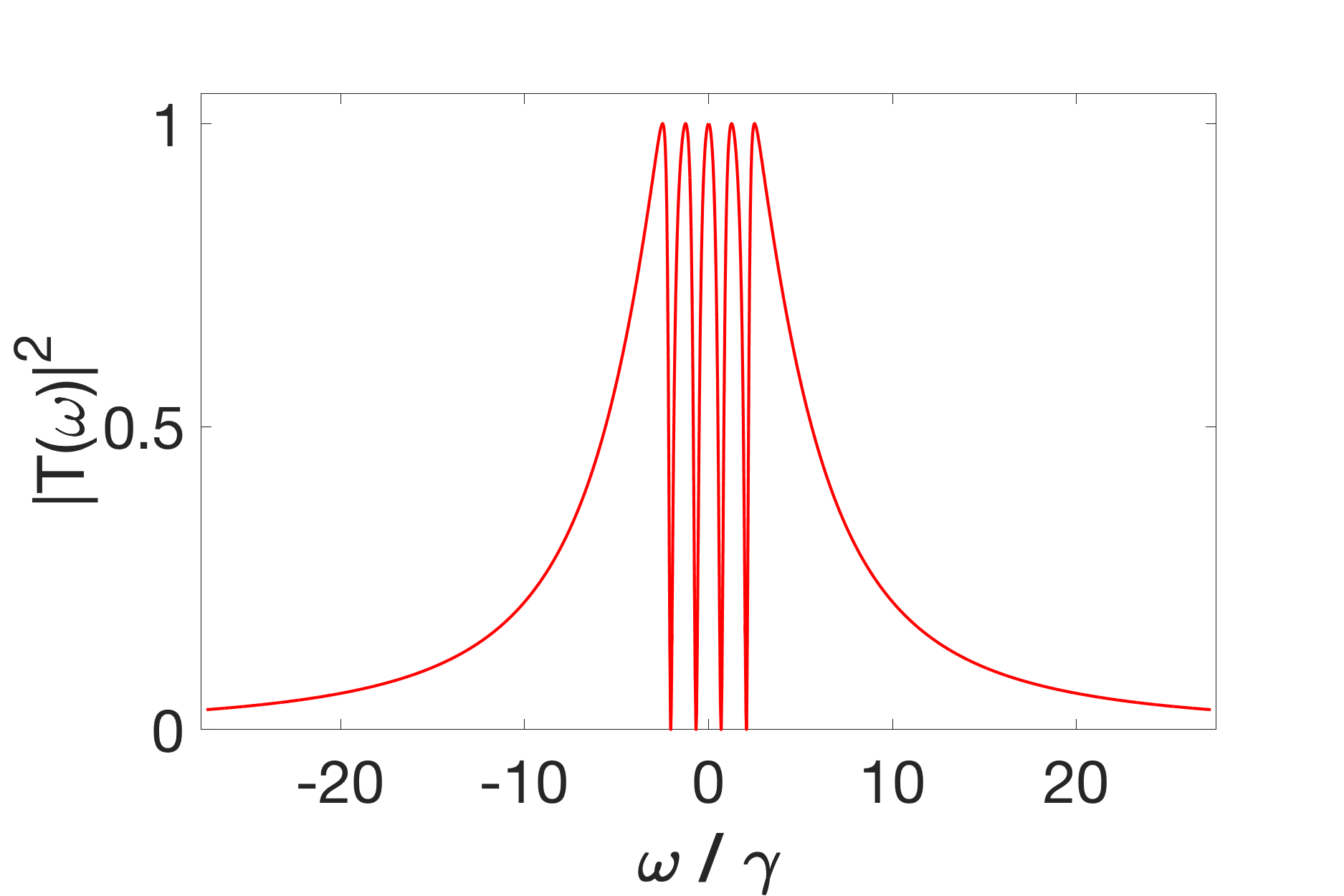}
            \caption[]%
            {{\small Transmission probability $|T(\omega)|^2$.}}    
            \label{parralelhomoT}
        \end{subfigure}
        \hfill
        \begin{subfigure}[b]{0.32\textwidth}  
            \centering 
            \includegraphics[width=1.1\textwidth]{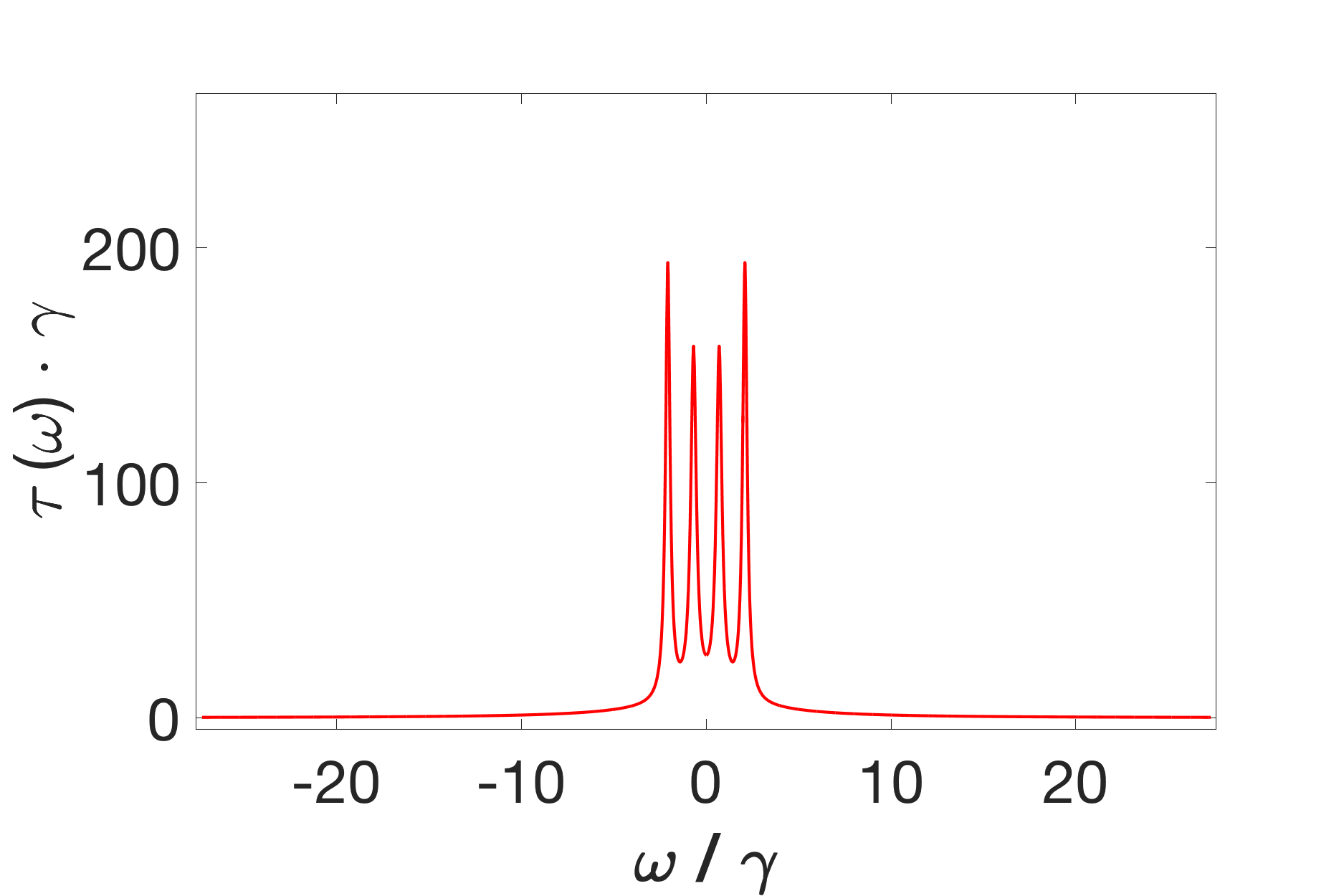}
            \caption[]%
            {{\small Group delay $\tau_g(\omega)$ in units of $1/\gamma$.}}    
            \label{parralelhomoG}
        \end{subfigure}
               \hfill
        \begin{subfigure}[b]{0.32\textwidth}  
            \centering 
            \includegraphics[width=1.1\textwidth]{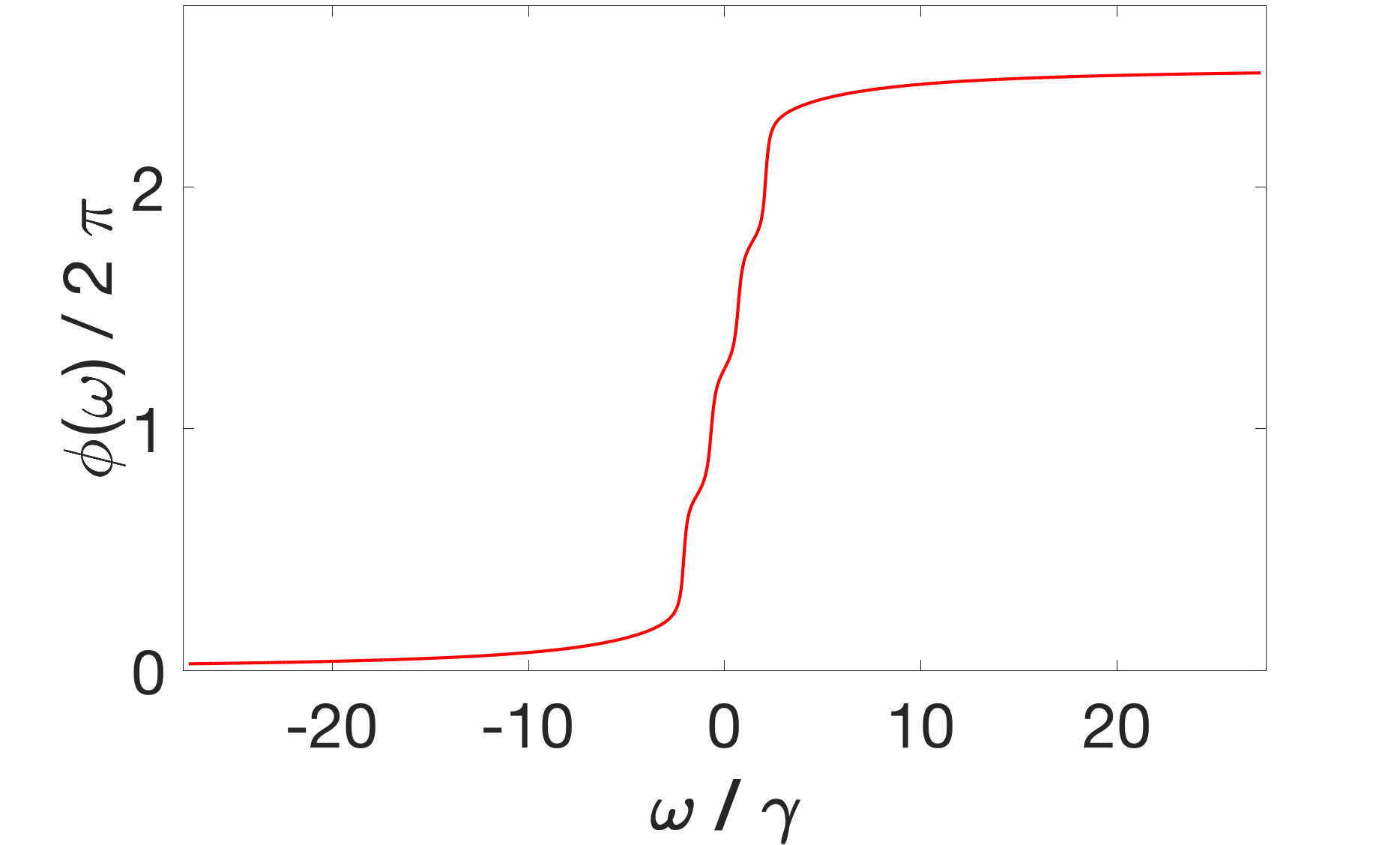}
            \caption[]%
            {{\small Phase $\phi(\omega) $ of $T(\omega)$ in units of $2\pi$.}}
            \label{parralelhomoP}
        \end{subfigure}
        \vskip\baselineskip
        \caption{\small Transmission probability, group delay, and transmission function phase for a parallel network with $N=5$ equally spaced discrete states with balanced decay rates to both continua ($\gamma=\Gamma$). Frequency is measured w.r.t. the average resonance frequency. We observe the five resonance frequencies $\omega_i$ each correspond to a perfectly transmitted frequency. The four frequencies of perfect refection correspond to solutions of $\sum_i\frac{1}{\Delta_i}=0$. The group delay is always largest for the highest and lowest frequency resonances (except in the large spacing limit, where they are of equal magnitude). The total change in phase of the transmission function is directly proportional to the number of discrete states.} 
        \label{parralelhomo}
    \end{figure*}

\bea\label{quantlangspectcorrk}
\,&\,\\
&\frac{i\Delta_i}{\sqrt{\gamma_i}}c_i (\omega) &= \sum\limits_j\frac{\sqrt{\gamma_j}(1+k)}{2} c_j(\omega) +a_{\rm in}(\omega)  +\sqrt{k}\,b_{\rm in}(\omega)\nonumber
\eea so that we observe strong correlations between discrete state amplitudes
\bea\label{quantlangspectcorrk2}
&\frac{i\Delta_i}{\sqrt{\gamma_i}}c_i (\omega) = \frac{i\Delta_{i'}}{\sqrt{\gamma_{i'}}}c_{i'} (\omega).
\eea

For non-degenerate states, Eq. \ref{quantlangspectcorrk2} means that a photon that is resonant with one state will only excite that state. (For infinitely-narrow discrete states, these correlations are satisfied trivially of course; $c_{i'}(\omega)$ is only non-zero at $\omega_{i'}$.) Eq. \ref{quantlangspectcorrk2} also indicates that the relative phase of discrete state amplitudes is frequency dependent; when $\omega_i < \omega < \omega_{i'}$, there is a relative phase of $\pi$ between $c_i$ and $c_{i'}$. (Outside of photo detection, this purely virtual coupling provides a possible alternative explanation for the destructive interference present in atoms along a fiber \cite{atomwire1, atomwire2} and multi-mode Fabry-Perot cavities \cite{multimodefabry}.) We can see destructive interference directly from the form of the reflection coefficient

\bea\label{reflectk}
R(\omega)=\frac{i-(k-1)\sum\limits_i \frac{\gamma_i}{2(\Delta_i)}}{i-(k+1)\sum\limits_i \frac{\gamma_i}{2(\Delta_i)}}
\eea from which we can determine $|T(\omega)|^2$ and the other quantities of interest. 

We can further specialize to the case of homogenous coupling where $\gamma_i=\gamma$ and $\Gamma_i=\Gamma$. This case is of interest for several reasons: most generally, this assumption directly follows from the first Markov approximation if the spacing between discrete states is small compared to the decays. It also simplifies the form of the correlations between discrete states $\Delta_i c_i(\omega)=\Delta_{i'} c_{i'}(\omega)$ \footnote{We now can perform a small sanity check. Considering the two continuum fields together, we define total input and output flux operators $J_{in} = \sqrt{\gamma} a_{in} + \sqrt{\Gamma} b_{in}$ and $J_{out} = \sqrt{\gamma} a_{out} + \sqrt{\Gamma} b_{out}$. Solving for the total outputs directly from (\ref{Nstatebound}), we find $J_{out} =\frac{i+ \frac{\gamma+\Gamma}{2}\,\sum_i \frac{1}{\Delta_i}}{i - \frac{\gamma+\Gamma}{2}\,\sum_i \frac{1}{\Delta_i}}\,J_{in}$. We can immediately see that $|J_{out}|=|J_{in}|$; photon flux is preserved through the system at every frequency, as it must be since there are no side channels present.}, as well as the form of the reflection coefficient 

\bea\label{reflecthomo}
R(\omega)=\frac{i-\frac{\Gamma-\gamma}{2}\sum\limits_i \frac{1}{\Delta_i}}{i-\frac{\Gamma+\gamma}{2}\sum\limits_i \frac{1}{\Delta_i}}.
\eea

\lettersection{Perfect transmission} From (\ref{reflectk}) and (\ref{reflecthomo}) we can see that in both cases there are $N$ frequencies of maximum transmission corresponding to each resonant frequency $\Delta_i=0$ with transmission probability $|T(\omega_i)|^2=\frac{4k}{(k+1)^2}$ and $|T(\omega_i)|^2=\frac{4\gamma\Gamma}{(\gamma+\Gamma)^2}$, respectively.  We also see $N-1$ frequencies of destructive interference corresponding to the $N-1$ solutions of $\sum\limits_{i}\frac{\gamma_i}{\Delta_i} =0$ (Fig. \ref{parralelk}) determined solely by the decays and resonances (and notably not by $k$). We similarly observe for the case of a network with homogenous decay rates $N-1$ frequencies of destructive interference corresponding to the solutions of $\sum\limits_{i}\frac{1}{\Delta_i} =0$ (Fig. \ref{parralelhomoT}). One might think that, in principle, a resonant frequency could coincide with a frequency of perfect reflection when $N>2$, in which they can annihilate. However, this only occurs when two discrete states are energetically degenerate, and since they couple to the same $1$D continuum, this is forbidden by unitarity; as we decrease the spacing between states, we see a resonant frequency and a frequency of destructive interference annihilate as a discrete state is forced to decouple from the system as the degeneracy becomes exact. 

In general, the condition for perfect transmission through a parallel network is that all the couplings be balanced ($\gamma=\Gamma$ or in the inhomogeneous uniformly unbalanced case, $k=1$). As we saw in the case for a completely arbitrary parallel network, we see that perfect transmission at some discrete state frequency $\omega_i$ not only requires balanced coupling $\gamma_i=\Gamma_i$, but also that all the other discrete states either be far away in frequency compared to the their decay rates ($|\omega_i-\omega_j|\gg\gamma_j,\Gamma_j$), or also be balanced ($\gamma_j=\Gamma_j$), or a mix of the two.

\lettersection{Spectral Bandwidth} Once we have the reflection coefficient, we can calculate the three quantities of interest. For all three cases we've discussed, we find that the spectral bandwidth is purely additive $\tilde{\Gamma} = \sum_i \frac{2\gamma_i\Gamma_i}{\gamma_i+\Gamma_i}$ and is completely independent of the spacing between discrete states \footnote{The uncertainty in frequency, as defined in \cite{vanenk2017} and calculated entropically from the spectral POVM \cite{spectralPOVM}, is in this case directly proportional to the bandwidth (and hence also independent of discrete state spacing).}.

\lettersection{Group Delay} We note that, unlike the simple model, the sharp peaks in the group delay (Fig. \ref{parralelhomoG}) correspond to frequencies of destructive interference and are greatest for the outermost frequencies of destructive interference despite all the frequencies of note in (Fig. \ref{parralelhomoT}) being completely destructive or constructive. We also observe that the relationship between the three quantities of interest discussed for the simple model is not present: here $\tau(\omega_i) > \tilde{\Gamma}^{-1} |T(\omega_i)|^2$ for each resonant frequency. We find that the phase of $T(\omega)$ increases by $\pi$ with each resonant frequency (\ref{parralelhomoP}). This provides a novel application for single-photon interferometry for resolving tightly-structured resonance structures and explains why, whereas the spectral bandwidth $\tilde{\Gamma}$ is independent of discrete state spacing, the group delay increases with close spacing; the same change in phase is occurring in a smaller spectral range so the magnitude of the group delay $|\tau_g(\omega)|=|\frac{d\phi(\omega)}{d\omega}|$ increases.

\vspace{-1em}
\subsection{Series Networks}

\begin{figure}[h!] 
	\includegraphics[width=\linewidth]{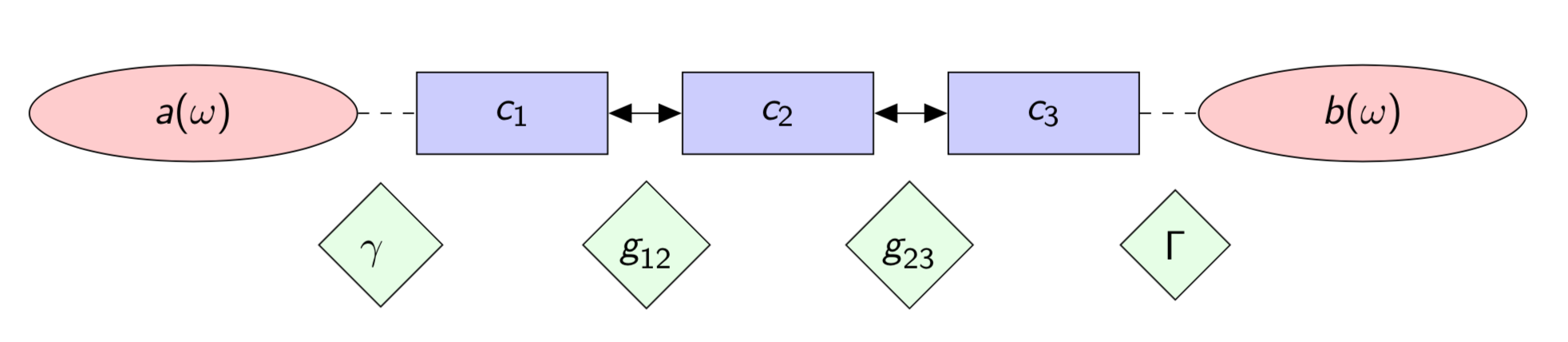} 
	\caption{A series network of three coherently coupled discrete states, each described by an operator $c_i$ and coupled to each other at rates $g_{ij}$. The first and last states are incoherently coupled to left (input) and right (output) continua $a$ and $b$ at rates $\gamma$ and $\Gamma$, respectively.}
	\label{seriesschem}
\end{figure}

\begin{center}
\begin{figure}[b]
\begin{tabular}{| c | c | c|}
\hline
$n $&$ a_n $&$ b_n $\\
\hline\hline
$0 $&$ 0 $&$ 1 $\\
\hline
$1$&$-\gamma $&$ \frac{\gamma}{2} -i\,\Delta_1 $\\
\hline
$2 $&$ g_{12}^2 $&$ -i\,\Delta_2 $\\
\hline
$\vdots$ & $\vdots$ & $\vdots$ \\
\hline
$N-1 $&$ g_{N-2 N-1}^2 $&$ -i\,\Delta_{N-1}$\\
\hline
$N  $&$ g_{N-1 N}^2 $&$ \frac{\Gamma}{2}-i\,\Delta_{N}$\\
\hline
$>N  $&$0 $&$0$\\
\hline
\end{tabular}
\caption{Wallis-Euler coefficients for $N>1$ discrete states in series. For a network with arbitrarily high-$N$, these can be used to generate transmission functions (that correctly encode the causal structure that is inherent to the network outside of the strong-coupling limit) using the Wallis-Euler recursion relations (\ref{WallisEuler}).}\label{table}
\end{figure}
\end{center}

        \begin{figure*}[t]
        \begin{subfigure}[b]{.475\textwidth}
            \centering
            \includegraphics[width=\textwidth]{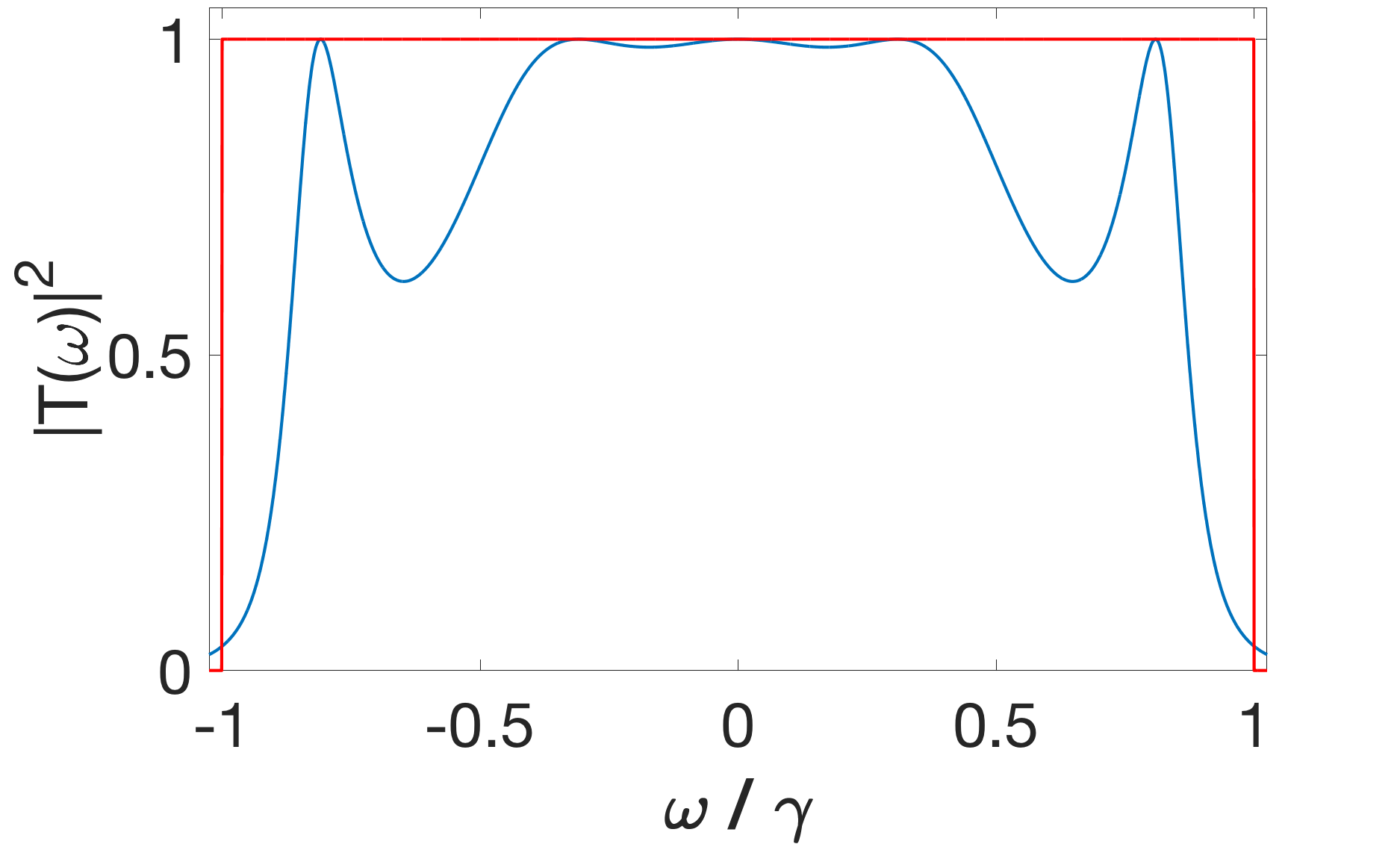}
            \caption[]%
            {{\small  $N=5$ discrete states, balanced decays $\gamma=\Gamma$}}    
            \label{CriticalOddBalanced}
        \end{subfigure}
        \hfill
                \begin{subfigure}[b]{.475\textwidth}   
            \centering 
            \includegraphics[width=\textwidth]{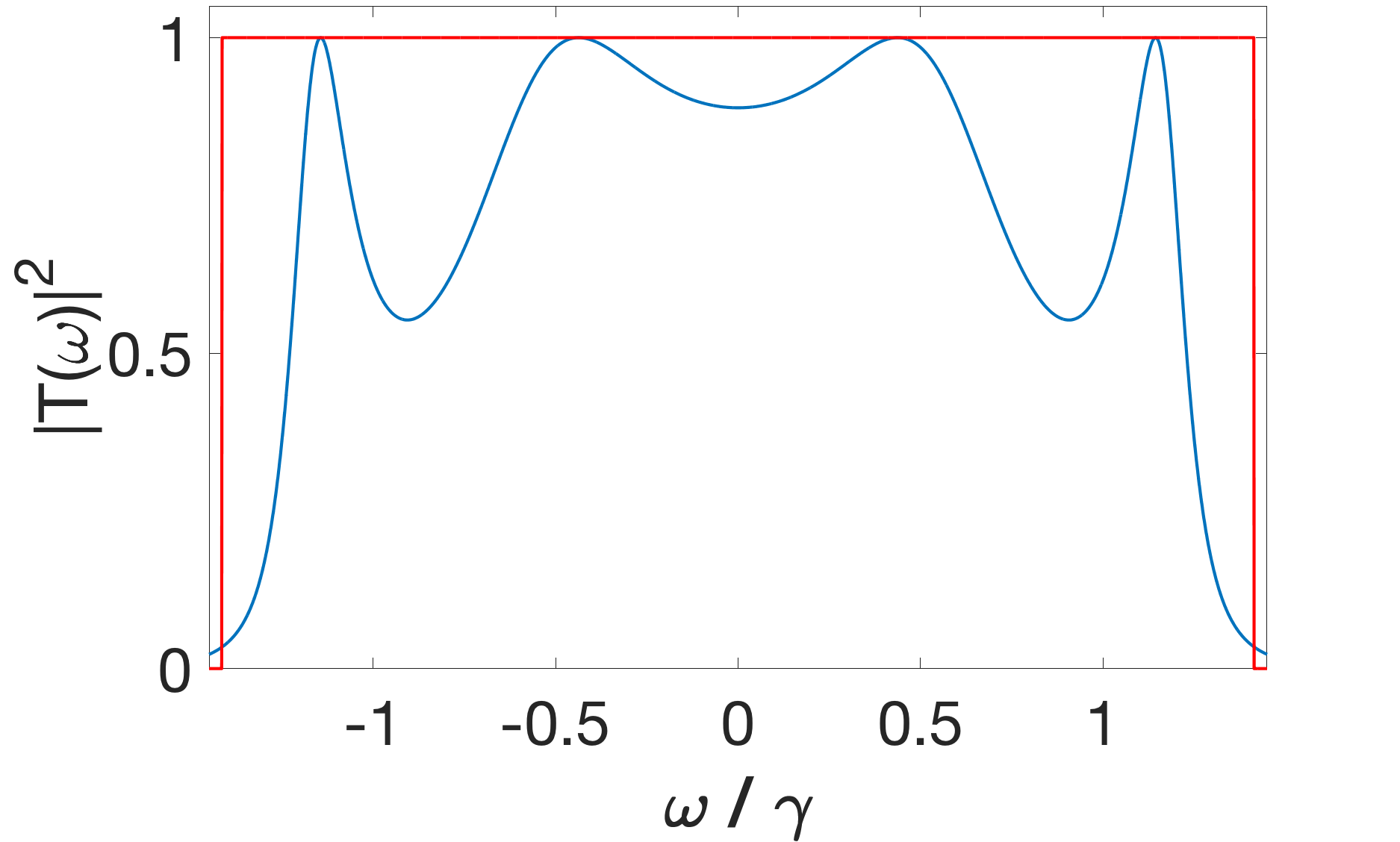}
            \caption[]%
            {{\small $N=5$ discrete states, unbalanced decays $2 \gamma=\Gamma$}}    
            \label{CriticalOddUnbalanced}
        \end{subfigure}
        \vskip\baselineskip
        \begin{subfigure}[b]{.475\textwidth}   
            \centering 
            \includegraphics[width=\textwidth]{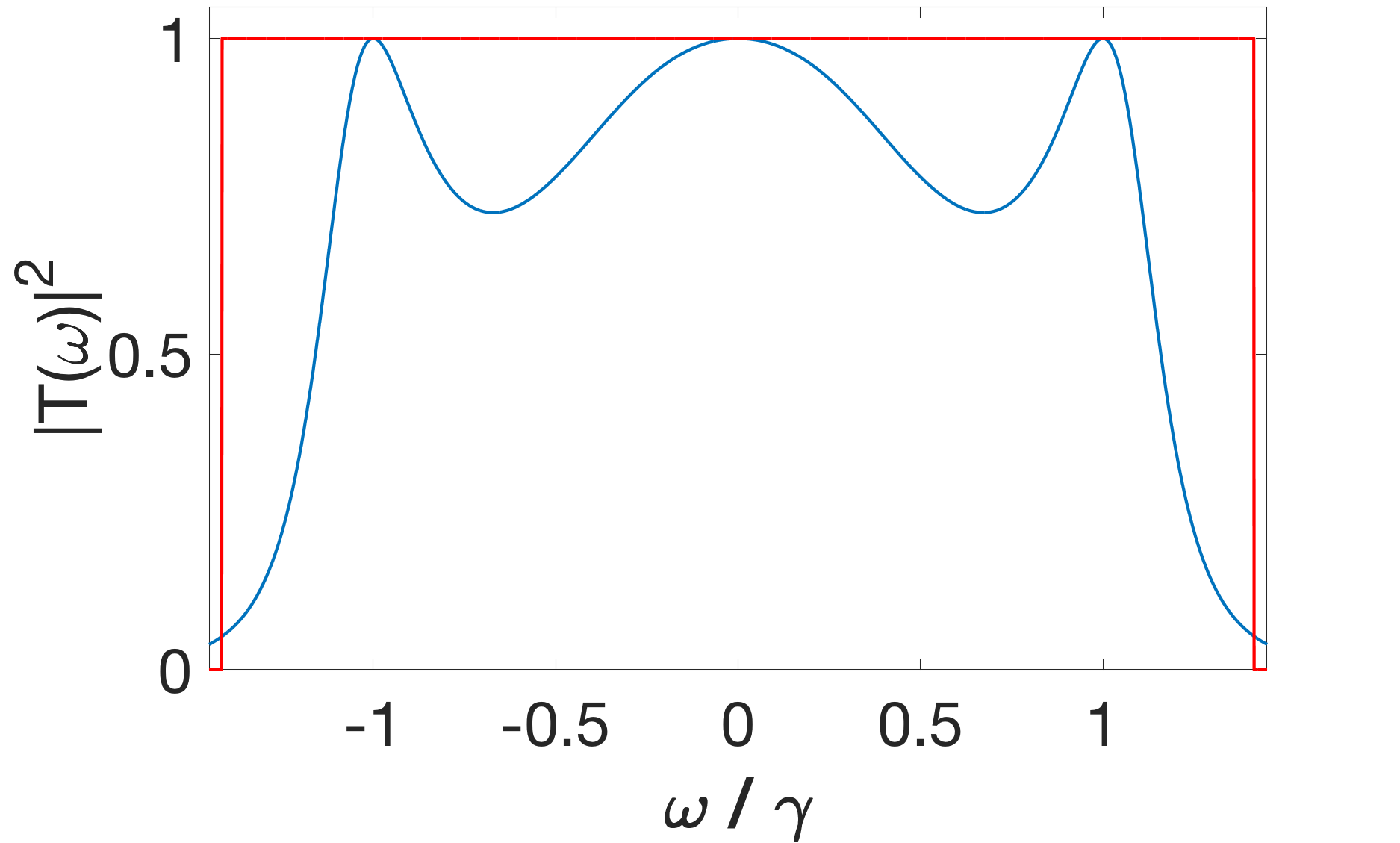}
            \caption[]%
            {{\small $N=4$ discrete states, unbalanced decays $2 \gamma=\Gamma$}}    
            \label{CriticalEvenUnbalanced}
        \end{subfigure}
         \quad\hfill
                         \begin{subfigure}[b]{.475\textwidth}  
            \centering 
            \includegraphics[width=\textwidth]{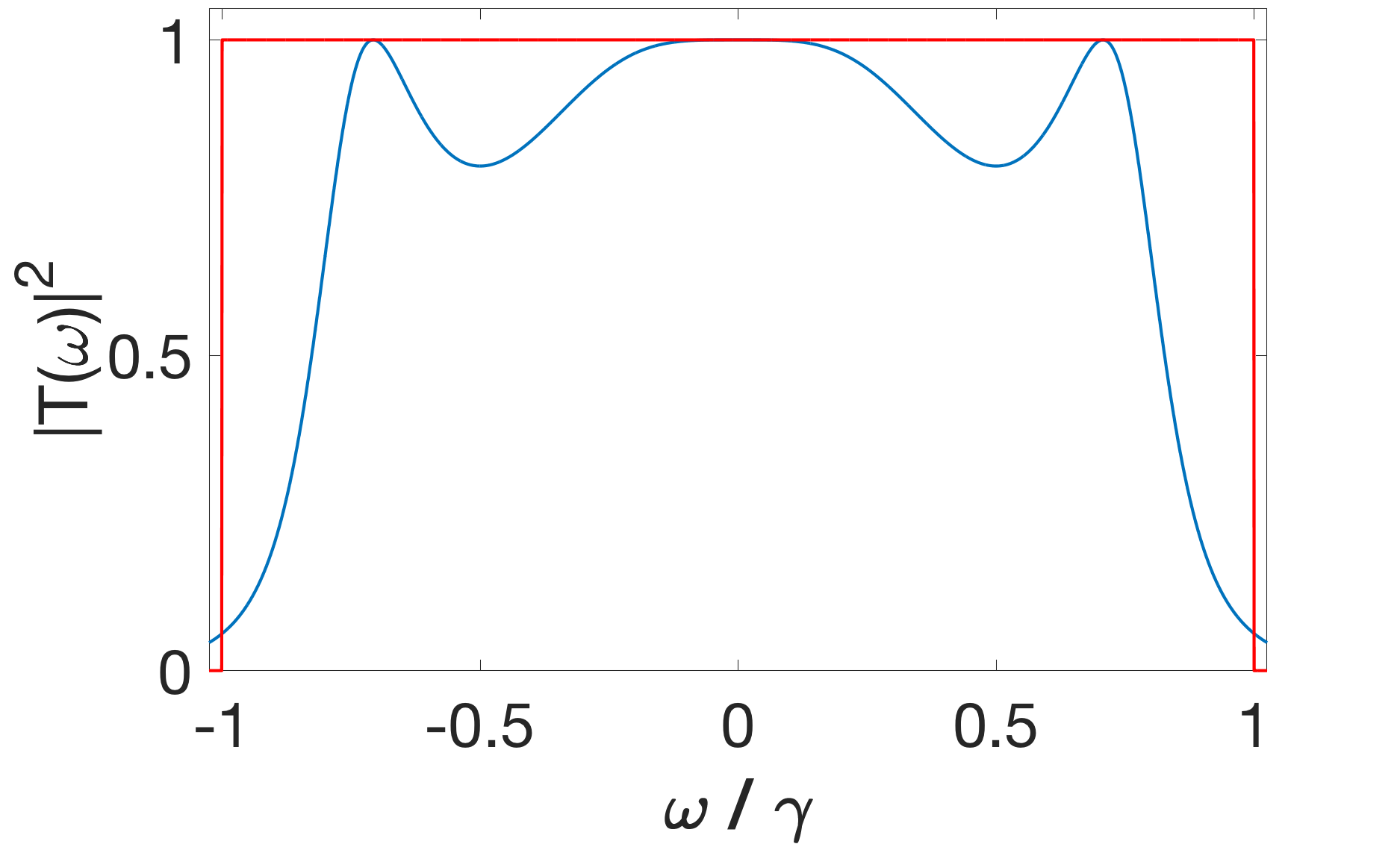}
            \caption[]%
            {{\small \small $N=4$ discrete states, balanced decays$\gamma=\Gamma$}}    
            \label{CriticalEvenBalanced}
                    \end{subfigure}
        \caption{\small Transmission probabilities for networks with $N$ discrete states in series with no relative detuning for the special case of homogenous critical coupling $g_{ij}\rightarrow g=\frac{\sqrt{\gamma\,\Gamma}}{2}$ (blue). For comparison, we plot a fictitious transmission probability (red) that is unity for $-2g\leq \omega \leq 2g$ and zero elsewhere. Frequencies are measured w.r.t. resonance. Transmission functions for networks with both balanced decays and unbalanced decays are plotted. Meeting both the balanced decay and critical coupling conditions ensures that the on-resonance transmission is both unity and maximally broadened, but are not necessary conditions for perfect transmission to occur at some frequency.} 
        \label{criticalpeaks}
    \end{figure*}

In general, we cannot diagonalize a completely arbitrary network in terms of a single set of parallel states; there may be a causal relationship embedded in the network structure; for example, for an atom in an s-state coupled via a photon to one or many p-states which subsequently decay to multiple d-states, we would have to diagonalize the p-states and d-states separately. A class of such systems are series quantum networks (Fig. \ref{seriesschem}), where only one state is coupled to each continuum, with the other states forming a chain between the two outer ones ($\gamma_{i>1} = 0$, $\Gamma_{i<N} = 0$, and $g_{ij} = 0$ for $j\neq i\pm1$). This provides a natural (and especially simple) model for energy transport and repeated spectral filtering (a series of Fabry-Perot cavities).  Furthermore, we can analytically determine the transmission function for arbitrary series networks, as we will now proceed to do.

\begin{figure*}[t]
        \begin{subfigure}[b]{0.475\textwidth}
            \centering
            \includegraphics[width=\textwidth]{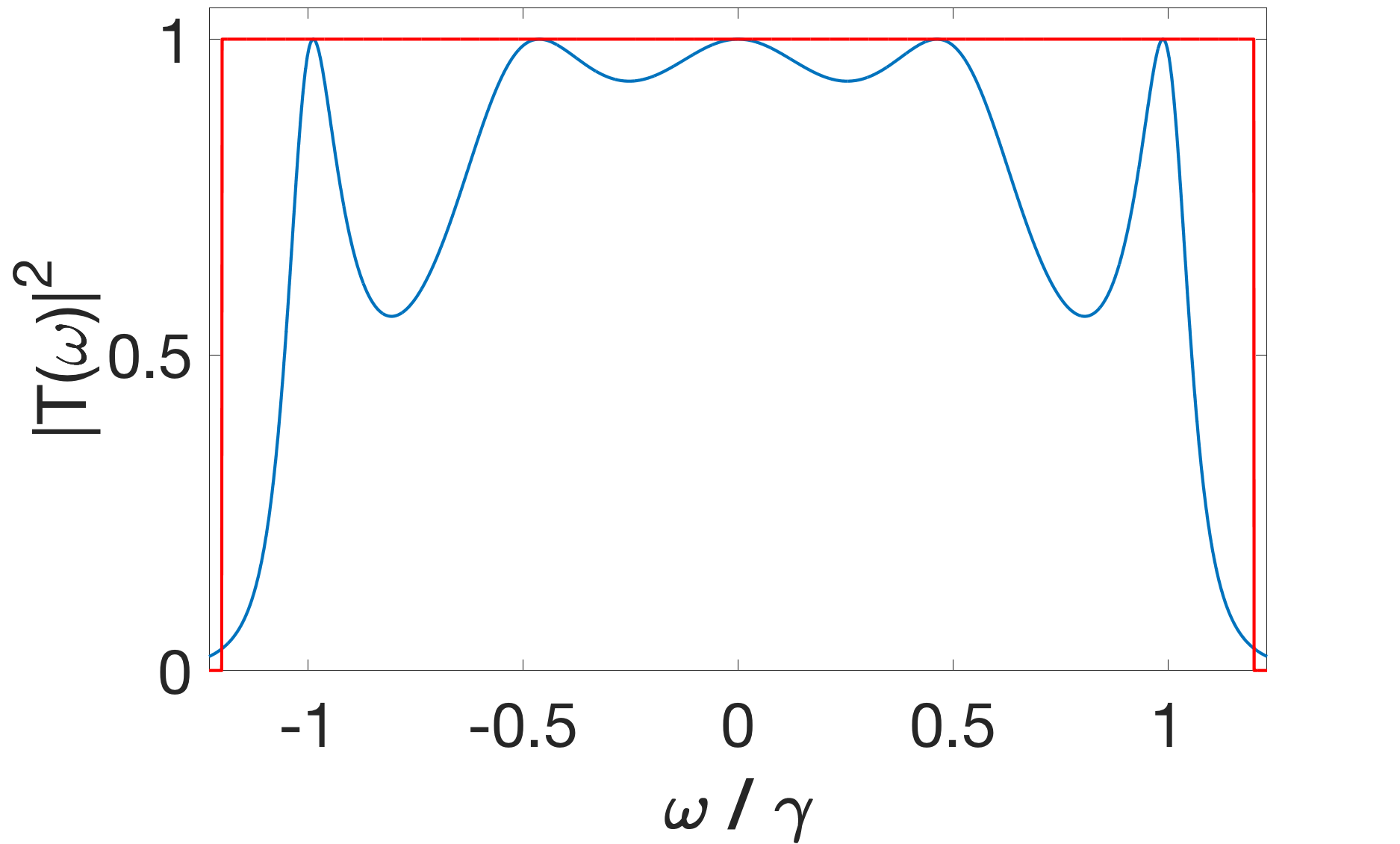}
            \caption[]%
            {{\small $N=5$ discrete states, over-coupled $g=\frac{6}{5}\frac{\sqrt{\gamma\Gamma}}{2}$}}    
            \label{OverOddBalanced}
        \end{subfigure}
        \hfill
                \begin{subfigure}[b]{0.475\textwidth}   
            \centering 
            \includegraphics[width=\textwidth]{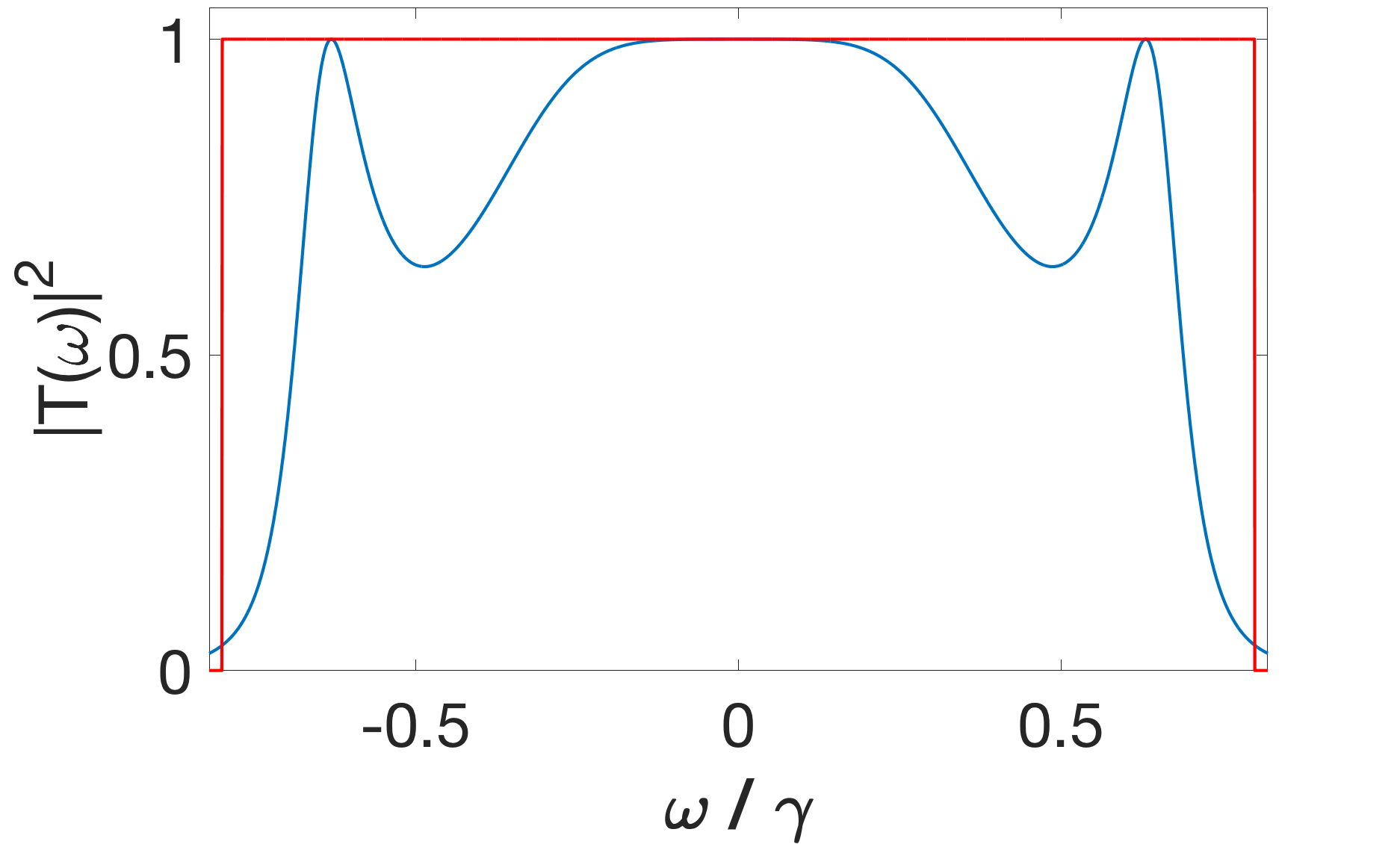}
            \caption[]%
            {{\small $N=5$ discrete states, under-coupled $g=\frac{4}{5}\frac{\sqrt{\gamma\Gamma}}{2}$}}    
            \label{UnderOddBalanced}
        \end{subfigure}
        \vskip\baselineskip
   \begin{subfigure}[b]{0.475\textwidth}  
            \centering 
            \includegraphics[width=\textwidth]{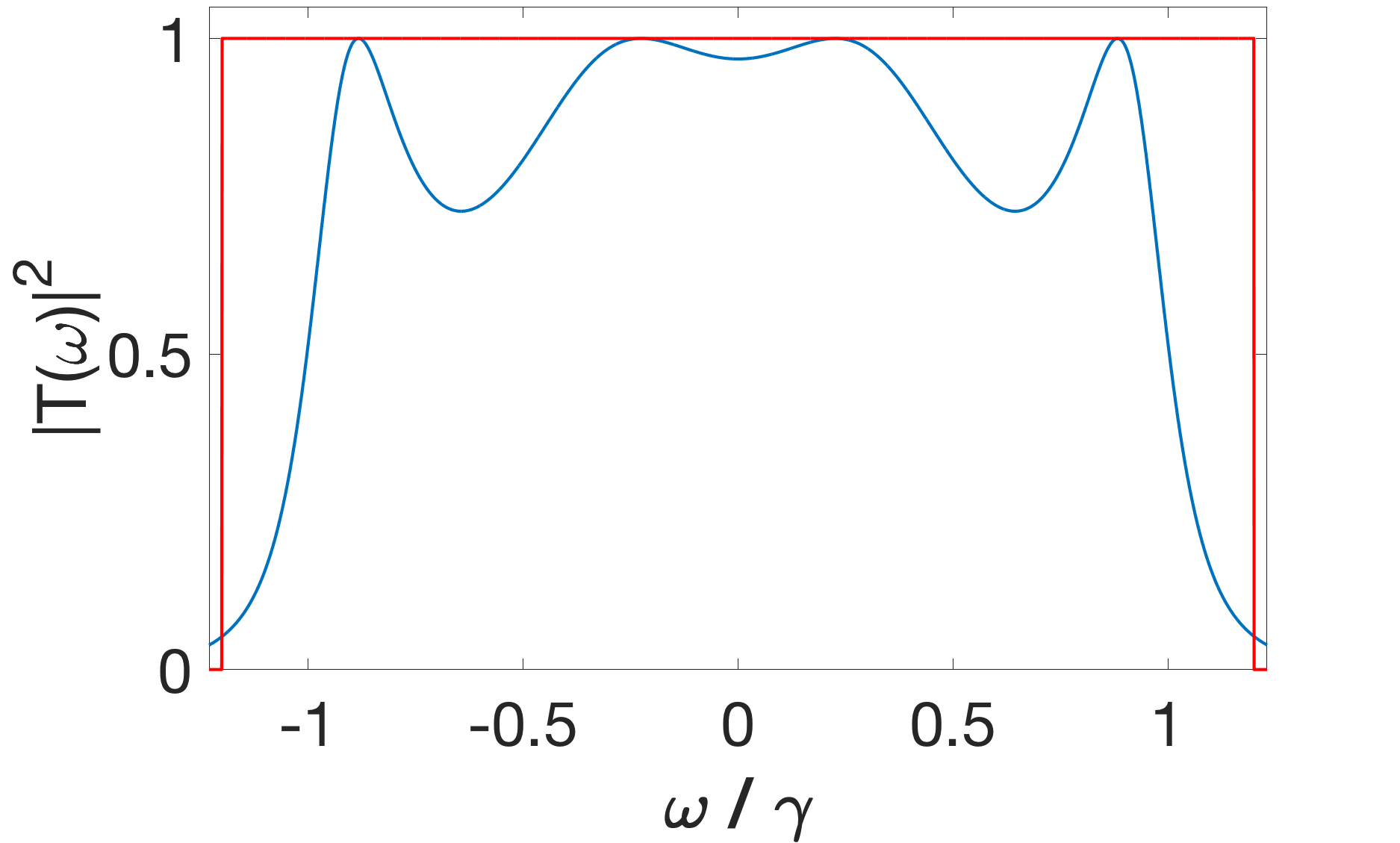}
            \caption[]%
            {{\small $N=4$ discrete states, over-coupled $g=\frac{6}{5}\frac{\sqrt{\gamma\Gamma}}{2}$}}    
            \label{OverEvenBalanced}
        \end{subfigure} 
        \quad\hfill
        \begin{subfigure}[b]{0.475\textwidth}   
            \centering 
            \includegraphics[width=\textwidth]{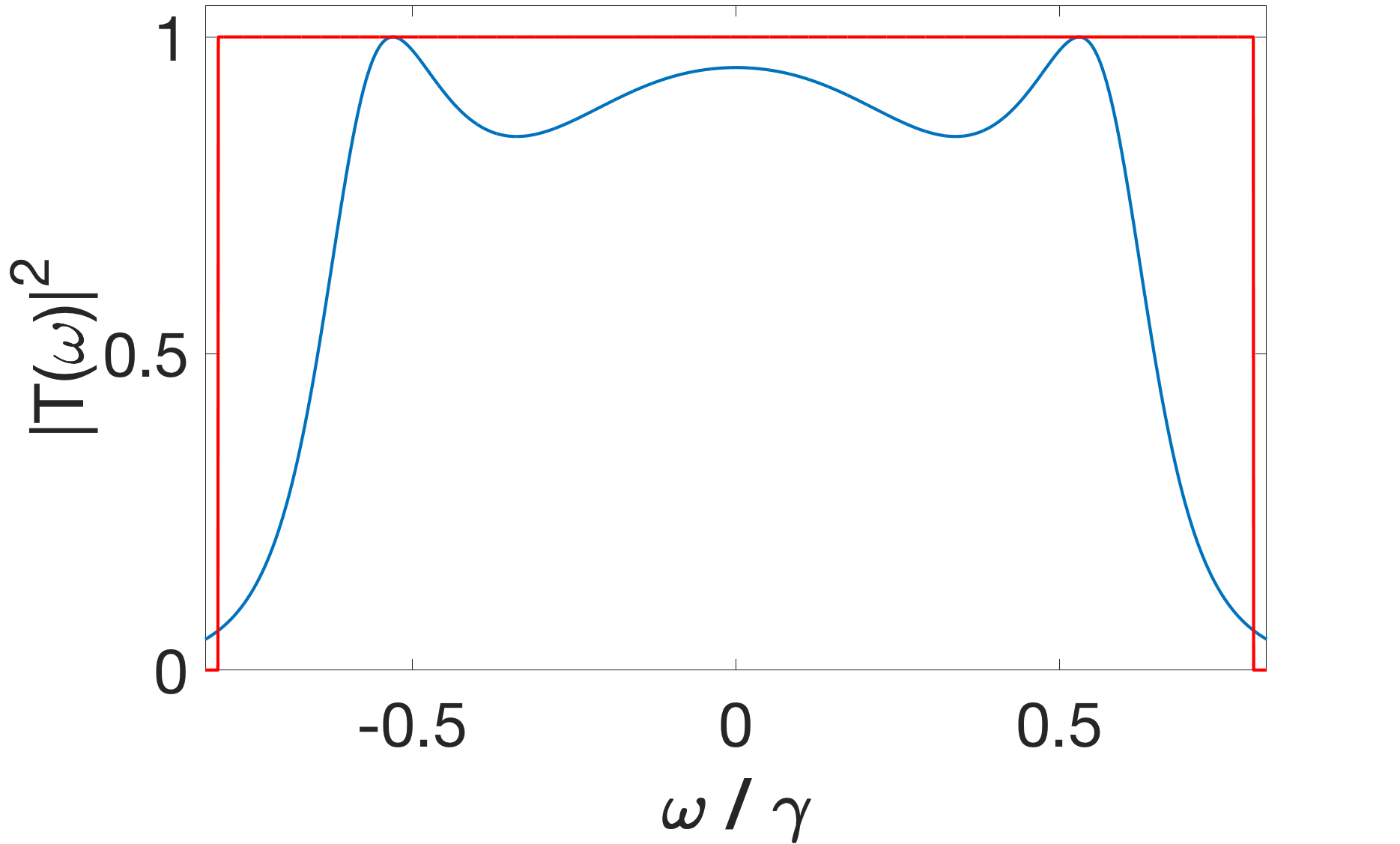}
            \caption[]%
            {{\small $N=4$ discrete states, under-coupled $g=\frac{4}{5}\frac{\sqrt{\gamma\Gamma}}{2}$}}    
            \label{UnderEvenBalanced}
        \end{subfigure}
                \caption{\small Transmission probabilities for networks with $N$ discrete states in series with no relative detuning for the special case of homogenous balanced decays $\gamma=\Gamma$ (blue). For comparison, we plot a fictitious transmission probability (red) that is unity for $-2g\leq \omega \leq 2g$ and zero elsewhere. Frequencies are measured w.r.t. resonance. Transmission functions for both over-coupled ($g>\frac{\sqrt{\gamma\Gamma}}{2}$) and under-coupled networks ($g<\frac{\sqrt{\gamma\Gamma}}{2}$) are plotted. A balanced over-coupled network will always have more peaks of perfect transmission than an under-coupled or critically-coupled one, though on-resonance transmission may not be a local maxima.}  
        \label{criticalpeaks}
    \end{figure*}

Since each state is only coupled to the two adjacent states, the system of $N$ equations of the form of (\ref{quantlangspect}) describing evolution of the system is solvable in a stepladder-type approach. Setting the expectation value of the second input continuum to zero, we solve for the $N$th state in terms of the $N-1$th. As we step up the ladder, we arrive at an expression for the reflection coefficient with the form of a generalized continued fraction

\bea\label{refseries}
R(\omega) &= 1-\cfrac{\gamma}{\frac{\gamma}{2}-i\Delta_1+\cfrac{g_{12}^2}{-i\Delta_2+\cfrac{g_{23}^2}{\dots+\cfrac{g_{N-1 N}^2}{\frac{\Gamma}{2}-i\Delta_N}}}}
\eea (we have dropped the subscripts on $\gamma_1$ and $\Gamma_N$). This equation looks difficult to analyze, but can be simplified using the Wallis-Euler recursion relations for continued fractions. We define two function $A_{n}$ and $B_{n}$ given by the following recurrence relations 

\bea\label{WallisEuler}
&B_{-1}=0, B_0=1, A_{-1}=1, A_0=b_0\nonumber \\ 
&A_n=b_n\,A_{n-1}+a_n\,A_{n-2}\,(n\geq1)\nonumber \\
&B_n=b_n\,B_{n-1}+a_n\,B_{n-2}\,(n\geq1)
\eea with coefficients $a_n$ and $b_n$ given in Fig. \ref{table}. The reflection coefficient is then given $R(\omega)=\frac{A_N}{B_N}$ for $N$ discrete states. (For $n<N$, the functions $A_n$ and $B_n$ have no clear physical meaning.) From this, we easily solve for conditions where $R(\omega)=0$ (perfect transmission).

\begin{figure}[t] 
	\includegraphics[width=\linewidth]{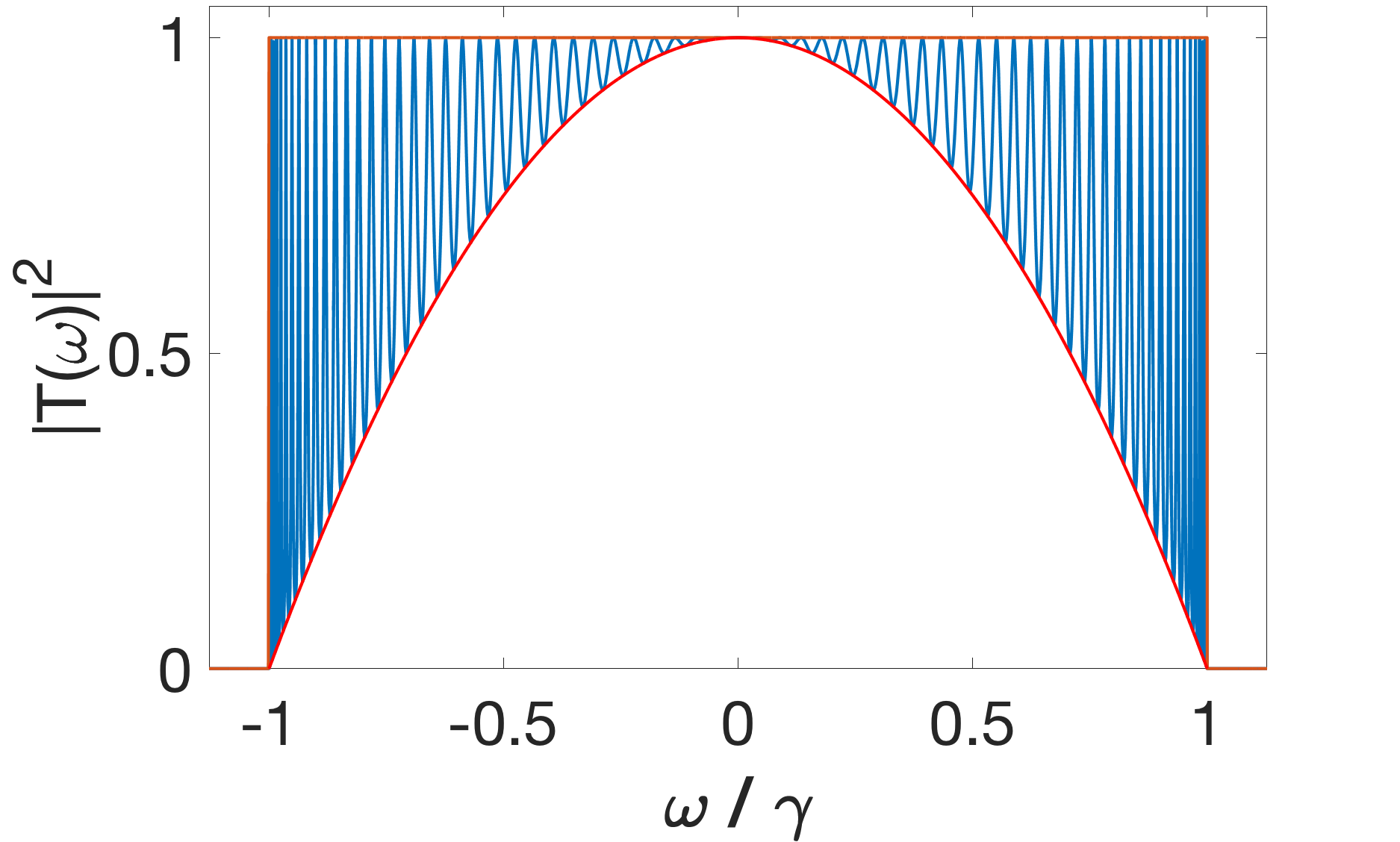} 
	\caption{Transmission probability (blue) for a series network with $N=70$ identical discrete states ($\omega_i=\omega_j$) with both balanced decays ($\gamma=\Gamma$) and uniform critical coupling ($g=\frac{\sqrt{\gamma\Gamma}}{2}$) conditions met. This results in a maximally-broadened on-resonance transmission. In the large-$N$ limit, the amplitude of this ideal (perfect and broadened) transmission function $|T(\omega)|$ is bounded below by a circle (parabola for $|T(\omega)|^2$) and above by a square (both red) with widths $2g$ and maxima of unity.}
	\label{CircleAndSquare}
\end{figure}

\lettersection{Perfect transmission} We begin by considering the unphysical but illuminating case of an infinite series of identical discrete states ($g_{i\,i+1}= g \,\forall  i$ and $\Delta_i = \Delta \,\forall i$) analytically. We find that the limit $\lim_{N\rightarrow\infty} R(\omega)$ only converges on-resonance for the special ``critical'' case \be
g=\frac{\sqrt{\gamma\,\Gamma}}{2}.
\ee 
Furthermore, the limit only converges to zero on resonance (perfect transmission) when we also have $\gamma=\Gamma$. This first condition corresponds to a series of discrete states coupled through their decays (e.g. Fabry-Perot cavities coupled through their evanescent fields), and the second condition corresponds to the same requirement of balanced decays we saw for parallel quantum networks.


While infinite series networks of discrete states are not realistic, these two conditions play different but important roles in all finite series networks where the discrete states are identical (though introducing relative detuning will change the critical values of $g$ and $\Gamma/\gamma$, as we will see). That there are two conditions can be explained thus; since $R(\omega)$ is in general complex, the condition $R(\omega)=0$ gives two constraint equations on the real and imaginary parts of $R(\omega)$. When $N$ is even [odd], the real part is an order $N$ [$N-1$] polynomial in the detunings $\Delta_i$, while the imaginary part is order $N-1$ [$N$]. In general, this means there are conditions for a minimum of $N-1$ frequencies of perfect transmission and we may find $N$ frequencies of perfect transmission only if the lower-order equation is satisfied trivially for all frequencies. When considering finite series networks of identical discrete states, these same two conditions appear in the constraint equations for perfect transmission \footnote{For $N=1$, the lower order equation is just the requirement that $\gamma=\Gamma$ and the higher order equation just requires $\Delta_1=0$ (on-resonance). For $N=2$, the lower order equation is $\gamma\Delta_2=\Gamma\Delta_1$ and the higher order equation is $g^2=\Delta_1 \Delta_2+\frac{\gamma\Gamma}{4}$. The lower order equation is only frequency independent when $\gamma=\Gamma$ and $\Delta_2=\Delta_1$. When the discrete states are degenerate, we find that $g\geq \frac{\sqrt{\gamma\,\Gamma}}{2}$ is the requirement for perfect transmission at one ($g= \frac{\sqrt{\gamma\,\Gamma}}{2}$) or two ($g> \frac{\sqrt{\gamma\,\Gamma}}{2}$) frequencies.}.

We now explore in detail the effects of these two conditions on series networks of identical discrete states with uniform coupling; first, consider fixing the coupling $g$ to be critical ($g=\frac{\sqrt{\gamma\Gamma}}{2}$). For odd $N$, we find that when the decays are balanced ($\gamma=\Gamma$), this ensures $N$ frequencies of perfect transmission  (Fig. \ref{CriticalOddBalanced}) with the resonance frequency at a local maxima of unity, and when the decays are unbalanced ($\gamma\neq\Gamma$), $N-1$ frequencies of perfect transmission (Fig. \ref{CriticalOddUnbalanced}) with the resonant frequency at a local minima. For even $N$, letting $g=\frac{\sqrt{\gamma\Gamma}}{2}$ always results in $N-1$ frequencies of perfect transmission (Fig. \ref{CriticalEvenUnbalanced}) with the on-resonant frequency at a local maxima. Here, having balanced decays broadens the the on-resonance maxima (Fig. \ref{CriticalEvenBalanced}), which will be desirable for detection of non-monochromatic photons (wave packets). 

Now we instead consider fixing the decays to be balanced ($\gamma=\Gamma$) and observe a switch in the on-resonance behavior; whereas above we had found for even $N$ the critical coupling condition was sufficient for on-resonance transmission to be at a local maxima, we now find that, for odd $N$ that the balanced decay condition results in $N$ peaks of unity transmission (Fig. \ref{OverOddBalanced}) with on-resonance transmission is at a local maxima. Here three peaks become degenerate to give $N-2$ frequencies of perfect transmission when $g\lesssim \frac{\sqrt{\gamma\Gamma}}{2}$ (Fig. \ref{UnderOddBalanced}). (This inequality rapidly becomes exact [$g< \frac{\sqrt{\gamma\Gamma}}{2}$] with increasing $N$.) Similarly, we find that for even $N$, the behavior of $T(\omega)$ depends strongly on the coupling, flipping between $N$ frequencies of perfect transmission with on-resonance transmission at a local minima for $g> \frac{\sqrt{\gamma\Gamma}}{2}$ (Fig. \ref{OverEvenBalanced}) and $N-2$ frequencies of perfect transmission with on-resonance transmission at a non-unity local maxima for $g< \frac{\sqrt{\gamma\Gamma}}{2}$ (Fig. \ref{UnderEvenBalanced}).

For both even and odd $N$, the width is increasingly determined by $g$ with increasing $N$, with the half-width asymptotically approaching $2g$. In the large-$N$ limit, we observe that the transmission function is asymptotically bounded between a circle and a square when both conditions for perfect transmission are met (Fig. \ref{CircleAndSquare}). We also observe that increasing $g$ past $\frac{\sqrt{\gamma\,\Gamma}}{2}$ while maintaining $\gamma=\Gamma$ induces Rabi splitting, with the $N$ or $N-1$ frequencies of perfect transmission spreading outwards (for odd and even $N$, respectively). For the special case of $N=2$, the presence of a second coupled discrete state resulting in a splitting and shift of the resonant frequency is in agreement with the effect seen in Ref.~\cite{Rosfjord2006} for an antireflective coating; perfect transmission is still possible provided $\gamma=\Gamma$ but the frequency that is perfectly transmitted is split in to two. 

\begin{figure}[b] 
	\includegraphics[width=\linewidth]{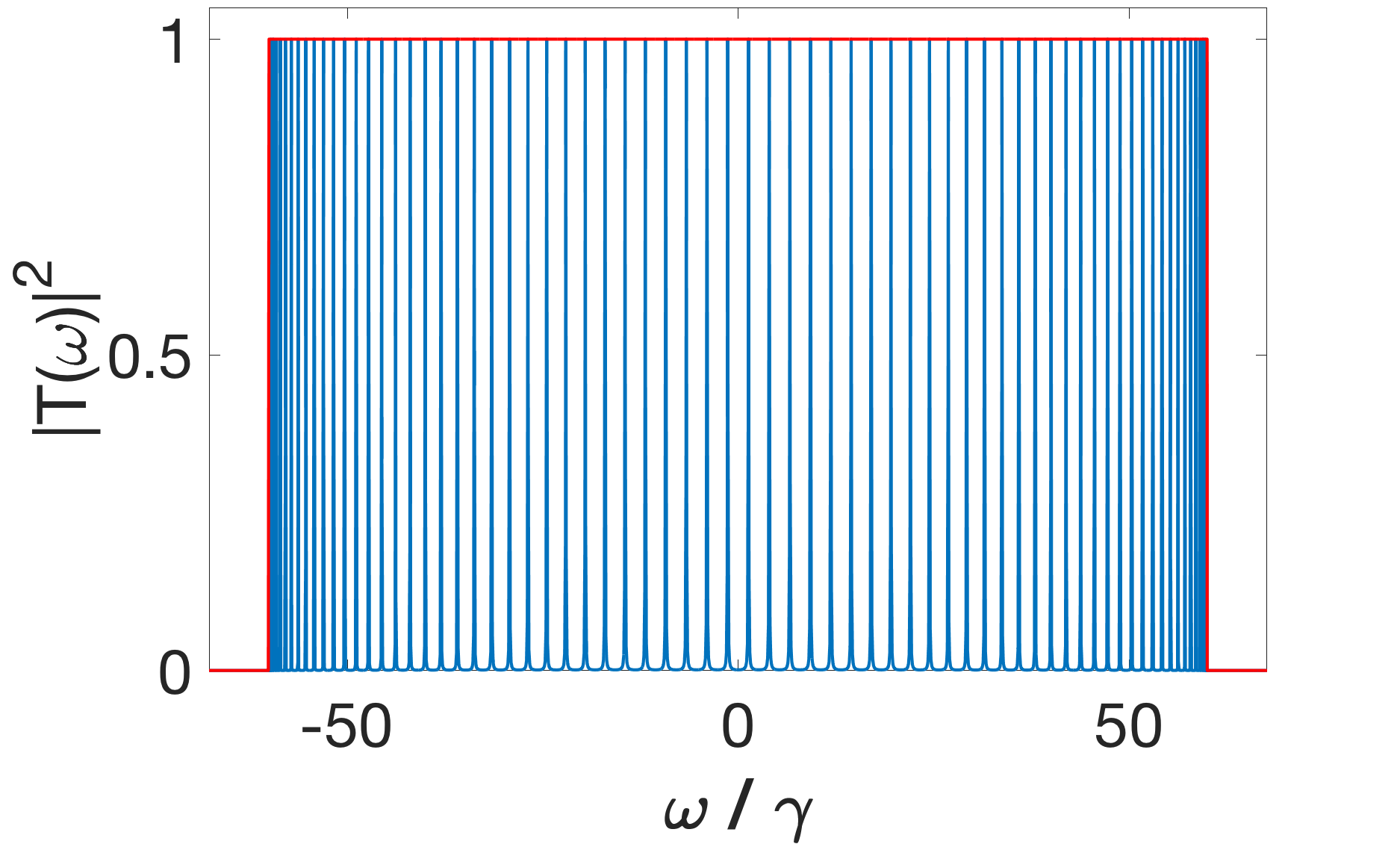} 
	\caption{Transmission probability for a series network with $N=70$ identical discrete states with balanced decays ($\gamma=\Gamma$) yielding perfect transmission in the strong-and-uniform coupling limit ($g\gg\frac{\sqrt{\gamma\Gamma}}{2}$). The result is a frequency-comb structure with $70$ frequencies of unity transmission between $-2g$ and $2g$, separated by $69$ regions of near-perfect reflection.} 
	\label{FrqeuencyComb}
\end{figure}

 In the high-$g$ limit, we start seeing a frequency-comb structure emerge (Fig. \ref{FrqeuencyComb}), with dips that approach perfect reflection. This is because in the strong-coupling limit, the causal ordering of the discrete states asymptotically disappears: the system becomes approximately diagonalizable as parallel modes with purely virtual coupling and the $N-1$ frequencies of perfect reflection from (\ref{reflectk}) manifest.

\begin{figure}[b] 
	\includegraphics[width=1\linewidth]{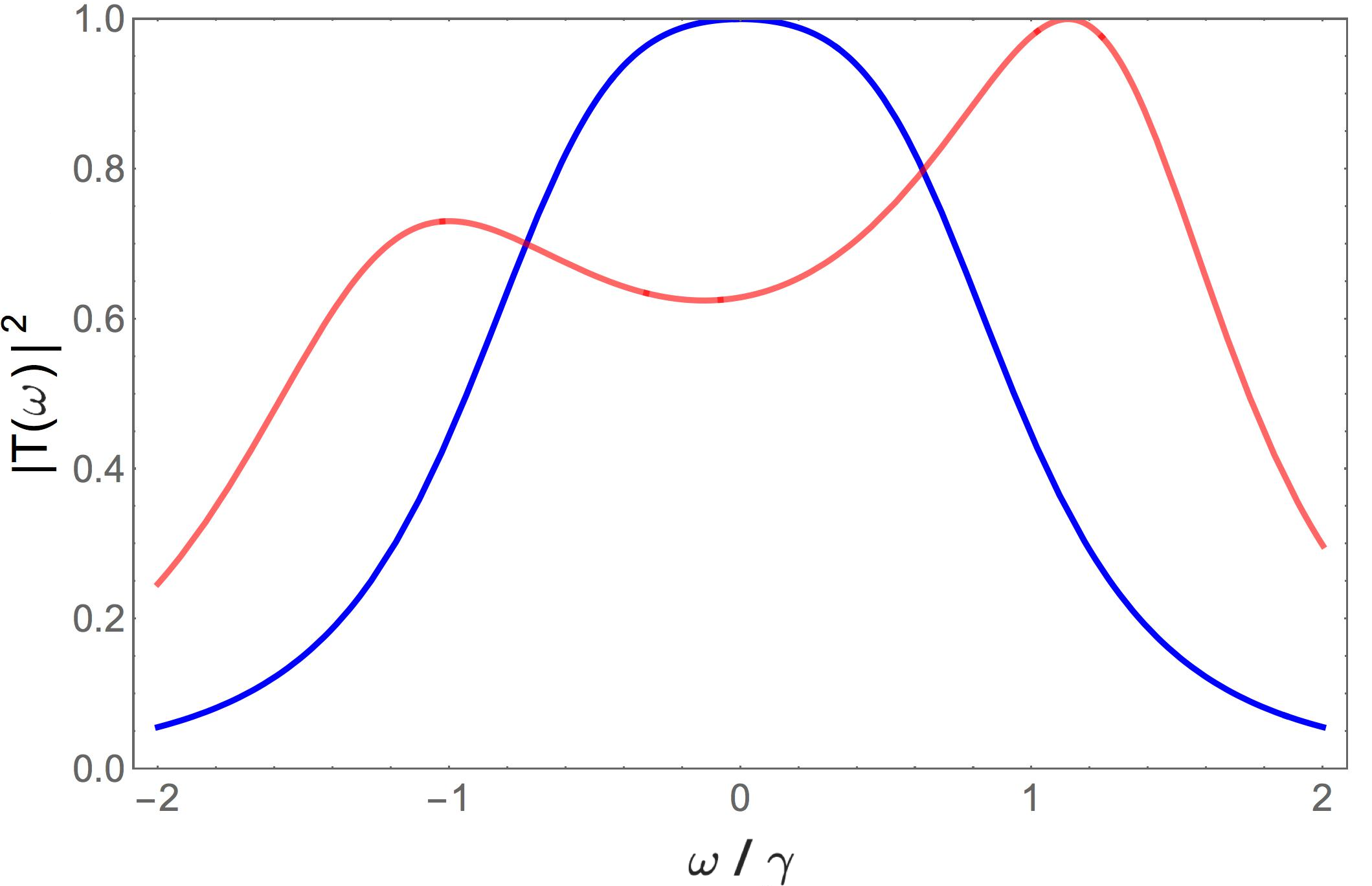} 
	\caption{Transmission probabilities for series networks with $N=2$ discrete states with no relative detuning (blue), and with non-zero relative detuning $\frac{\omega_2-\omega_1}{\gamma}=\frac{3}{4}$ (red). In both cases, the decays are not balanced ($\frac{\Gamma}{\gamma}=2$) but the couplings are chosen such that perfect transmission at a frequency is achieved.}
	\label{detunedstates}
\end{figure}

We now consider the effects of introducing relative detunings between discrete states as, generally, a quantum network will not be comprised of identical states. Still there are sufficient degrees of freedom in (\ref{refseries}) such that, by tuning the parameters, perfect transmission at \emph{some} frequencies is always possible. This is even true when $\gamma\neq\Gamma$ (for $N>1$): considering only two discrete states in series, we find that perfect transmission occurs at a frequency $\omega=\frac{\omega_1+\omega_2}{2} + \frac{\Gamma\omega_1-\gamma\omega_2}{\Gamma-\gamma}$ when $g_{12}=\sqrt{\frac{\gamma\Gamma}{4} + \left(\frac{\Gamma\omega_1-\gamma\omega_2}{\Gamma-\gamma}\right)^2 - \left(\frac{\omega_1-\omega_2}{2}\right)^2}$ (Fig. \ref{detunedstates}). For two detuned discrete states in series with balanced decays ($\gamma=\Gamma$), perfect transmission is impossible as the critical value of $g$ is infinite; the transmission efficiency asymptotically becomes perfect in the strong-coupling limit. However, this a special case and is not true for larger numbers ($N>2$) discrete states. Critically, given a series network of $N$ discrete states with arbitrary relative detunings, we can \emph{always} find at least one set of parameters (couplings and decay rates) such that $N-1$ frequencies are perfectly transmitted.

\begin{figure}[t] 
	\includegraphics[width=\linewidth]{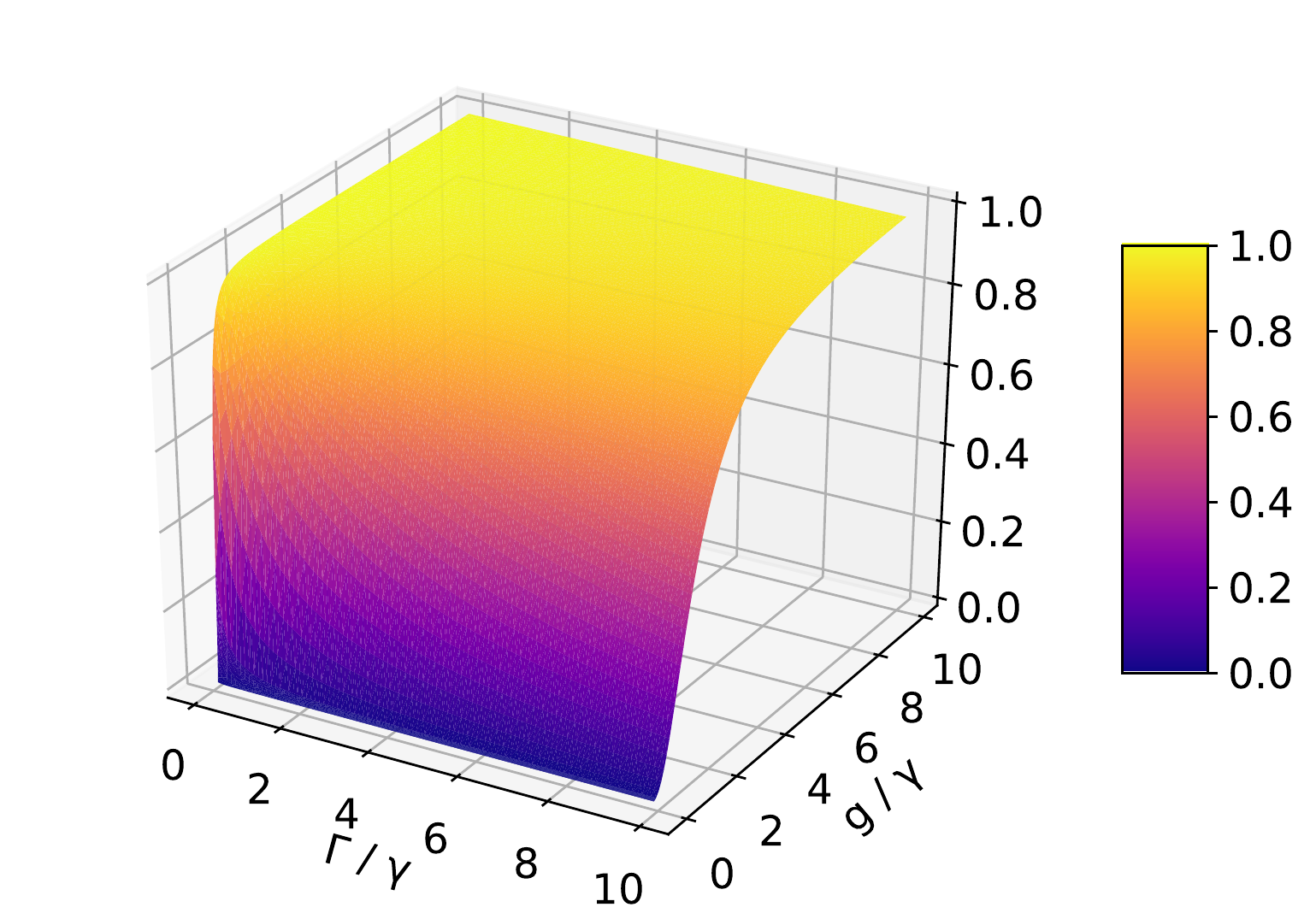} 
	\caption{Normalized spectral bandwidth $\tilde{\Gamma}/\tilde{\Gamma}_{\rm max}$ for a series network with $N=2$ discrete states and no relative detuning. Here the maximum bandwidth, given by that of a single discrete state $\tilde{\Gamma}_{\rm max}=\frac{2\gamma\Gamma}{\gamma+\Gamma}$, is reach in the strong coupling limit. The balanced decay condition does not effect bandwidth as an increase in decay rates just scales the strong coupling limit regime ($g\gg\frac{\sqrt{\gamma\Gamma}}{2}$).}
	\label{tildeGammaN1}
\end{figure}

\begin{figure}[t] 
	\includegraphics[width=\linewidth]{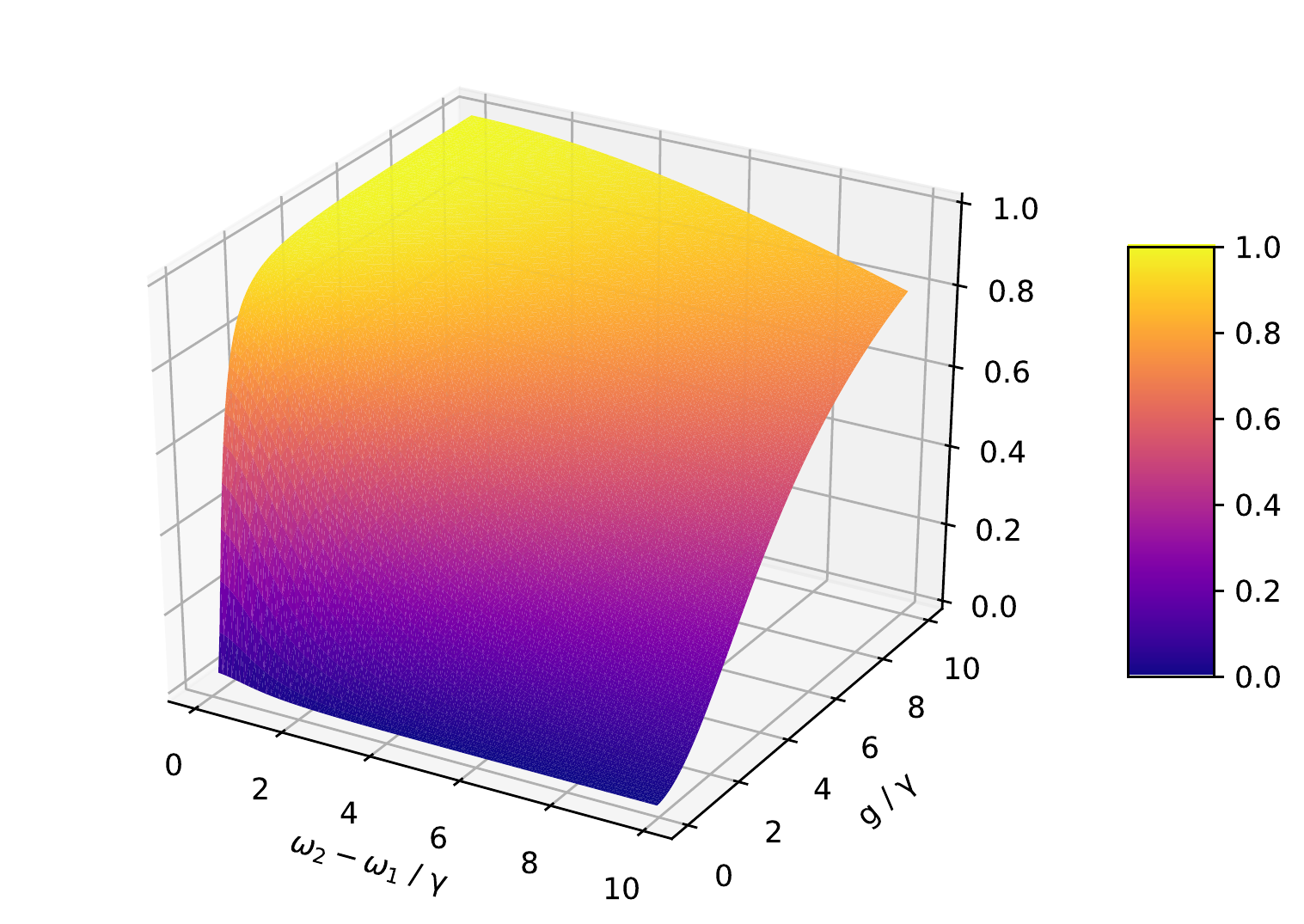} 
	\caption{Normalized spectral bandwidth $\tilde{\Gamma}/\tilde{\Gamma}_{\rm max}$ for a series network with $N=2$ discrete states with relative detuning $\omega_2-\omega_1$. We still observe $\tilde{\Gamma}/\tilde{\Gamma}_{\rm max} = 1$, but at a higher coupling strength for greater detunings. }
	\label{N1RelativeBandwidthDetuning}
\end{figure}

 \lettersection{Spectral Bandwidth} Once we have the form of the transmission function, we can calculate the spectral bandwidth for these systems. The spectral bandwidth decreases with additional discrete states and is strictly bounded above by the single discrete state bandwidth $\tilde{\Gamma}\leq\frac{2\gamma\Gamma}{\gamma+\Gamma}$. For discrete states without detuning, equality is reached in the strong-coupling limit of $g\gg\frac{\sqrt{\gamma\Gamma}}{2}$ but independently of whether $\gamma=\Gamma$ (Fig. \ref{tildeGammaN1}). Introducing detuning between discrete states lowers the bandwidth, but equality with the upper limit still occurs for sufficiently strong coupling (Fig. \ref{N1RelativeBandwidthDetuning}); the strong coupling is able to better mask the discrepancy between discrete state frequencies as the dressed states become the more physical description and as the system more strongly resembles a parallel network. (In this limit, each dressed state is coupled to the continua at reduced decays so that the total bandwidth is still bounded by $\tilde{\Gamma}\leq\frac{2\gamma\Gamma}{\gamma+\Gamma}$ with $\gamma$ and $\Gamma$ the two original decay rates for the series network.)

\begin{figure}[ht]
        \begin{subfigure}[b]{0.475\textwidth}
            \centering
            \includegraphics[width=.9\textwidth]{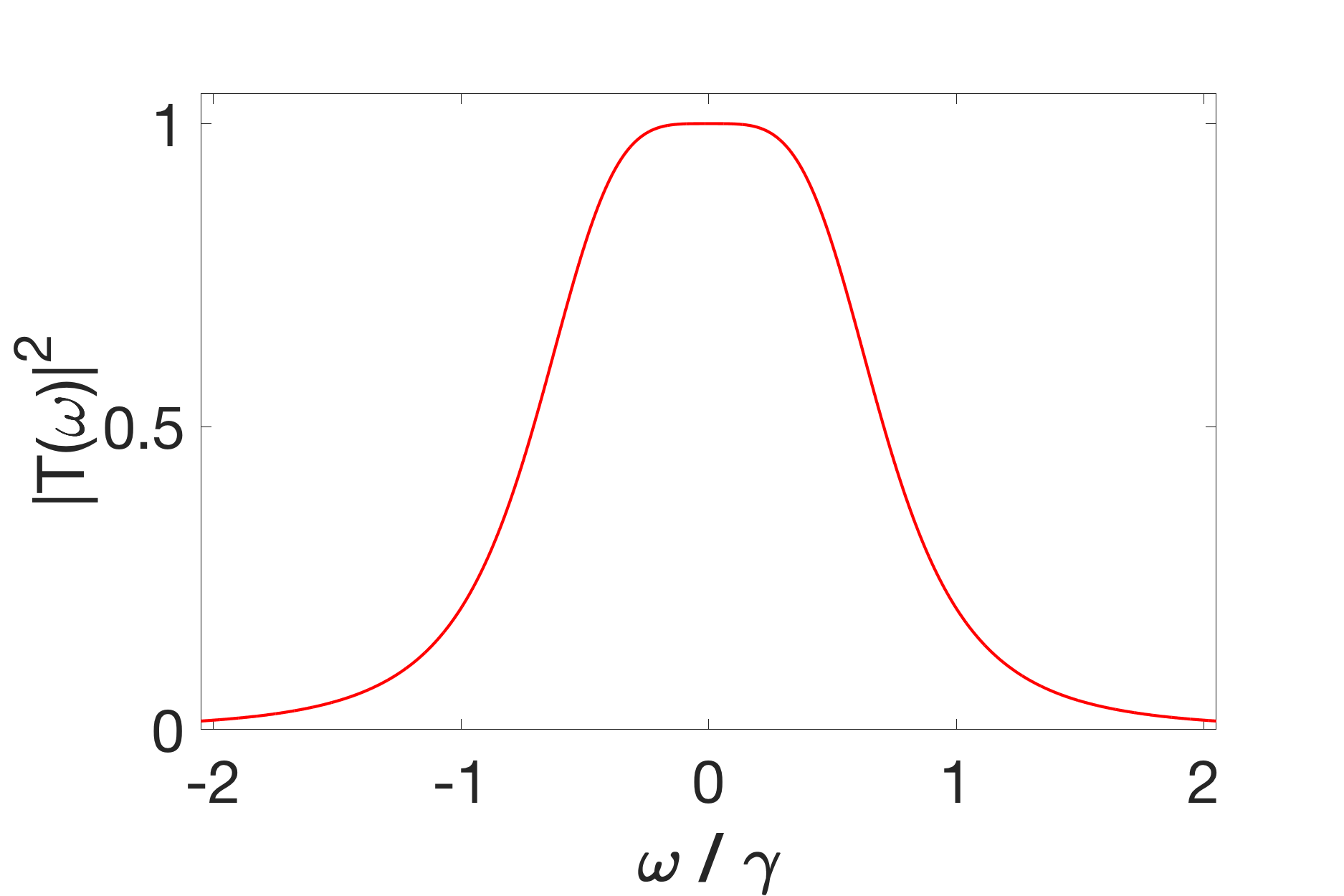}
            \caption[]%
    {{\small Transmission probability, critically coupled $g=\frac{\sqrt{\gamma\Gamma}}{2}$}}    
            \label{2discretestatesSeriesTrans}
        \end{subfigure}
        \hfill
        \begin{subfigure}[b]{0.475\textwidth}  
            \centering 
            \includegraphics[width=.9\textwidth]{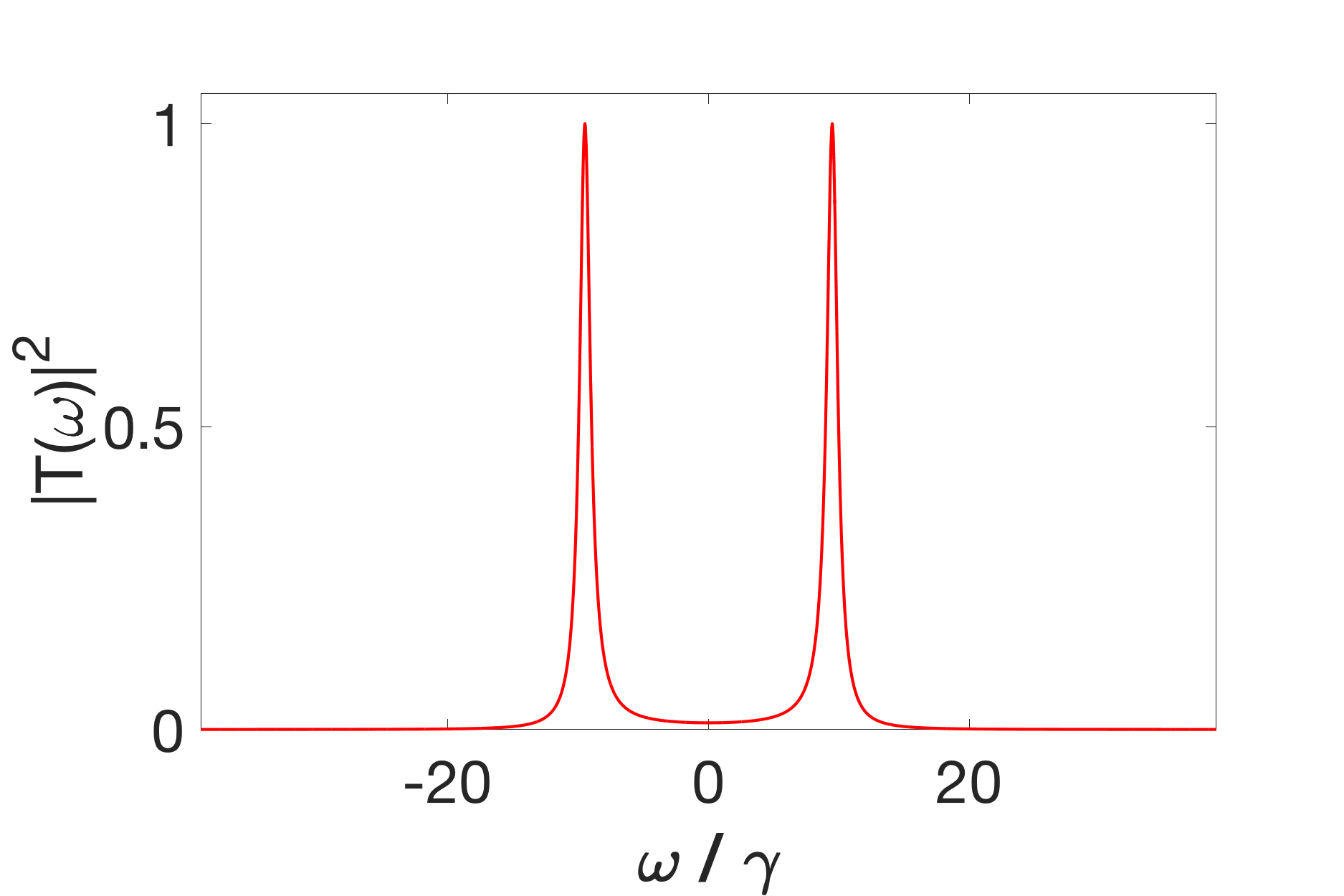}
            \caption[]%
            {{\small Transmission probability, over-coupled $g=18 \frac{\sqrt{\gamma\Gamma}}{2}$}}    
            \label{2discretestatesSeriesTransOver}
        \end{subfigure}
        \begin{subfigure}[b]{0.475\textwidth}   
            \centering 
            \includegraphics[width=.9\textwidth]{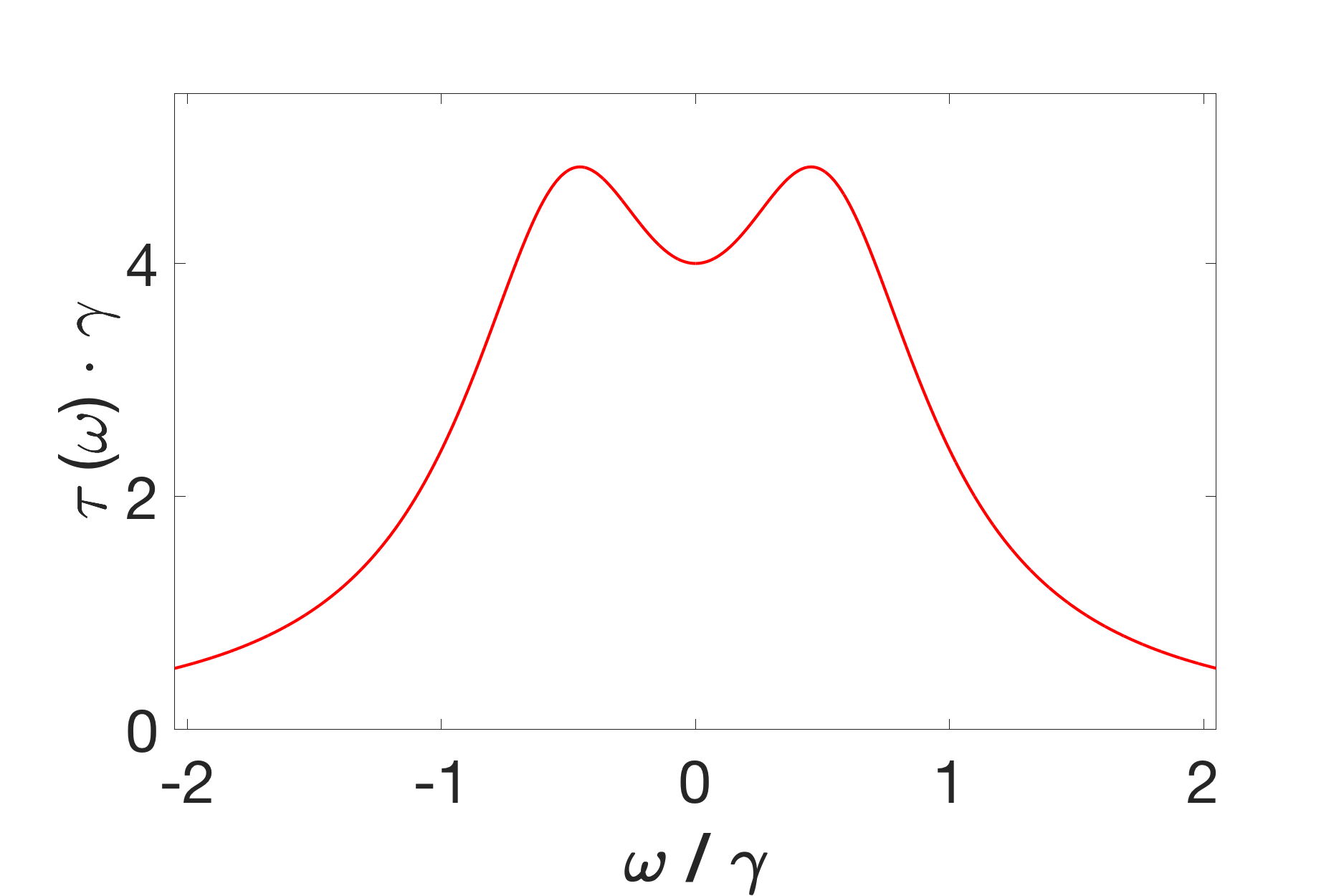}
            \caption[]%
            {{\small Group delay, critically coupled $g=\frac{\sqrt{\gamma\Gamma}}{2}$}}    
            \label{2discretestatesSeriesTau}
        \end{subfigure}
        \quad\hfill
        \begin{subfigure}[b]{0.475\textwidth}   
            \centering 
            \includegraphics[width=.9\textwidth]{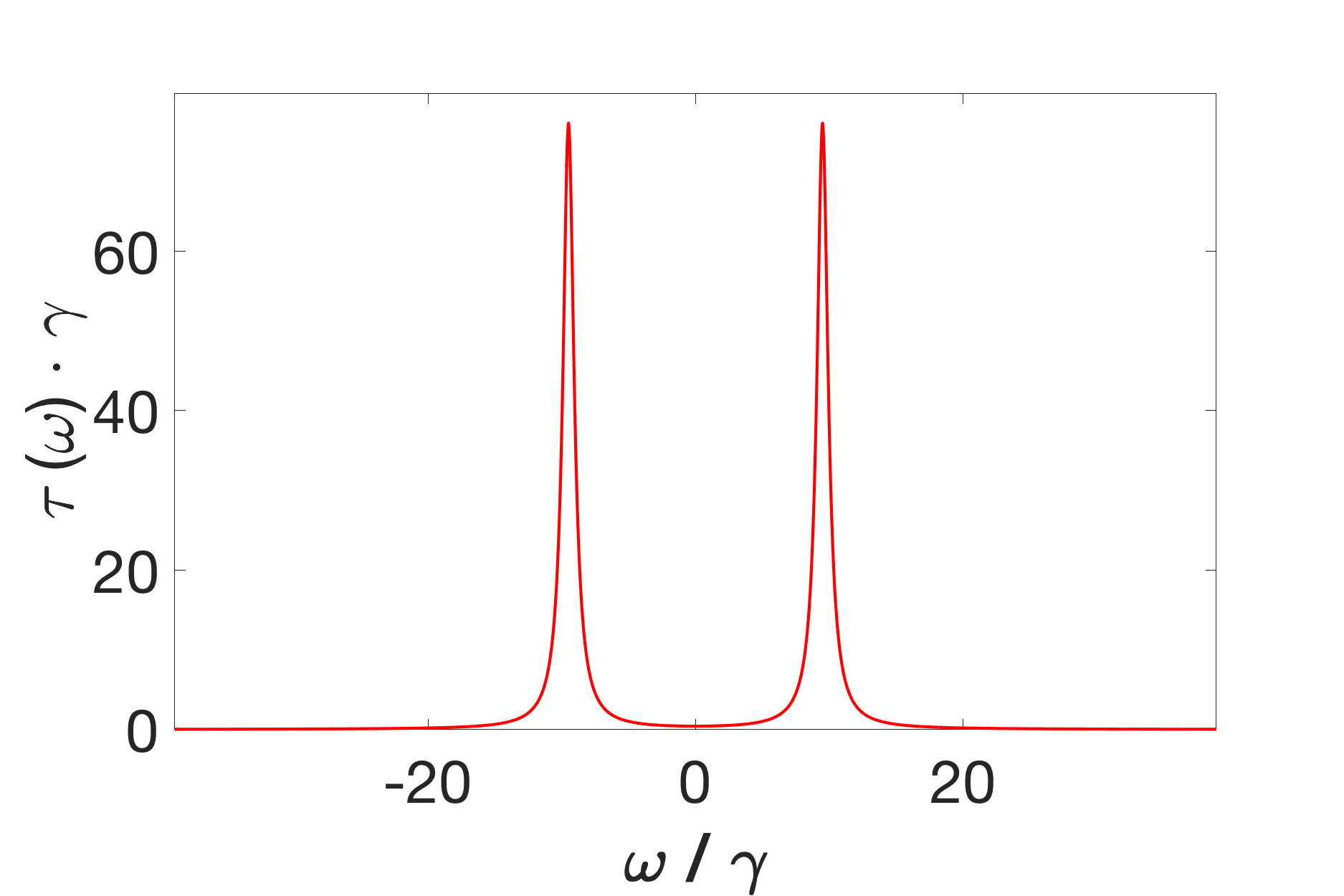}
            \caption[]%
            {{\small Group delay, over-coupled $g=18 \frac{\sqrt{\gamma\Gamma}}{2}$}}    
            \label{2discretestatesSeriesTauOver}
        \end{subfigure}
        \caption{\small Transmission probabilities and group delays for both over-coupled and critically coupled series networks of $N=2$ discrete states with balanced decay rates ($\gamma=\Gamma$) and no relative detuning. A small change in the sharpness of the transmission function's peaks can have a large effect on the magnitude of the group delay.} 
        \label{2discfig}
    \end{figure}

\begin{figure*}[t]
        \begin{subfigure}[b]{0.475\textwidth}   
            \centering 
            \includegraphics[width=.9\textwidth]{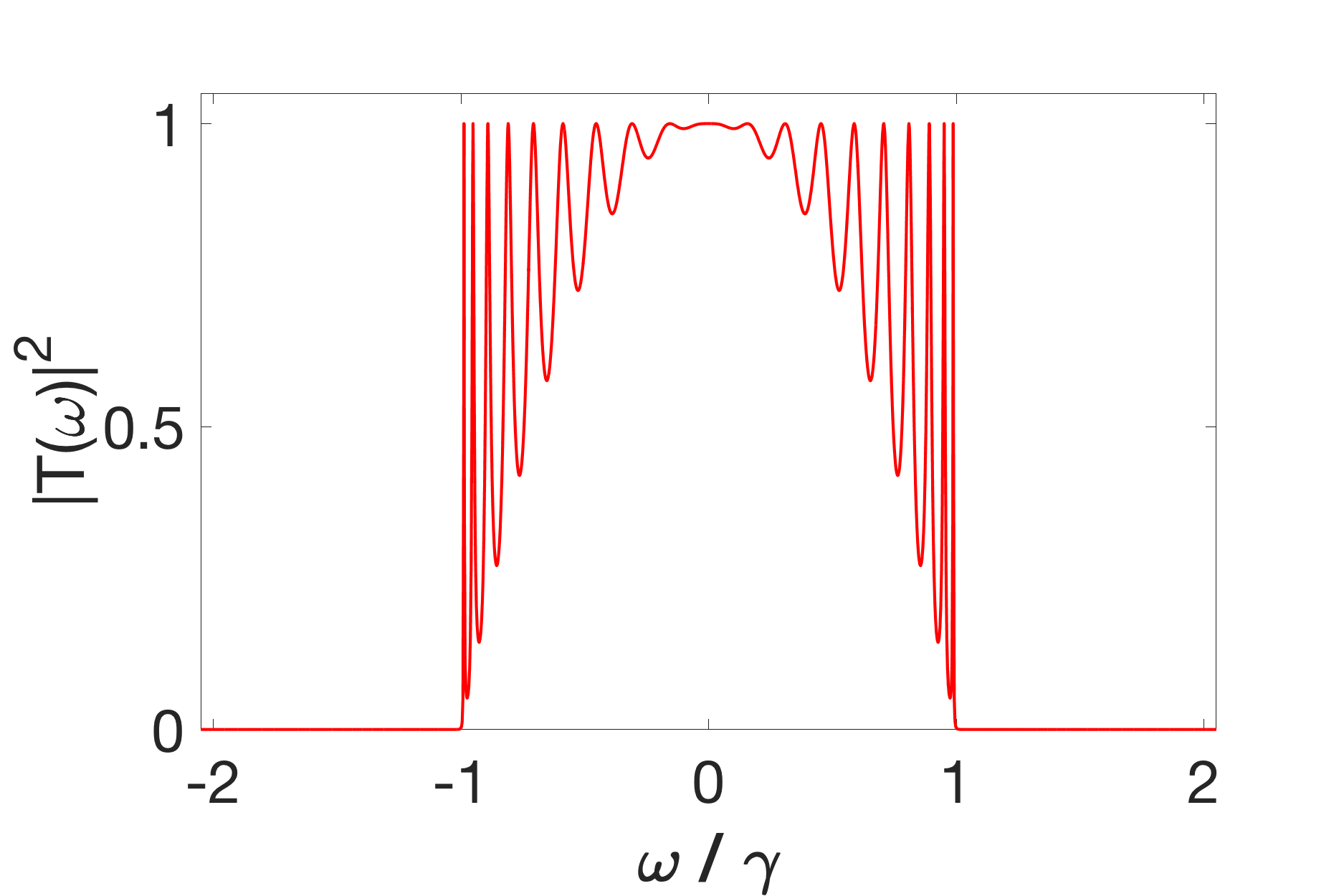}
            \caption[]%
    {{\small Transmission probability, no relative detuning}}    
            \label{20discretestatesSeriesTrans}
        \end{subfigure}  \hfill
        \begin{subfigure}[b]{0.475\textwidth}   
            \centering 
            \includegraphics[width=.9\textwidth]{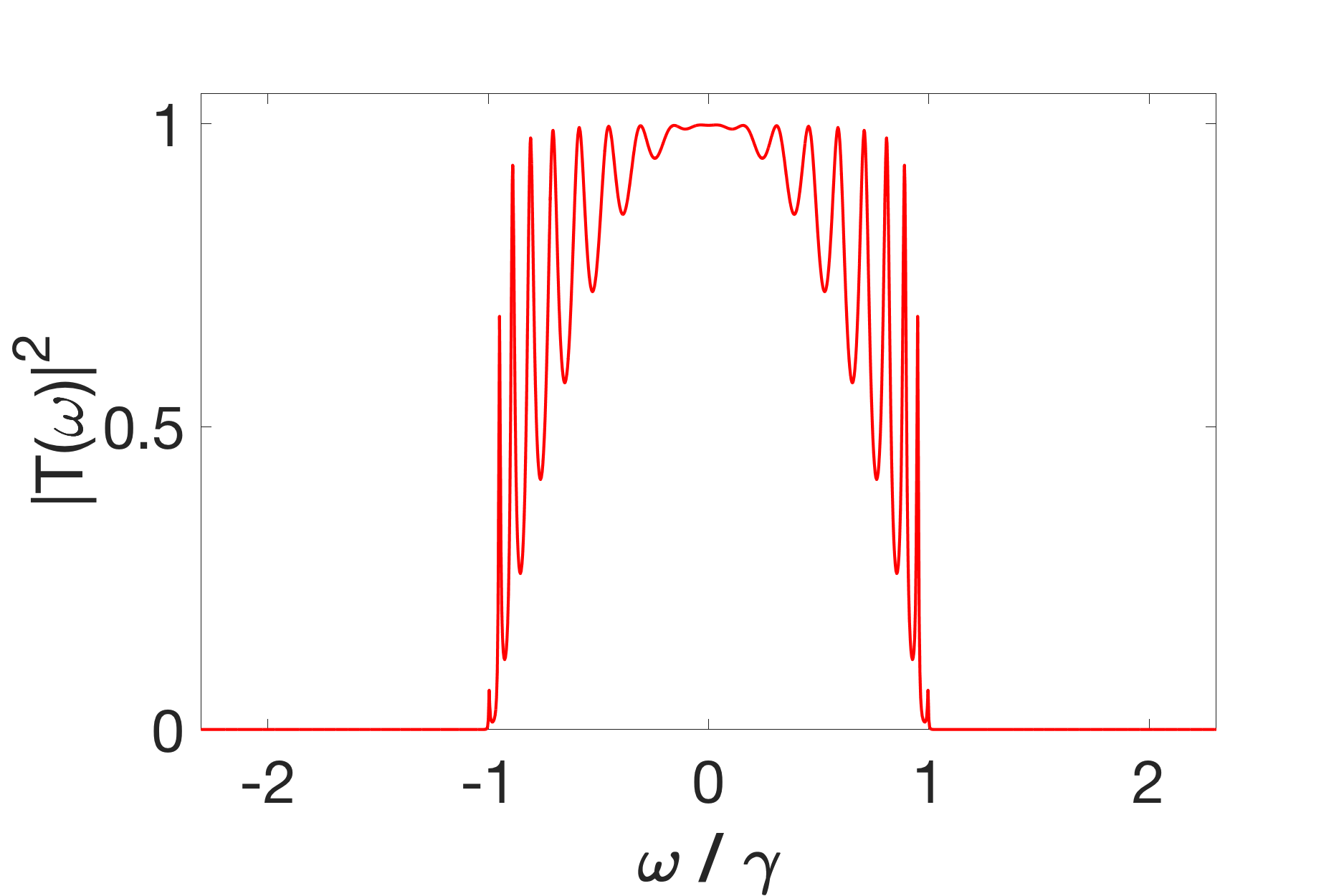}
            \caption[]%
    {{\small Transmission probability, detuning $\omega_i - \omega_{i+1} = \frac{1}{50}\gamma$}}    
            \label{DetunedN20Trans}
        \end{subfigure}
                \begin{subfigure}[b]{0.475\textwidth}   
            \centering 
            \includegraphics[width=.9\textwidth]{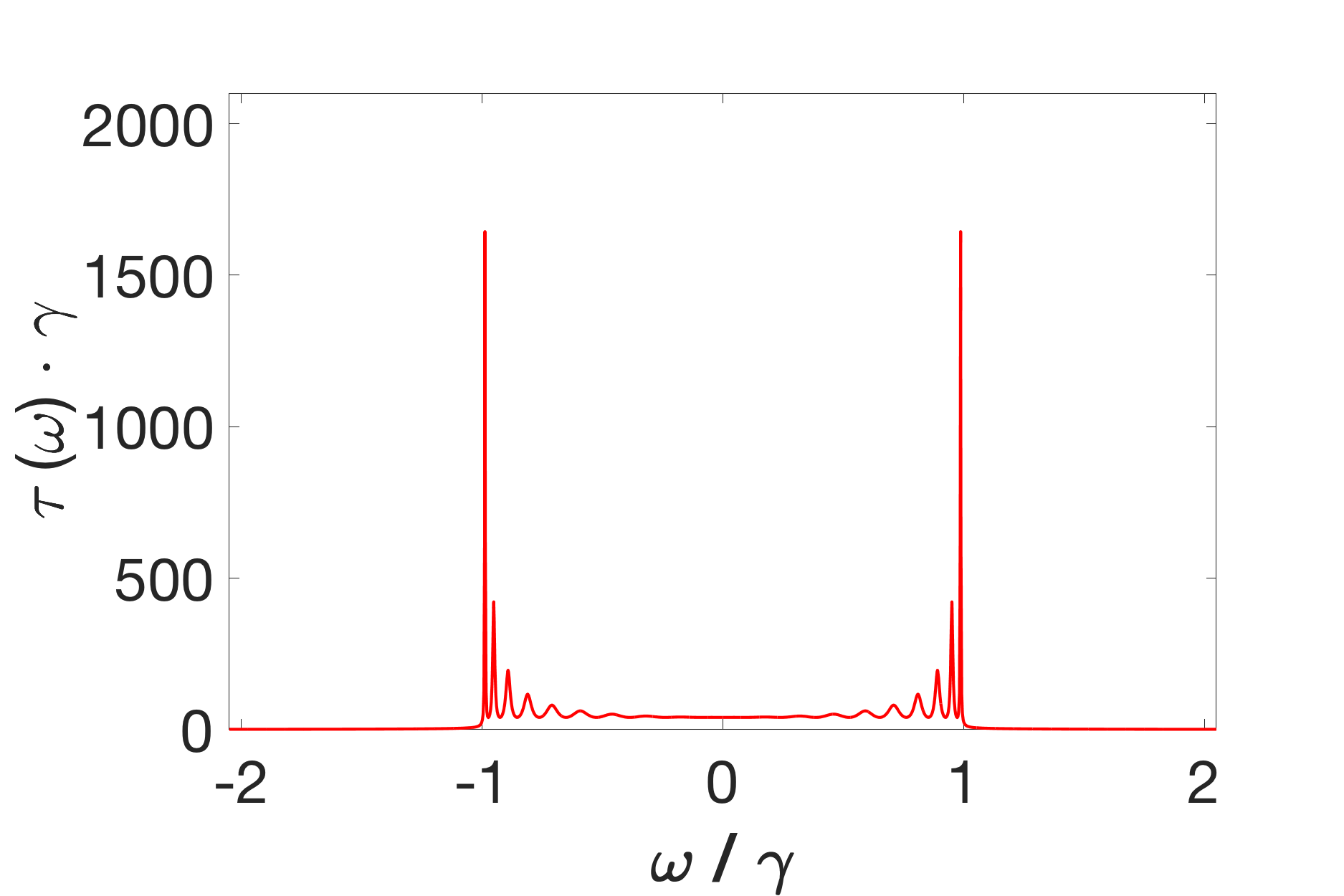}
            \caption[]%
    {{\small Group delay, no relative detuning}}    
            \label{20discretestatesSeriesTau}
        \end{subfigure}
	\quad
        \begin{subfigure}[b]{0.475\textwidth}   
            \centering 
            \includegraphics[width=.9\textwidth]{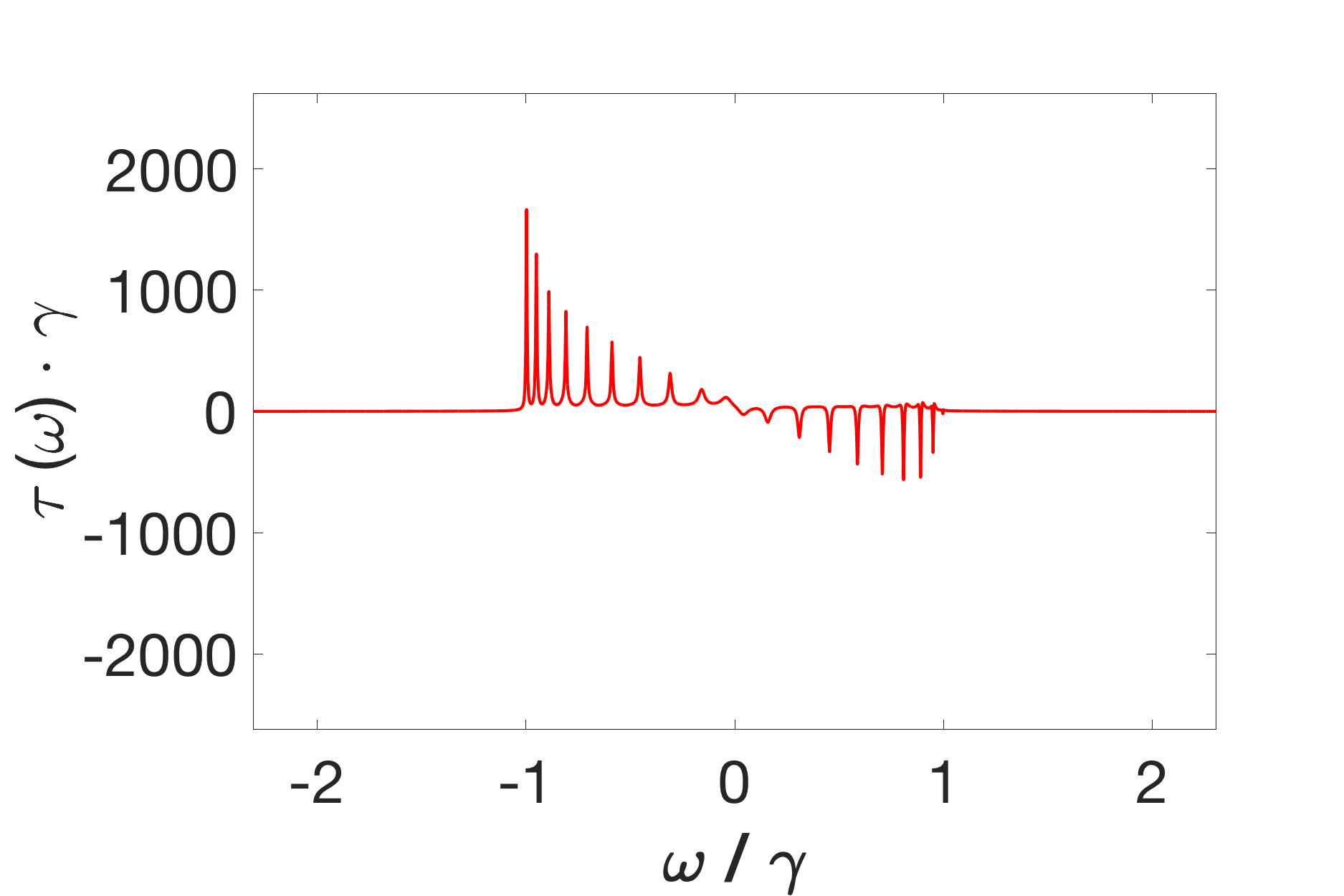}
            \caption[]%
    {{\small Group delay, detuning $\omega_i - \omega_{i+1} = \frac{1}{50}\gamma$}}     
            \label{DetunedN20Tau}
        \end{subfigure}
        \caption{\small Transmission probabilities and group delays for series networks with $N=20$ discrete states, balanced decay rates, critical coupling, and varied relative detuning. Again, the group delay is of the largest magnitude near the edge of the transmission function ($\pm2g$ in the high $N$ limit). A relative detuning may result in a negative group delay.} 
        \label{N20}
    \end{figure*}

 \lettersection{Group Delay} Lastly, we consider the group-delay for series networks, which increases with both $g$ and $N$ but for different reasons. As $g$ increases, the peaks of the transmission function sharpen so that the phase changes more rapidly. This results in an increased group delay (Fig. \ref{2discretestatesSeriesTrans}-d). As $N$ increase, we observe (as we did for parallel networks) that the peaks in the group delay are not of uniform magnitude, even when the transmission function itself is rather flat (Fig. \ref{20discretestatesSeriesTrans}). Instead, the frequencies of maximum delay are those closest to $\pm 2g$ (Fig. \ref{20discretestatesSeriesTau}) increase the most with $N$. This is because the oscillations in the transmission function are most dense here, resulting in a rapid change in transmission function phase and thus a larger group delay. On the contrary, the group delay is relatively flat in the center of the transmission window (Fig. \ref{20discretestatesSeriesTau}) so that a spectrally-narrow (compared to $g$) input photon pulse will not be significantly dispersed. (That is, the maximum dispersion-induced jitter $\mathcal{T}_g$ from (\ref{maxdisp}) will be very small.)

We can also consider the effect of detunings on the frequency-dependent group delay. We observe the group delay can be negative for series networks with detuning (Fig. \ref{DetunedN20Tau}), with an asymmetric structure that depends on the ordering of the detunings (except in the strongly coupled limit). Again, the magnitude of the group delay depends on the spacing of discrete states, increasing when the resonant frequencies are closely spaced.

  \newpage
\clearpage

\onecolumngrid

    \newpage
\clearpage

\twocolumngrid

\subsection{Hybrid Networks}

 \begin{figure}[ht] 
	\includegraphics[width=1\linewidth]{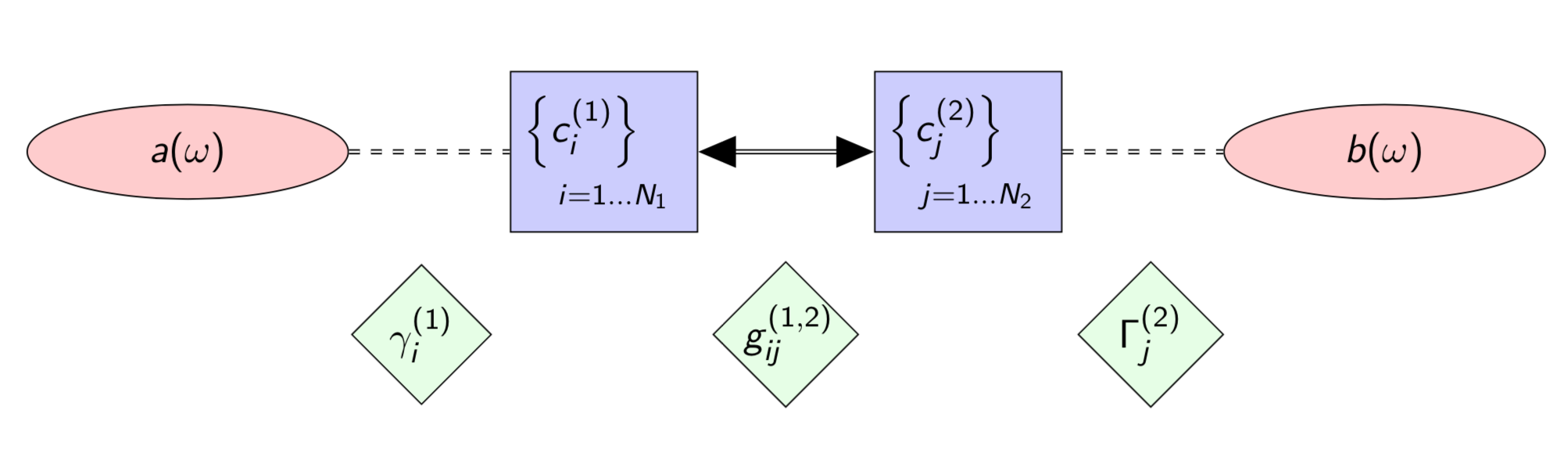} 
	\caption{A hybrid network of two manifolds, each with $N_k$ discrete states described by an operator $c^{(k)}_i (\omega)$. The discrete states are coherently coupled at rates $g^{(1,2)}_{ij}$ and incoherently coupled to left (input) and right (output) continua $a$ and $b$ with decay rates $\gamma^{(1)}_i$ and $\Gamma^{(2)}_j$, respectively.}	\label{hybridschem}
\end{figure}

We now begin to approach the case of a general two-sided quantum network. The fully general problem is intractable analytically, but luckily there are several simplifications we can make that correspond to the network representing realistic photo detecting systems. To illustrate, consider the case of two parallel networks in series: each discrete state connected to each discrete state in the other manifold (but not necessarily at the same rate) with each of the two manifolds of purely virtually coupled discrete states coupled to their own continuum.  One could imagine generating different networks from this one by removing a coupling $g_{ij}$ between discrete states, permuting which discrete states are disconnected, removing an additional coupling, permuting, and so on. However, this is unphysical: two discrete states in parallel cannot be prevented from coupling to the same discrete state except by selection rules. But the discrete states are also coupled to the same continuum (which has been already made $1$D in effect), so they must satisfy the same selection rules. The same argument applies in reverse: no discrete state within a manifold can individually stop being coupled to a continuum without the rest of the discrete states doing so as well. And we can similarly apply it to manifolds embedded in a larger network away from a continuum: the requirement that all states satisfy the same selection rules is strong. This means we can ignore partially connected networks. It also gives us a helpful way to organize discrete states: into manifolds of discrete states (which are purely virtually coupled to each other after diagonalization) that are all coupled to the same set of discrete or continuum states.

We can now focus on a very large class of quantum networks, hybrid systems consisting of manifolds in series (Fig. \ref{hybridschem}). It will be helpful to denote couplings between discrete states $i$ and $j$ or manifolds $k$ and $\ell$ as $g_{ij}^{(k,\ell)}$, and denote decay rates and discrete states within a manifold with a superscript ($\gamma_i^{(k)}$, $\Gamma_i^{(k)}$, and $c_i^{(k)}(\omega)$, with $\gamma_i^{(k)}=0$ for $k>1$ and $\Gamma_i^{(k)}=0$ for $k<M$, where $M$ is the number of manifolds). This allows us to rewrite the general equation (\ref{quantlangspect}) in a more explicit form

 \begin{widetext}
\bea\label{quantlangspectexp}
\,&\,\\
-i\Delta_i c_i^{(k)} (\omega)= &- \sum\limits_{j=1}^{N_k}\frac{\sqrt{\gamma_i^{(k)}\gamma_j^{(k)}} + \sqrt{\Gamma_i^{(k)}\Gamma_j^{(k)}}}{2} c_j^{(k)}(\omega)  - i \sum\limits_{j=1}^{N_{k-1}} g_{ij}^{(k-1,k)} c_j^{(k-1)}- i \sum\limits_{j=1}^{N_{k+1}}g_{ij}^{(k,k+1)} c_j^{(k+1)}-\sqrt{\gamma_i^{(k)}}a_{\rm in}(\omega)  -\sqrt{\Gamma_i^{(k)}}b_{\rm in}(\omega)\nonumber
\eea \end{widetext} where the superscripts denote labels of manifolds and we've implicitly defined $g_{ij}^{(0,1)} = g_{ij}^{(N,N+1)} = 0 \,\forall i,j$. We denote the number of discrete states in each manifold $N_k$ such that $\sum\limits_{k=1}^M N_k = N$. In general (\ref{quantlangspectexp}) is still only numerically solvable but we can now note two cases that yield analytic solutions.

\begin{figure}[h!]
        \begin{subfigure}[b]{0.475\textwidth}
            \centering
            \includegraphics[width=\textwidth]{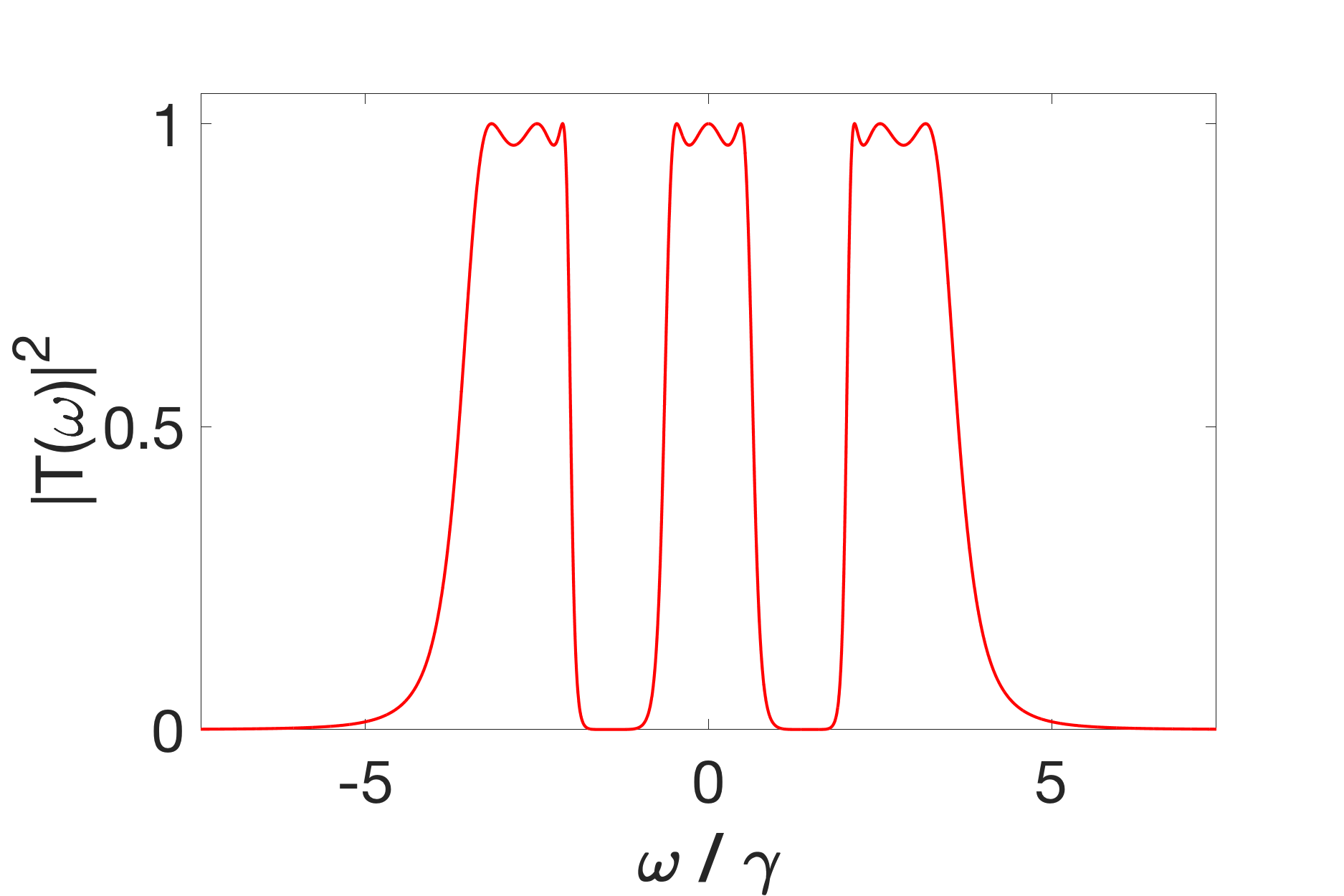}
            \caption[]%
            {{\small Transmission probability }}    
            \label{HybridTrans333}
        \end{subfigure}
        \vskip\baselineskip
        \begin{subfigure}[b]{0.475\textwidth}  
            \centering 
            \includegraphics[width=\textwidth]{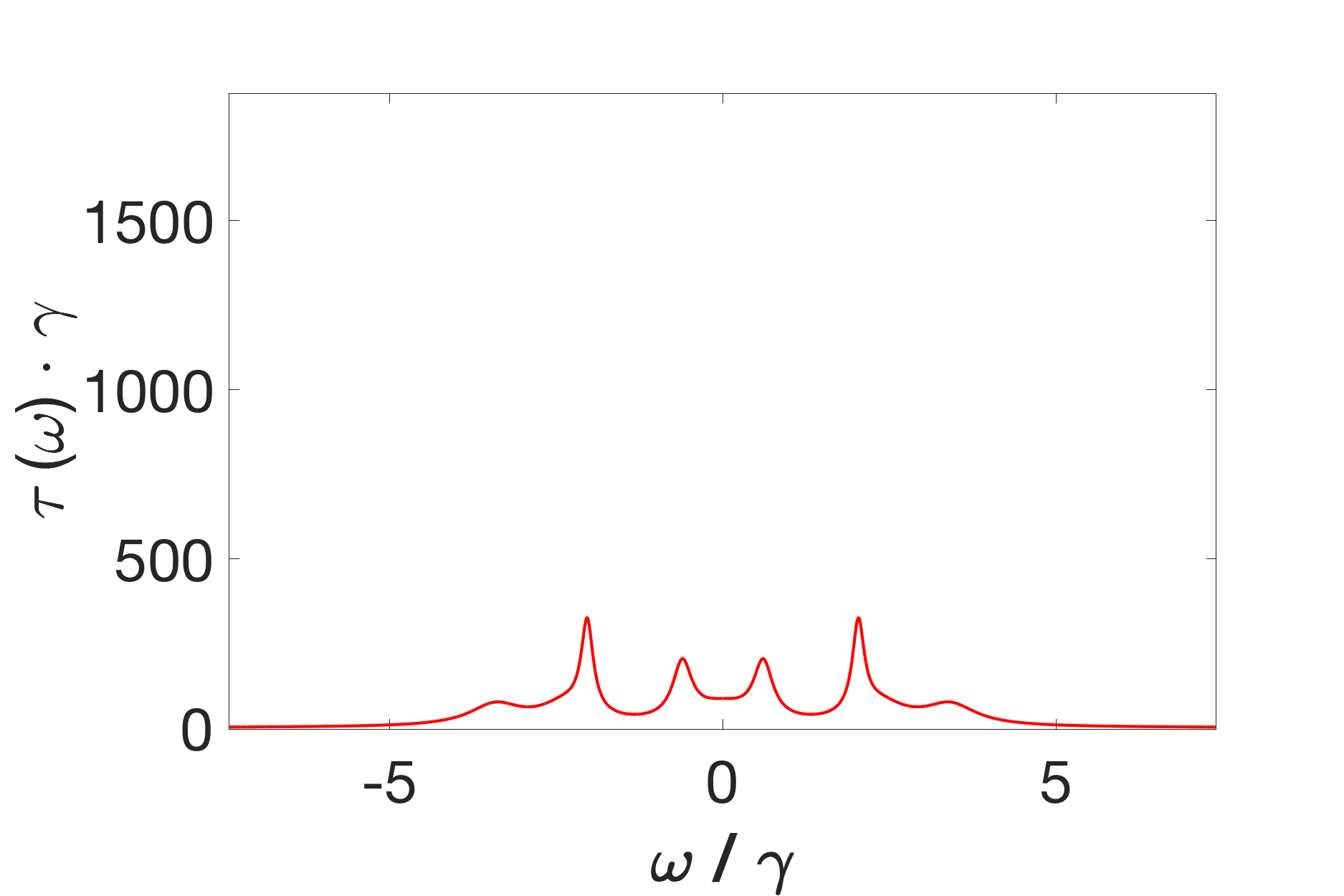}
            \caption[]%
            {{\small Group delay}}    
            \label{HybridTau333}
        \end{subfigure}
\caption{\small Transmission probability and group delay for a hybrid network consisting of $3$ manifolds in series, each with $3$ discrete states in series with balanced decays and uniform critical coupling. Within each manifold, the discrete states have frequency spacing by $2.5\gamma$. Using the same special conditions as for series network, perfect transmission is achieved for $9$ frequencies, with the multi-layered structure of $|T(\omega)|^2$ encoding the manifold structure.} 
        \label{hybridplot333}
    \end{figure}

The first is the case of critical coupling between members of each adjacent manifold and, additionally, uniformly unbalanced decays: we first define an effective decay rate for internal couplings within the system so that $g_{ij}^{(k,\ell)}= \frac{\sqrt{\gamma_i^{(k)}\Gamma_j^{(\ell)}}}{2}$ (effectively specializing to the critical coupling case for series networks), and then consider the special case of $\Gamma_i^{(k)}/\gamma_i^{(k)}=k^{(k)}\,\forall i,k$ (inhomogeneous decays that are uniformly unbalanced within each manifold). This leads to a reflection coefficient of the form

\bea\label{refseriesc1}
&\\
R(\omega) &= 1-\cfrac{2 h^{(1)}}{h^{(1)}-i+\cfrac{k^{(1)} h^{(1)}h^{(2)}}{-i+\cfrac{k^{(2)} h^{(2)}h^{(3)}}{\dots+\cfrac{k^{(N-1)} h^{(N-1)}h^{(N)}}{\sqrt{k^{(N)}}h^{(N)}-i}}}}\nonumber
\eea where we have defined a new function $h^{(k)} = \sum\limits_{i=1}^{N_k} \frac{\gamma_i^{(k)}}{2\Delta_i^{(k)}}$. (The appearance of a lone $\sqrt{k^{(N)}}$ at the end of (\ref{refseriesc1}) is due to the final purely virtual coupling to the output continuum, since we've absorbed the rest of the decay rate into the function $h^{(k)}$.) This function encodes the zeroes and singularities we found for parallel networks, which previously gave rise to frequencies of constructive interference and completely destructive interference. 

While (\ref{refseriesc1}) is tractable, it is of limited applicability to real systems. More relevant is the second solvable case of homogenous coupling and decays within manifolds: $g_{ij}^{(k,\ell)}= g^{(k,\ell)}$, $\gamma_i^{(k)}=\gamma^{(k)}$, and $\Gamma_i^{(k)}=\Gamma^{(k)}$. This leads to a reflection coefficient of the form

\bea\label{refseriesc2}
& \\
R(\omega) &= 1-\cfrac{\gamma f^{(1)}}{\frac{\gamma}{2}f^{(1)}-i+\cfrac{(g^{(1,2)})^2 f^{(1)}f^{(2)}}{-i+\cfrac{(g^{(2,3)})^2 f^{(2)} f^{(3)}}{\dots+\cfrac{(g^{(N-1, N)})^2 f^{(N-1)} f^{(N)}}{\frac{\Gamma}{2} f^{(N)}-i}}}}\nonumber
\eea

where again we have defined a new function $f^{(k)} = \sum\limits_{i=1}^{N_k}  \frac{1}{\Delta_i^{(k)}}$. This provides a nice model for multi-mode systems in series (for instance, a linear network of identical multimode optical cavities).

        We see in both (\ref{refseriesc1}) and (\ref{refseriesc2}) a combination of the structures we observed in  (\ref{reflectk}) and (\ref{reflecthomo}) for parallel networks and (\ref{refseries}) for series networks: correlations in amplitudes between discrete states within a given manifold manifest via a function $h^{(k)}$ or $f^{(k)}$ with $N_k$ poles and $N_k-1$ zeroes (potentially perfectly transmitted and reflected frequencies, depending on the hybrid network's resonance structure and couplings), and causal ordering of the manifolds manifests in a continued fraction structure. This latter property makes them easily analyzable using the Wallis-Euler recurrence relations, allowing us to find $R(\omega)$ and from there $T(\omega)$ and the other quantities of interest.

\begin{figure}[t]
        \begin{subfigure}[b]{0.475\textwidth}
            \centering
            \includegraphics[width=\textwidth]{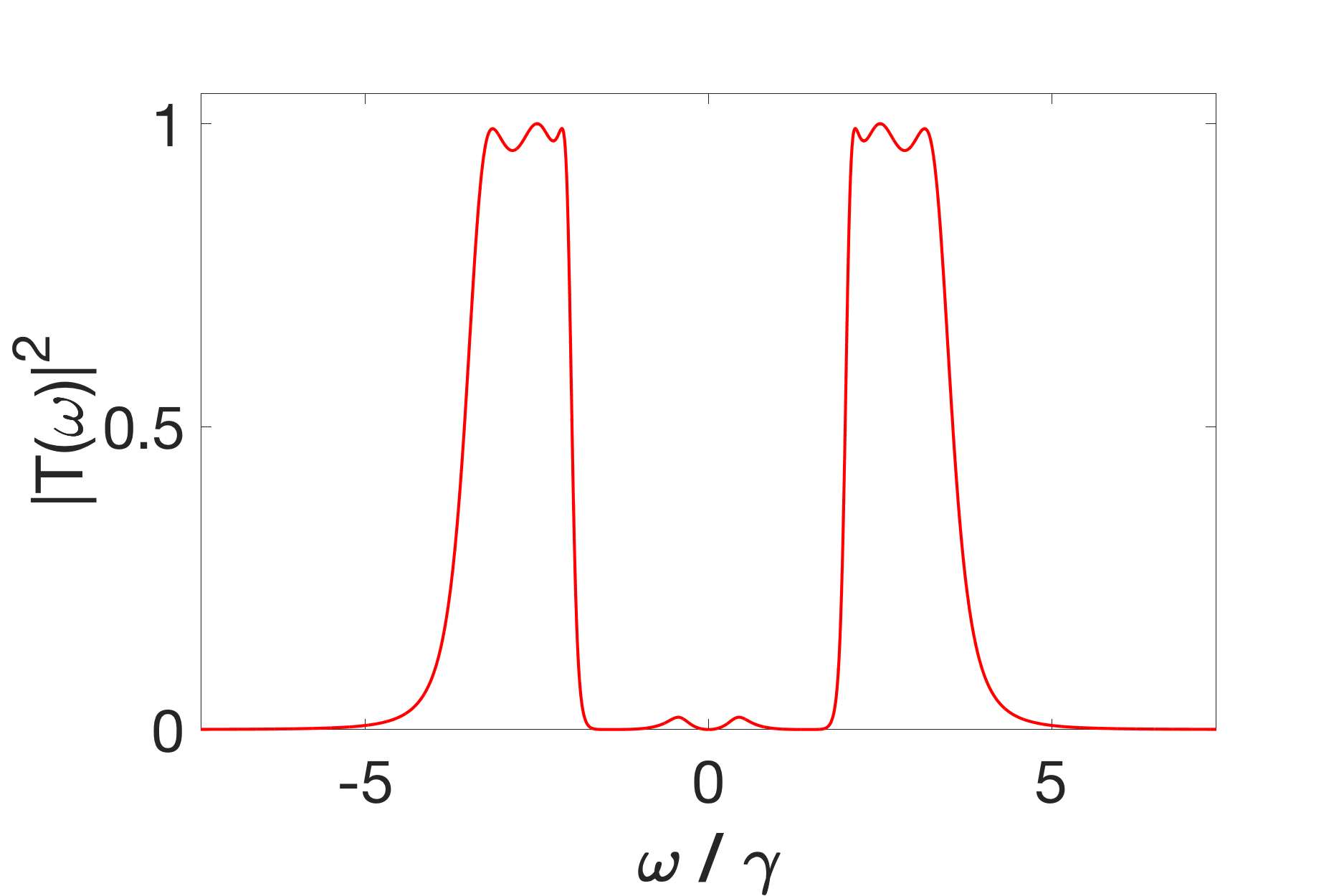}
            \caption[]%
            {{\small Transmission probability}}    
            \label{HybridTrans233}
        \end{subfigure}
        \vskip\baselineskip
        \begin{subfigure}[b]{0.475\textwidth}  
            \centering 
            \includegraphics[width=\textwidth]{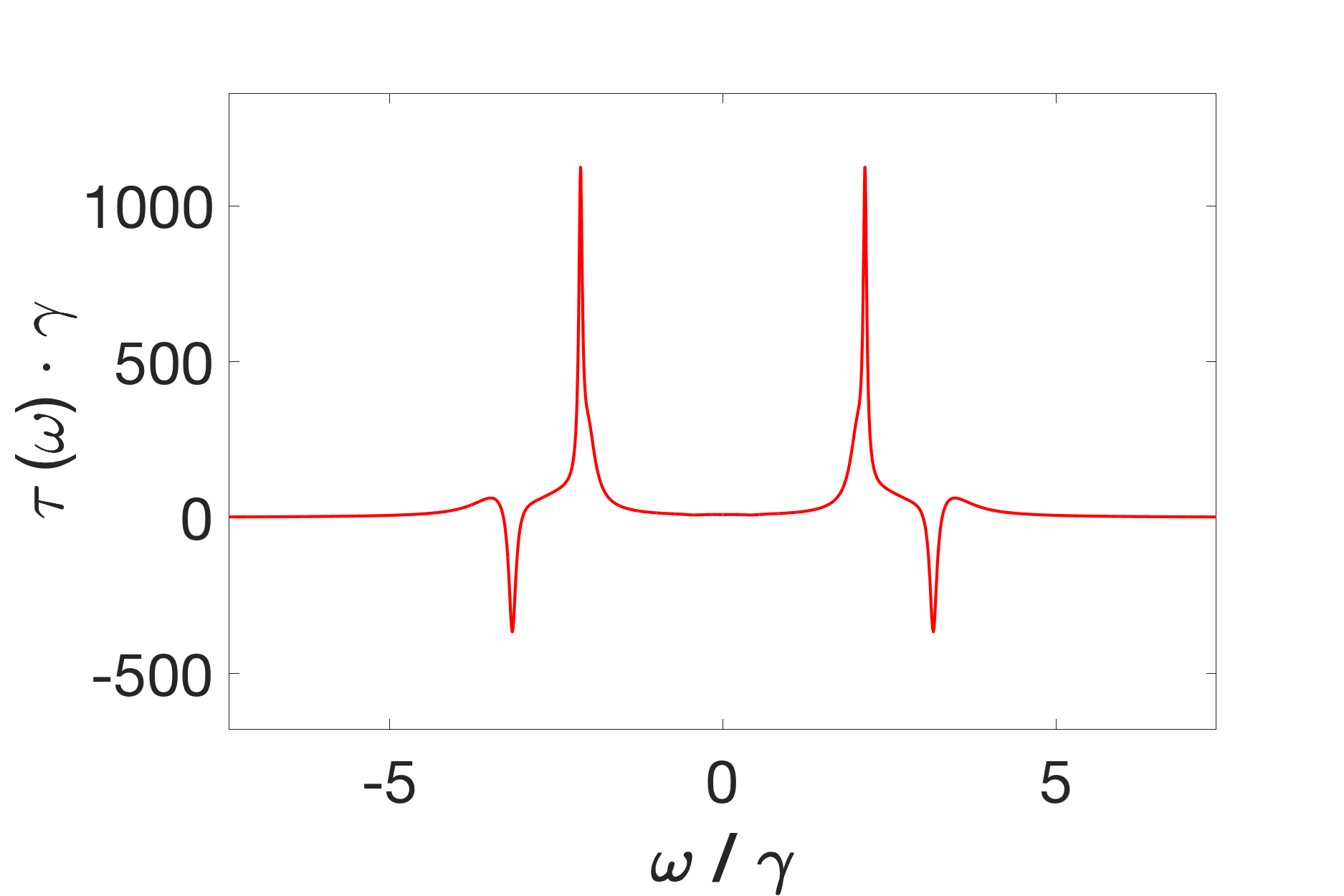}
            \caption[]%
            {{\small Group delay}}    
            \label{HybridTau233}
        \end{subfigure}
        \caption{\small Transmission probability and group delay for a hybrid network consisting of $3$ manifolds in series, the first with $2$ discrete states and the second and third with $3$ discrete states, with balanced decays and uniform coupling $g=\frac{\sqrt{\gamma\Gamma}}{2}$ (the critical value for networks without detuning). Within the first manifold, the discrete states have frequency spacing $5\gamma$. Within the second and third manifolds, the discrete states are detuned by $2.5\gamma$, resulting in relative detuning between the manifolds. Since the couplings are no longer tuned to the critical parameters for a detuned system, perfect transmission is only achieved at $6$ frequencies.} 
        \label{hybridplot233}
\end{figure}

\lettersection{Perfect transmission} Focusing on the case of homogenous coupling, we can use the same trick of examining the convergence of an infinite series of identical manifolds around one of resonant frequencies $\omega_i$ to find the two critical conditions as we did for series networks, which again are $\gamma=\Gamma$ and $g=\frac{\sqrt{\gamma\Gamma}}{2}$. We also observe that the form of $T(\omega)$ for hybrid systems exhibits a combination of features of series and parallel networks  (Fig. \ref{HybridTrans333} and \ref{HybridTrans233}). This results in layers of structure, with the small dips and peaks corresponding to intra-manifold structure layered on top of the larger dips and peaks of the inter-manifold structure. 

When at least one discrete state in each manifold have the same resonance, letting $\gamma=\Gamma$ and $g=\frac{\sqrt{\gamma\Gamma}}{2}$ ensures perfect transmission at $M$ frequencies. For a general hybrid network, the number of peaks of unity (perfect transmission) of $T(\omega)$ are bounded above by $M \,Min\{N_k\}\leq N$, with $M$ the number of manifolds each with $N_k$ discrete states. Here the latter equality is reached for networks that are either completely in parallel or completely in series, with critical parameters in either case \footnote{The bound is somewhat stronger for critically coupled networks, with the number of perfectly transmitted frequencies bounded by $(M-\frac{1}{2}(1+(-1)^M))\,Min\{N_k\}$ instead. This is due to the on-resonance broadening that occurs for critically coupled systems, which prevents the splitting of one peak to two when the number of manifolds $M$ is even.}.

\lettersection{Spectral Bandwidth} For hybrid networks where the first and last manifold have the same number of discrete states $N_1=N_M$, the spectral bandwidth is bounded above by the parallel network bandwidth for that number of discrete states $\tilde{\Gamma}\leq \frac{N_1 2\gamma\Gamma}{\gamma+\Gamma}$ with equality reached in the strong coupling limit. (When the first and last manifold have different number of discrete states $N_1\neq N_M$, the bandwidth is bounded above by $\frac{2\gamma\Gamma}{\gamma+\Gamma}(Min\{N_1,N_M\}+X)$ for a homogeneously decaying network with $X$ a network-dependent number that is always less than $1/2$.) 

\lettersection{Group Delay} We observe the same structural properties of the group delay (Fig. \ref{HybridTau333}) as we did for other networks; the frequencies with the largest delays are those where oscillations in the transmission function are most dense. We find that networks with non-identical manifolds can give rise to group delays that are not strictly positive (Fig. \ref{HybridTau233}). 


\begin{figure}[t]
        \begin{subfigure}[b]{0.475\textwidth}
            \centering
            \includegraphics[width=\textwidth]{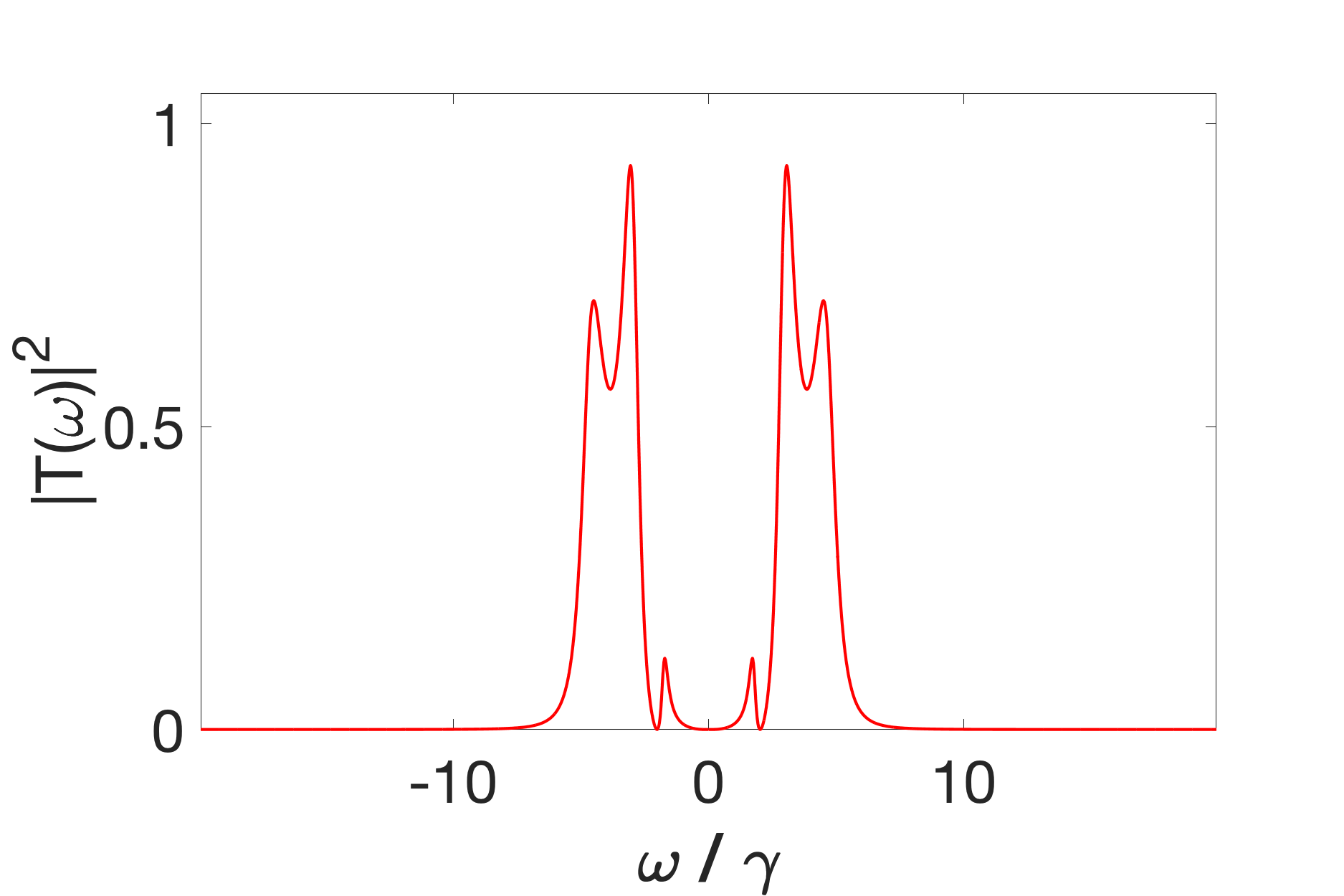}
            \caption[]%
            {{\small Transmission probability }}    
            \label{HybridTrans223}
        \end{subfigure}
        \vskip\baselineskip
        \begin{subfigure}[b]{0.475\textwidth}  
            \centering 
            \includegraphics[width=\textwidth]{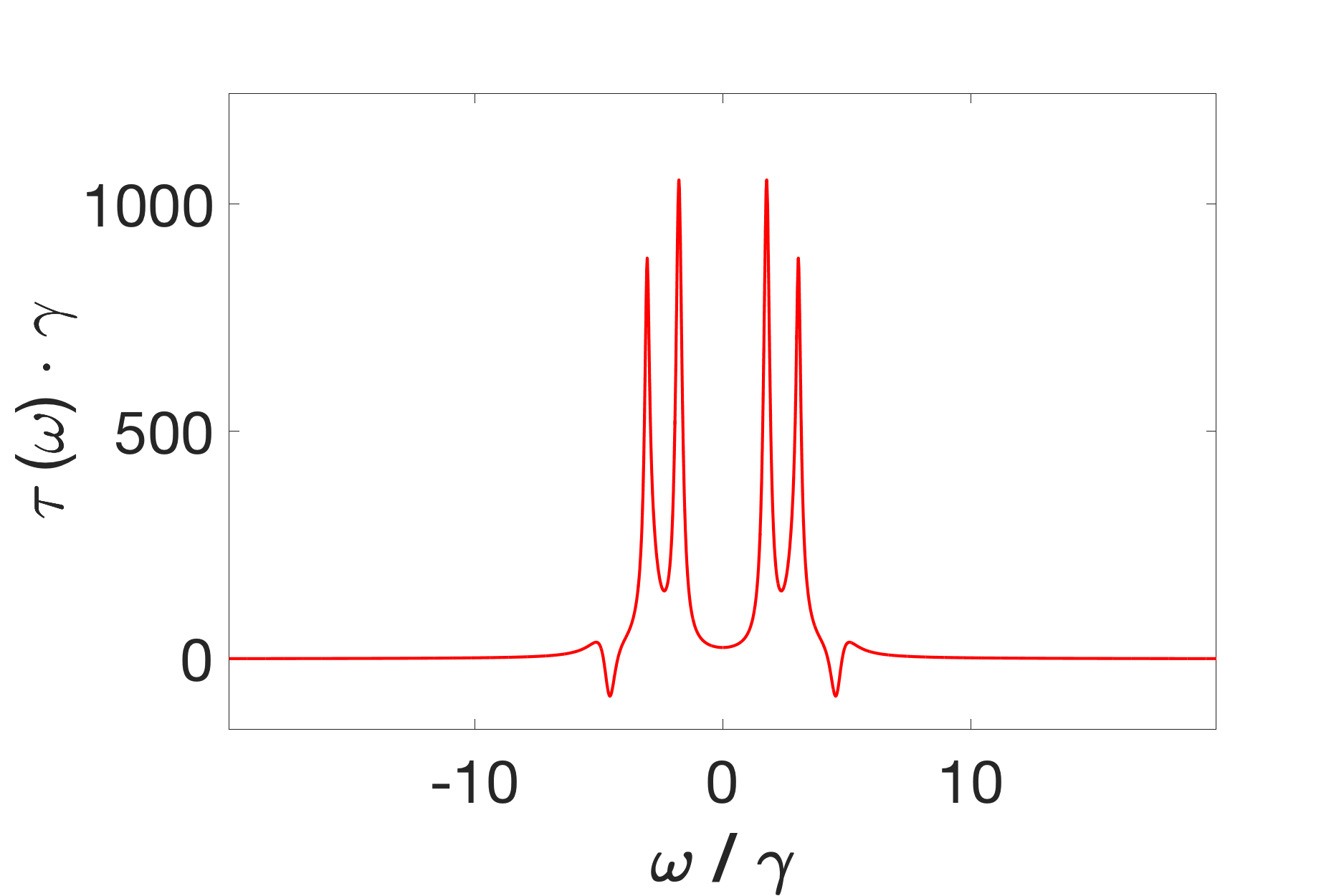}
            \caption[]%
            {{\small Group delay}}    
            \label{HybridTau223}
        \end{subfigure}
        \caption{\small Transmission probability and group delay for a hybrid network consisting of $3$ manifolds in series, the first two with $2$ discrete states and the third with $3$ discrete states, with balanced decays $\gamma=\Gamma$ and uniform coupling $g=\sqrt{\gamma\Gamma}$. Within the first and second manifolds, the discrete states are spaced by $2.5\gamma$. Within the third manifold, the discrete states are spaced by $7\gamma$. Now no frequencies are perfectly transmitted, and finding couplings such that perfect transmission is achieved becomes less trivial. \vspace{1em}} 
        \label{hybridplot223}
    \end{figure}

\section{Other Networks}

\subsection{General Two-Sided Networks}

We now begin to extrapolate from the above analyses to a larger class of two-sided networks. Both the series networks and hybrid networks we've discussed have the key property of asymptotic irrelevance of the causal ordering of discrete states in the strong coupling limit. This means that the asymptotically strong-coupling behavior of these networks is entirely determined by the properties of fully parallel networks (\ref{RGen}), and we can make several statements that will apply to any network that has the same behavior (that is, a network that resembles (\ref{RGen}) in the strong coupling limit). From the structure of (\ref{RGen}) alone, we can bound the total number of peaks and troughs of $|T(\omega)|^2$; we find there are at most $2N-1$ peaks and $2N-2$ troughs in the strong coupling limit and since the number of maxima never decreases with increasing $g$, we can extrapolate these bounds to weakly coupled systems as well. In transitioning from $2N-1$ peaks or constructive interference to $N$ peaks of perfect transmission, we observe a pair-wise merging of peaks (with one extra peak leftover when $N$ is odd). We also find that the number of dips of zero transmission (perfect reflection) is at most $\sum\limits_{k=1}^{M} (N_k - 1) \leq N-1$ where the latter equality is only reached for a completely parallel network or in the strong coupling limit.

We suspect that the above argument for upper bounds applies more generally than just to the specific networks consisting of manifolds in series; that is, there is a large class of networks that reduce to a parallel description in the strong coupling limit \footnote{It's worth reiterating here that \emph{all} networks reduce to (\ref{RGen}) in the very-weak coupling limit $2g_{ij}\ll\sqrt{\gamma_i\gamma_j}+\sqrt{\Gamma_i\Gamma_j} \,\forall i,j$}. However, we cannot generalize from the above analysis to a fully arbitrary two-sided network as there also are networks with different topologies. For instance, we can consider networks with loops of discrete states that exist outside the main chain of manifolds that connect the two continua. Since it is not necessary for a photon to pass through the loop of discrete states to make it through the network, these networks behave differently in the strong coupling limit. We can make a further distinction between disconnected loops (dead-ends where photons have to back track) and connected loops (chains of manifolds that provide an alternate route to the output continua). It is possible that connected loops behave more like the loop-free networks discussed above in the strong-coupling limit, but with their specific loop-structure encoded in $T(\omega)$ in unexpected ways.

\subsection{Additional Continua}

  \begin{figure}[h] 
	\includegraphics[width=.6\linewidth]{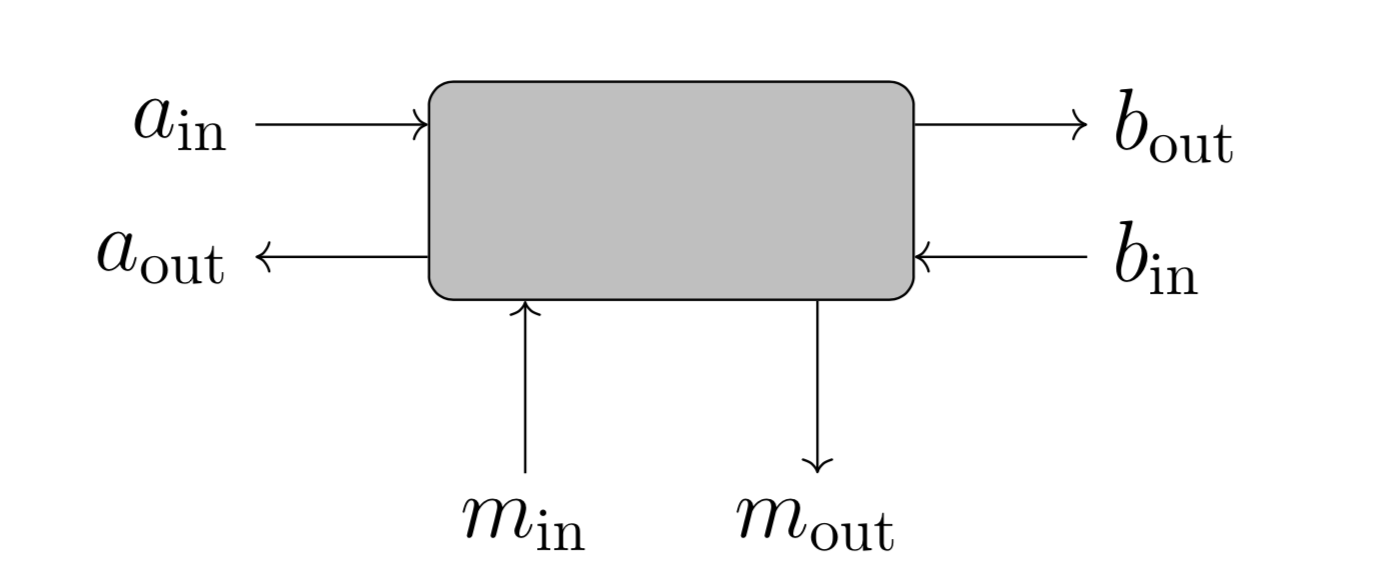} 
	\caption{Input-output field operators for a black box network with one additional side channel. This will give rise to losses that make perfect transmission impossible, as well as dark counts if the side channel contains excitations (i.e. thermal).}
	\label{3ports}
\end{figure}

We have thus far only considered quantum networks with two continnua, and have shown that perfect transmission through a general network structure is possible. We will briefly analyze more general multi-port quantum networks (Fig. \ref{3ports}) to illustrate how they lead to inefficiencies and dark counts. 

Introducing a third continuum coupled to our network of discrete states at rates $\mu_i$, we rewrite (\ref{quantlangspect}) in the form 

\begin{widetext}\bea\label{quantlangspectside}
-i\Delta_i c_i (\omega)= - \sum\limits_j\,\left(\frac{\sqrt{\gamma_i\,\gamma_j} + \sqrt{\Gamma_i\,\Gamma_j}+\sqrt{\mu_i\,\mu_j} }{2} + i g_{i\,j}\right) c_j(\omega)-\sqrt{\gamma_i}\,b_{in}(\omega)  -\sqrt{\Gamma_i}\,a_{in}(\omega)-\sqrt{\mu_i}\,m_{in}(\omega),
\eea\end{widetext} where we have introduced a new input field annihilation operator $m_{in}(\omega)$ for the additional continuum (satisfying the canonical commutation relations), satisfying the same form of input-output relations as in (\ref{Nstatebound}) 

\bea
m_{out}(\omega) &= m_{in} (\omega)+ \sum_i\,\sqrt{\mu_i}\,c_i (\omega)\label{out fields:3}.
\eea

When there are side channels and all decay rates in the system are homogenous ($\mu_i=\mu$, $\gamma_i=\gamma$, and $\Gamma_i=\Gamma$), it is impossible to achieve perfect transmission at any input frequency without adding additional excitations (active filtering). To see this, consider taking expectation values and imposing $m_{in}(\omega)\rightarrow0$ over all frequencies. Flux conservation requires that, for a perfectly transmitted frequency $\omega'$, we also have $m_{out}(\omega')\rightarrow0$. From (\ref{out fields:3}), we see the only way to achieve this is for $\sum_i c_i(\omega') \rightarrow 0$ but from (\ref{Nstatebound}), we see that this results in $a_{out}(\omega') = a_{in}(\omega')$ and $b_{out}(\omega') = b_{in}(\omega')$---the frequency $\omega'$ is instead perfectly reflected by the system.

It is still possible that side channels with inhomogeneous coupling $\mu_i$ could yield a system that perfectly transmit some light at a frequency $\omega'$ satisfying both conditions $a_{out}(\omega') = a_{in}(\omega')-\sum_i\,\sqrt{\gamma_i}c_i(\omega') = 0$ and $\sum_i\,\sqrt{\mu_i} c_i(\omega')=0$. After all, the $c_i(\omega')$ with resonant frequencies above and below $\omega'$ generically differ in phase by $\pi$ and could in principle cancel out. However, this would require \emph{incredible} fine-tuning of the system. For a uniformly inhomogeneous parallel network such as (\ref{reflectk}) with a side channel, we can easily see that this approach fails even in the low-loss limit near resonance (where it would be most likely to succeed): we find $m_{out}(\omega_j) \approx \frac{2\mu_j}{\gamma_j+\Gamma_j} a_{in}(\omega_j)$. In general, output channels outside your control lead to loss.

Similarly, we can determine from (\ref{quantlangspectside}) and (\ref{out fields:3}) that, in general, input channels outside your control lead to extra dark counts; photons that are in initially populated side channel can end up in the monitored continuum. If the side channel is at a finite temperature, there will be thermal photons that could populate the system. The probability of this vanishes for $k_b T\ll \hbar \omega_i \,\forall i$, but will generally be non-negligible for high-temperature systems. The specific contribution depends on the specific form of the full transfer matrix, which now will include a function $D_m(\omega)$ that governs the probability of thermal excitations in the side channel to end up in the monitored continuum. 

Of course, the internal mode $\bin$ may be occupied by thermal excitations as well, but the contribution to dark counts by these will depend strongly on the amplification mechanism. Furthermore, if only frequencies such that $|T(\omega_i)|=1$ are amplified, we find that thermal excitations $\bin$ do not contribute at all to dark counts as they always leak out of the system, ending up in the continuum mode $\aout$ populated by reflected input photons.

\section{Conclusions}

We have studied the behavior of coherent quantum networks and have found that they provide a diverse structure of transmission functions for modeling the first stage of single photon detectors (transmission of a single excitation from the input continuum, through the system, to a monitored output continuum). Inefficiencies and dark counts can be modeled through the incorporation of additional continua (side channels). While we do not find fundamental limits to transmission efficiency, spectral bandwidth, or frequency-dependent group delay across all the studied networks (series, parallel, and hybrid), we do find that some networks are better suited to certain applications than others, as we will now discuss.

 \lettersection{Perfect Transmission} Often the most important metric for a photo detector is photo detection efficiency, which for a quantum network is tantamount to ensuring $|T(\omega_i)|=1$ is achieved for some frequency or frequencies $\omega_i$. The main conclusions of this paper are as follows:
 \begin{itemize}
 \item[(i)] Ensuring the decay rates to the input and output continua are balanced ($\gamma=\Gamma$) guarantees perfect transmission for at least one frequency in almost all quantum networks without loops, side channels, and detunings  [including all not covered by (ii)]. 
  \item[(ii)] Ensuring the couplings between manifolds are uniform and critical ($g=\frac{\sqrt{\gamma\Gamma}}{2}$) guarantees perfect transmission for at least one frequency in almost all quantum networks without loops, side channels, and detunings [including all not covered by (i)].
  \item[(iii)] For an arbitrarily detuned quantum network without loops and side channels, we are \emph{always} able to find conditions for the couplings and decay rates such that perfect transmission occurs for at least one frequency.
 \end{itemize}

That finding conditions such that perfect transmission $|T(\omega_i)|=1$ is always possible indicates that perfect photo detection is possible in a wide variety of physical systems. 

 
 
 
If one additionally wants a broadened transmission spectrum at a \emph{particular} perfectly transmitted frequency---so that a broad range of frequencies is detected almost perfectly---we similarly find a variety of ways to accomplish it. One way is to use a parallel network distributed over a small range of states. However, this will result in a number of dark states which will not be detected at all (spectral hole-burning). Instead, it is better to use a series network that meets both the critical coupling and balanced decay conditions, resulting in a maximally broadened on resonance transmission function as seen in Fig. \ref{CircleAndSquare}.

\lettersection{Spectral Bandwidth} We do not find fundamental limits to the minimum or maximum bandwidth of a network $\tilde{\Gamma}$ (or, conversely, to the interaction time between a network and incident light $\tilde{\Gamma}^{-1}$), but that it is generally proportional to the decay rates to both the input and monitored continua ($\tilde{\Gamma}\propto \frac{\gamma\Gamma}{\gamma+\Gamma}$, here assumed to be homogenous across states). We also find that some network structures are more suited to high-bandwidth applications than others; for networks with equivalent decay rates, we generally find that $\tilde{\Gamma}_{\rm series} \leq \tilde{\Gamma}_{\rm simple}\leq \tilde{\Gamma}_{\rm parallel}$. For a series network, equality with the upper limit  of $\tilde{\Gamma}_{\rm simple}=\frac{2\gamma\Gamma}{\gamma+ \Gamma}$ is reached only in the strong-coupling limit, and the lower limit ($\tilde{\Gamma}=0$) corresponds to a completely de-coupled or infinitely-detuned network (so that a photon can never pass through). For a parallel network, the bandwidth is always given $\tilde{\Gamma}_{\rm parallel} = \sum_i = \frac{2\gamma_i\Gamma_i}{\gamma_i+\Gamma_i}$ regardless of detuning, so the lower limit simply corresponds to a single discrete state (reproducing the simple model). Unlike both series and hybrid networks where the spectral bandwidth decreases with detuning, the spectral bandwidth of a parallel network is independent of detuning. This makes parallel networks the better candidate for implementation of broadband single photon detection where the frequencies that need to be detected are distinct (so that spectral hole-burning is a non-issue). 

Considering hybrid networks, we find that their bandwidths are bounded above and below by parallel and series networks, respectively; given a series network with the same coupling strengths and manifold-number, and a parralel network with the same manifold resonance structure and decay rates, the bandwidth of a Hybrid network is bounded $\tilde{\Gamma}_{\rm series} \leq \tilde{\Gamma}_{\rm hybrid}\leq \tilde{\Gamma}_{\rm parallel}$. The upper limit for a hybrid network is approached only in the limit of strong coupling, and the lower limit is approached when there is only a single discrete state in each manifold. 

 \lettersection{Group Delay} The maximum magnitude of frequency-dependent group delay $\tau(\omega)$ increases with both the couplings between discrete states and the density of oscillations in the transmission function $T(\omega)$. In particular, the bounded-box structure of a uniformly coupled series networks yields large delays near $\omega=\pm 2g$. We observe that a negative group delay may occur for series networks with relative detuning between the discrete states. (For more on negative group delay, see Ref.~\cite{solli2002}.) For general networks, the group delay can vary immensely over the range of frequencies where $|T(\omega)|^2$ is non-negligible, with the effect being strongest for hybrid networks (where the resonance structures can be most dense). This means we can expect dispersion effects to be substantial in many photo detection platforms. For applications such as frequency discriminating delay-lines, we may expect hybrid networks to be the best performing candidate as they allow for the finest control of group delay structure.

 \lettersection{Tradeoffs} For an arbitrary series or hybrid network with arbitrary decay rates, perfect transmission requires a particular choice of the couplings. This critical value is generically of order $\sqrt{\frac{\gamma\Gamma}{2}}$. However, the spectral bandwidth is maximized when $g\gg\sqrt{\frac{\gamma\Gamma}{2}}$. Furthermore, the magnitude and location of the maximum group delays depends strongly on the coupling $g$ (and especially so in the high-$N$ limit). So for series and hybrid networks, there is a clear tradeoff between efficient transmission and the spectral bandwidth (which saturates in the high-$g$ limit) \footnote{Of course, this assumes a large spectral bandwidth is a desired. There is always an obvious tradeoff between the spectral bandwidth and the amount one learns about the incoming light; the more frequencies that can be detected efficiently, the less a single click tells you about the incoming light. Whether a high bandwidth or a low bandwidth is preferred will ultimately depend on the application of a SPD.}, with the group delay changing as well. We contrast this with the case of parallel networks where the three quantities are completely independent: perfect detection requires balanced decays ($\gamma=\Gamma$), the bandwidth can be directly scaled by scaling both decay rates together, and the frequency-dependent group delay $\tau(\omega)$ alone depends on the detuning between discrete states, which is an independent quantity. So for photo detectors where all three quantities must be determinable independently, parallel networks form the clear contenders. The exception to this are situations where a negative group delay is required, which parallel networks never exhibit. Then the use of a hybrid or series network is unavoidable, as are their accompanying tradeoffs.

\lettersection{Final Remarks} We have studied a variety of quantum networks to uncover tradeoffs and limits fundamental to single photon detection that may arise in the first step of photo detection (filtering). The spectral bandwidth is the only main quantity of interest where there appear to be fundamental limits for particular classes of networks (parallel, series, and hybrid), but even these scale with the relevant decay rates for the networks. Given the freedom to adjust coupling strengths and decay rates, arbitrary group delay, transmission efficiency, and spectral bandwidth are attainable for any network so long as they are attained individually (together there may be tradeoffs for series and hybrid networks).

We do not find fundamental limitations to dark counts and losses in this analysis: in general uncontrolled input channels lead to dark counts and uncontrolled output channels lead to loss. These can always be mitigated by cooling the system ($k_b T\ll\hbar\omega_i\,\forall i$) and reducing coupling to side channels ($\mu_i,\,\nu_i\ll\gamma_i+\Gamma_i \,\forall i$). 

Other sources of noise, such as from signal amplification (for more on this specifically, see Ref.~\cite{proppamp}), classical parameter fluctuations, as well as the noise inherent in a non-ideal quantum measurement (as described by an arbitrary photo detection POVM) can also be included in our model to give a fully quantum description of the entire single photon detection process \cite{Propp2}.



This work is supported by funding from
DARPA under
Contract No. W911NF-17-1-0267.
\bibliography{networks}

\begin{thebibliography}{79}%
\makeatletter
\providecommand \@ifxundefined [1]{%
 \@ifx{#1\undefined}
}%
\providecommand \@ifnum [1]{%
 \ifnum #1\expandafter \@firstoftwo
 \else \expandafter \@secondoftwo
 \fi
}%
\providecommand \@ifx [1]{%
 \ifx #1\expandafter \@firstoftwo
 \else \expandafter \@secondoftwo
 \fi
}%
\providecommand \natexlab [1]{#1}%
\providecommand \enquote  [1]{``#1''}%
\providecommand \bibnamefont  [1]{#1}%
\providecommand \bibfnamefont [1]{#1}%
\providecommand \citenamefont [1]{#1}%
\providecommand \href@noop [0]{\@secondoftwo}%
\providecommand \href [0]{\begingroup \@sanitize@url \@href}%
\providecommand \@href[1]{\@@startlink{#1}\@@href}%
\providecommand \@@href[1]{\endgroup#1\@@endlink}%
\providecommand \@sanitize@url [0]{\catcode `\\12\catcode `\$12\catcode
  `\&12\catcode `\#12\catcode `\^12\catcode `\_12\catcode `\%12\relax}%
\providecommand \@@startlink[1]{}%
\providecommand \@@endlink[0]{}%
\providecommand \url  [0]{\begingroup\@sanitize@url \@url }%
\providecommand \@url [1]{\endgroup\@href {#1}{\urlprefix }}%
\providecommand \urlprefix  [0]{URL }%
\providecommand \Eprint [0]{\href }%
\providecommand \doibase [0]{http://dx.doi.org/}%
\providecommand \selectlanguage [0]{\@gobble}%
\providecommand \bibinfo  [0]{\@secondoftwo}%
\providecommand \bibfield  [0]{\@secondoftwo}%
\providecommand \translation [1]{[#1]}%
\providecommand \BibitemOpen [0]{}%
\providecommand \bibitemStop [0]{}%
\providecommand \bibitemNoStop [0]{.\EOS\space}%
\providecommand \EOS [0]{\spacefactor3000\relax}%
\providecommand \BibitemShut  [1]{\csname bibitem#1\endcsname}%
\let\auto@bib@innerbib\@empty
\bibitem [{\citenamefont {Young}\ \emph {et~al.}(2018)\citenamefont {Young},
  \citenamefont {Sarovar},\ and\ \citenamefont {L{\'e}onard}}]{young2018}%
  \BibitemOpen
  \bibfield  {author} {\bibinfo {author} {\bibfnamefont {S.~M.}\ \bibnamefont
  {Young}}, \bibinfo {author} {\bibfnamefont {M.}~\bibnamefont {Sarovar}}, \
  and\ \bibinfo {author} {\bibfnamefont {F.}~\bibnamefont {L{\'e}onard}},\
  }\href@noop {} {\bibfield  {journal} {\bibinfo  {journal} {Phys. Rev. A}\
  }\textbf {\bibinfo {volume} {97}},\ \bibinfo {pages} {033836} (\bibinfo
  {year} {2018})}\BibitemShut {NoStop}%
\bibitem [{\citenamefont {Sunter}\ and\ \citenamefont
  {Berggren}(2018)}]{sunter2018}%
  \BibitemOpen
  \bibfield  {author} {\bibinfo {author} {\bibfnamefont {K.~A.}\ \bibnamefont
  {Sunter}}\ and\ \bibinfo {author} {\bibfnamefont {K.~K.}\ \bibnamefont
  {Berggren}},\ }\href@noop {} {\bibfield  {journal} {\bibinfo  {journal}
  {Appl. Opt}\ }\textbf {\bibinfo {volume} {57}},\ \bibinfo {pages} {4872}
  (\bibinfo {year} {2018})}\BibitemShut {NoStop}%
\bibitem [{\citenamefont {Gemmell}\ \emph {et~al.}(2017)\citenamefont
  {Gemmell}, \citenamefont {Hills}, \citenamefont {Bradshaw}, \citenamefont
  {Rawlings}, \citenamefont {Green}, \citenamefont {Heath}, \citenamefont
  {Tsimvrakidis}, \citenamefont {Dobrovolskiy}, \citenamefont {Zwiller},
  \citenamefont {Dorenbos}, \citenamefont {Crook},\ and\ \citenamefont
  {Hadfield}}]{gemmell2017}%
  \BibitemOpen
  \bibfield  {author} {\bibinfo {author} {\bibfnamefont {N.~R.}\ \bibnamefont
  {Gemmell}}, \bibinfo {author} {\bibfnamefont {M.}~\bibnamefont {Hills}},
  \bibinfo {author} {\bibfnamefont {T.}~\bibnamefont {Bradshaw}}, \bibinfo
  {author} {\bibfnamefont {T.}~\bibnamefont {Rawlings}}, \bibinfo {author}
  {\bibfnamefont {B.}~\bibnamefont {Green}}, \bibinfo {author} {\bibfnamefont
  {R.~M.}\ \bibnamefont {Heath}}, \bibinfo {author} {\bibfnamefont
  {K.}~\bibnamefont {Tsimvrakidis}}, \bibinfo {author} {\bibfnamefont
  {S.}~\bibnamefont {Dobrovolskiy}}, \bibinfo {author} {\bibfnamefont
  {V.}~\bibnamefont {Zwiller}}, \bibinfo {author} {\bibfnamefont {S.~N.}\
  \bibnamefont {Dorenbos}}, \bibinfo {author} {\bibfnamefont {M.}~\bibnamefont
  {Crook}}, \ and\ \bibinfo {author} {\bibfnamefont {R.~H.}\ \bibnamefont
  {Hadfield}},\ }\href@noop {} {\bibfield  {journal} {\bibinfo  {journal}
  {Supercond. Sci. Technol.}\ }\textbf {\bibinfo {volume} {30}},\ \bibinfo
  {pages} {11LT01} (\bibinfo {year} {2017})}\BibitemShut {NoStop}%
\bibitem [{\citenamefont {{Dewitt}}\ \emph {et~al.}(1965)\citenamefont
  {{Dewitt}}, \citenamefont {{Blandin}},\ and\ \citenamefont
  {{Cohen-Tannoudji}}}]{Glauber}%
  \BibitemOpen
  \bibinfo {editor} {\bibfnamefont {C.}~\bibnamefont {{Dewitt}}}, \bibinfo
  {editor} {\bibfnamefont {A.}~\bibnamefont {{Blandin}}}, \ and\ \bibinfo
  {editor} {\bibfnamefont {C.}~\bibnamefont {{Cohen-Tannoudji}}},\ eds.,\
  \href@noop {} {\emph {\bibinfo {title} {Quantum Optics and Electronics}}}\
  (\bibinfo {year} {1965})\BibitemShut {NoStop}%
\bibitem [{\citenamefont {{Mandel}}\ and\ \citenamefont
  {{Wolf}}(1995)}]{MandelWolf}%
  \BibitemOpen
  \bibfield  {author} {\bibinfo {author} {\bibfnamefont {L.}~\bibnamefont
  {{Mandel}}}\ and\ \bibinfo {author} {\bibfnamefont {E.}~\bibnamefont
  {{Wolf}}},\ }\href@noop {} {\emph {\bibinfo {title} {{Optical Coherence and
  Quantum Optics}}}},\ \bibinfo {edition} {2nd}\ ed.\ (\bibinfo  {publisher}
  {Cambridge University Press},\ \bibinfo {year} {1995})\BibitemShut {NoStop}%
\bibitem [{\citenamefont {Kelley}\ and\ \citenamefont
  {Kleiner}(1964)}]{kelley1964}%
  \BibitemOpen
  \bibfield  {author} {\bibinfo {author} {\bibfnamefont {P.~L.}\ \bibnamefont
  {Kelley}}\ and\ \bibinfo {author} {\bibfnamefont {W.~H.}\ \bibnamefont
  {Kleiner}},\ }\href@noop {} {\bibfield  {journal} {\bibinfo  {journal} {Phys.
  Rev.}\ }\textbf {\bibinfo {volume} {136}},\ \bibinfo {pages} {A316} (\bibinfo
  {year} {1964})}\BibitemShut {NoStop}%
\bibitem [{\citenamefont {Scully}\ and\ \citenamefont
  {Lamb}(1969)}]{scully1969}%
  \BibitemOpen
  \bibfield  {author} {\bibinfo {author} {\bibfnamefont {M.~O.}\ \bibnamefont
  {Scully}}\ and\ \bibinfo {author} {\bibfnamefont {W.~E.}\ \bibnamefont
  {Lamb}},\ }\href@noop {} {\bibfield  {journal} {\bibinfo  {journal} {Phys.
  Rev.}\ }\textbf {\bibinfo {volume} {179}},\ \bibinfo {pages} {368} (\bibinfo
  {year} {1969})}\BibitemShut {NoStop}%
\bibitem [{\citenamefont {Yurke}\ and\ \citenamefont
  {Denker}(1984)}]{yurke1984}%
  \BibitemOpen
  \bibfield  {author} {\bibinfo {author} {\bibfnamefont {B.}~\bibnamefont
  {Yurke}}\ and\ \bibinfo {author} {\bibfnamefont {J.~S.}\ \bibnamefont
  {Denker}},\ }\href@noop {} {\bibfield  {journal} {\bibinfo  {journal} {Phys.
  Rev. A}\ }\textbf {\bibinfo {volume} {29}},\ \bibinfo {pages} {1419}
  (\bibinfo {year} {1984})}\BibitemShut {NoStop}%
\bibitem [{\citenamefont {Ueda}(1999)}]{ueda1999}%
  \BibitemOpen
  \bibfield  {author} {\bibinfo {author} {\bibfnamefont {M.}~\bibnamefont
  {Ueda}},\ }\href@noop {} {\bibfield  {journal} {\bibinfo  {journal} {Quantum
  Opt.: J. Eur. Opt. Soc. B}\ }\textbf {\bibinfo {volume} {1}},\ \bibinfo
  {pages} {131} (\bibinfo {year} {1999})}\BibitemShut {NoStop}%
\bibitem [{\citenamefont {Schuster}\ \emph {et~al.}(2005)\citenamefont
  {Schuster}, \citenamefont {Wallraff}, \citenamefont {Blais}, \citenamefont
  {Frunzio}, \citenamefont {Huang}, \citenamefont {Majer}, \citenamefont
  {Girvin}, \citenamefont {Schoelkopf},\ and\ \citenamefont
  {{RJ}}}]{schuster2005}%
  \BibitemOpen
  \bibfield  {author} {\bibinfo {author} {\bibfnamefont {D.~I.}\ \bibnamefont
  {Schuster}}, \bibinfo {author} {\bibfnamefont {A.}~\bibnamefont {Wallraff}},
  \bibinfo {author} {\bibfnamefont {A.}~\bibnamefont {Blais}}, \bibinfo
  {author} {\bibfnamefont {L.}~\bibnamefont {Frunzio}}, \bibinfo {author}
  {\bibfnamefont {R.-S.}\ \bibnamefont {Huang}}, \bibinfo {author}
  {\bibfnamefont {J.}~\bibnamefont {Majer}}, \bibinfo {author} {\bibfnamefont
  {S.~M.}\ \bibnamefont {Girvin}}, \bibinfo {author} {\bibnamefont
  {Schoelkopf}}, \ and\ \bibinfo {author} {\bibnamefont {{RJ}}},\ }\href@noop
  {} {\bibfield  {journal} {\bibinfo  {journal} {Phys. Rev. Lett.}\ }\textbf
  {\bibinfo {volume} {94}},\ \bibinfo {pages} {123602} (\bibinfo {year}
  {2005})}\BibitemShut {NoStop}%
\bibitem [{\citenamefont {Clerk}\ \emph {et~al.}(2010)\citenamefont {Clerk},
  \citenamefont {Devoret}, \citenamefont {Girvin}, \citenamefont {Marquardt},\
  and\ \citenamefont {Schoelkopf}}]{clerk2010}%
  \BibitemOpen
  \bibfield  {author} {\bibinfo {author} {\bibfnamefont {A.~A.}\ \bibnamefont
  {Clerk}}, \bibinfo {author} {\bibfnamefont {M.~H.}\ \bibnamefont {Devoret}},
  \bibinfo {author} {\bibfnamefont {S.~M.}\ \bibnamefont {Girvin}}, \bibinfo
  {author} {\bibfnamefont {F.}~\bibnamefont {Marquardt}}, \ and\ \bibinfo
  {author} {\bibfnamefont {R.~J.}\ \bibnamefont {Schoelkopf}},\ }\href@noop {}
  {\bibfield  {journal} {\bibinfo  {journal} {Rev. Mod. Phys.}\ }\textbf
  {\bibinfo {volume} {82}},\ \bibinfo {pages} {1155} (\bibinfo {year}
  {2010})}\BibitemShut {NoStop}%
\bibitem [{\citenamefont {Propp}\ and\ \citenamefont {van
  Enk}(2019)}]{proppamp}%
  \BibitemOpen
  \bibfield  {author} {\bibinfo {author} {\bibfnamefont {{\relax Tz}.~B.}\
  \bibnamefont {Propp}}\ and\ \bibinfo {author} {\bibfnamefont {S.~J.}\
  \bibnamefont {van Enk}},\ }\href@noop {} {\bibfield  {journal} {\bibinfo
  {journal} {Opt. Express}\ }\textbf {\bibinfo {volume} {27}},\ \bibinfo
  {pages} {23454} (\bibinfo {year} {2019})}\BibitemShut {NoStop}%
\bibitem [{\citenamefont {Helmer}\ \emph {et~al.}(2009)\citenamefont {Helmer},
  \citenamefont {Mariantoni}, \citenamefont {Solano},\ and\ \citenamefont
  {Marquardt}}]{helmer2009}%
  \BibitemOpen
  \bibfield  {author} {\bibinfo {author} {\bibfnamefont {F.}~\bibnamefont
  {Helmer}}, \bibinfo {author} {\bibfnamefont {M.}~\bibnamefont {Mariantoni}},
  \bibinfo {author} {\bibfnamefont {E.}~\bibnamefont {Solano}}, \ and\ \bibinfo
  {author} {\bibfnamefont {F.}~\bibnamefont {Marquardt}},\ }\href@noop {}
  {\bibfield  {journal} {\bibinfo  {journal} {Phys. Rev. A}\ }\textbf {\bibinfo
  {volume} {79}},\ \bibinfo {pages} {052115} (\bibinfo {year}
  {2009})}\BibitemShut {NoStop}%
\bibitem [{\citenamefont {Nehra}\ \emph {et~al.}(2017)\citenamefont {Nehra},
  \citenamefont {Chang}, \citenamefont {Beling},\ and\ \citenamefont
  {Pfister}}]{photonnumber2017}%
  \BibitemOpen
  \bibfield  {author} {\bibinfo {author} {\bibfnamefont {R.}~\bibnamefont
  {Nehra}}, \bibinfo {author} {\bibfnamefont {C.-H.}\ \bibnamefont {Chang}},
  \bibinfo {author} {\bibfnamefont {A.}~\bibnamefont {Beling}}, \ and\ \bibinfo
  {author} {\bibfnamefont {O.}~\bibnamefont {Pfister}},\ }\href@noop {}
  {\bibfield  {journal} {\bibinfo  {journal} {arXiv preprint arXiv:1708.09015}\
  } (\bibinfo {year} {2017})}\BibitemShut {NoStop}%
\bibitem [{\citenamefont {Migdall}\ \emph {et~al.}(2003)\citenamefont
  {Migdall}, \citenamefont {Castelletto},\ and\ \citenamefont
  {Ware}}]{migdall2003}%
  \BibitemOpen
  \bibfield  {author} {\bibinfo {author} {\bibfnamefont {A.~L.}\ \bibnamefont
  {Migdall}}, \bibinfo {author} {\bibfnamefont {S.}~\bibnamefont
  {Castelletto}}, \ and\ \bibinfo {author} {\bibfnamefont {M.~J.}\ \bibnamefont
  {Ware}},\ }in\ \href@noop {} {\emph {\bibinfo {booktitle} {Quantum
  {Information} and {Computation}}}},\ Vol.\ \bibinfo {volume} {5105}\
  (\bibinfo  {publisher} {International Society for Optics and Photonics},\
  \bibinfo {year} {2003})\ pp.\ \bibinfo {pages} {294--303}\BibitemShut
  {NoStop}%
\bibitem [{\citenamefont {Combes}\ \emph {et~al.}(2017)\citenamefont {Combes},
  \citenamefont {Kerckhoff},\ and\ \citenamefont {Sarovar}}]{combes2017}%
  \BibitemOpen
  \bibfield  {author} {\bibinfo {author} {\bibfnamefont {J.}~\bibnamefont
  {Combes}}, \bibinfo {author} {\bibfnamefont {J.}~\bibnamefont {Kerckhoff}}, \
  and\ \bibinfo {author} {\bibfnamefont {M.}~\bibnamefont {Sarovar}},\
  }\href@noop {} {\bibfield  {journal} {\bibinfo  {journal} {Adv. Phys: X}\
  }\textbf {\bibinfo {volume} {2}},\ \bibinfo {pages} {784} (\bibinfo {year}
  {2017})}\BibitemShut {NoStop}%
\bibitem [{\citenamefont {Fan}\ \emph {et~al.}(2010)\citenamefont {Fan},
  \citenamefont {Kocaba{\c{s}}},\ and\ \citenamefont {Shen}}]{fan2010}%
  \BibitemOpen
  \bibfield  {author} {\bibinfo {author} {\bibfnamefont {S.}~\bibnamefont
  {Fan}}, \bibinfo {author} {\bibfnamefont {{\c{S}}.~E.}\ \bibnamefont
  {Kocaba{\c{s}}}}, \ and\ \bibinfo {author} {\bibfnamefont {J.-T.}\
  \bibnamefont {Shen}},\ }\href@noop {} {\bibfield  {journal} {\bibinfo
  {journal} {Phys. Rev. A}\ }\textbf {\bibinfo {volume} {82}},\ \bibinfo
  {pages} {063821} (\bibinfo {year} {2010})}\BibitemShut {NoStop}%
\bibitem [{\citenamefont {Caneva}\ \emph {et~al.}(2015)\citenamefont {Caneva},
  \citenamefont {Manzoni}, \citenamefont {Shi}, \citenamefont {Douglas},
  \citenamefont {Cirac},\ and\ \citenamefont {Chang}}]{caneva2015}%
  \BibitemOpen
  \bibfield  {author} {\bibinfo {author} {\bibfnamefont {T.}~\bibnamefont
  {Caneva}}, \bibinfo {author} {\bibfnamefont {M.~T.}\ \bibnamefont {Manzoni}},
  \bibinfo {author} {\bibfnamefont {T.}~\bibnamefont {Shi}}, \bibinfo {author}
  {\bibfnamefont {J.~S.}\ \bibnamefont {Douglas}}, \bibinfo {author}
  {\bibfnamefont {J.~I.}\ \bibnamefont {Cirac}}, \ and\ \bibinfo {author}
  {\bibfnamefont {D.~E.}\ \bibnamefont {Chang}},\ }\href@noop {} {\bibfield
  {journal} {\bibinfo  {journal} {New J. Phys.}\ }\textbf {\bibinfo {volume}
  {17}},\ \bibinfo {pages} {113001} (\bibinfo {year} {2015})}\BibitemShut
  {NoStop}%
\bibitem [{\citenamefont {Xu}\ and\ \citenamefont {Fan}(2015)}]{xu2015}%
  \BibitemOpen
  \bibfield  {author} {\bibinfo {author} {\bibfnamefont {S.}~\bibnamefont
  {Xu}}\ and\ \bibinfo {author} {\bibfnamefont {S.}~\bibnamefont {Fan}},\
  }\href@noop {} {\bibfield  {journal} {\bibinfo  {journal} {Phys. Rev. A}\
  }\textbf {\bibinfo {volume} {91}},\ \bibinfo {pages} {043845} (\bibinfo
  {year} {2015})}\BibitemShut {NoStop}%
\bibitem [{\citenamefont {O’Sullivan}\ \emph {et~al.}(2012)\citenamefont
  {O’Sullivan}, \citenamefont {Mirhosseini}, \citenamefont {Malik},\ and\
  \citenamefont {Boyd}}]{osullivan2012}%
  \BibitemOpen
  \bibfield  {author} {\bibinfo {author} {\bibfnamefont {M.~N.}\ \bibnamefont
  {O’Sullivan}}, \bibinfo {author} {\bibfnamefont {M.}~\bibnamefont
  {Mirhosseini}}, \bibinfo {author} {\bibfnamefont {M.}~\bibnamefont {Malik}},
  \ and\ \bibinfo {author} {\bibfnamefont {R.~W.}\ \bibnamefont {Boyd}},\
  }\href@noop {} {\bibfield  {journal} {\bibinfo  {journal} {Opt. Express}\
  }\textbf {\bibinfo {volume} {20}},\ \bibinfo {pages} {24444} (\bibinfo {year}
  {2012})}\BibitemShut {NoStop}%
\bibitem [{\citenamefont {Bouchard}\ \emph {et~al.}(2018)\citenamefont
  {Bouchard}, \citenamefont {Valencia}, \citenamefont {Brandt}, \citenamefont
  {Fickler}, \citenamefont {Huber},\ and\ \citenamefont
  {Malik}}]{Bouchard2018}%
  \BibitemOpen
  \bibfield  {author} {\bibinfo {author} {\bibfnamefont {F.}~\bibnamefont
  {Bouchard}}, \bibinfo {author} {\bibfnamefont {N.~H.}\ \bibnamefont
  {Valencia}}, \bibinfo {author} {\bibfnamefont {F.}~\bibnamefont {Brandt}},
  \bibinfo {author} {\bibfnamefont {R.}~\bibnamefont {Fickler}}, \bibinfo
  {author} {\bibfnamefont {M.}~\bibnamefont {Huber}}, \ and\ \bibinfo {author}
  {\bibfnamefont {M.}~\bibnamefont {Malik}},\ }\href@noop {} {\bibfield
  {journal} {\bibinfo  {journal} {Opt. Express}\ }\textbf {\bibinfo {volume}
  {26}},\ \bibinfo {pages} {31925} (\bibinfo {year} {2018})}\BibitemShut
  {NoStop}%
\bibitem [{\citenamefont {Fontaine}\ \emph {et~al.}(2019)\citenamefont
  {Fontaine}, \citenamefont {Ryf}, \citenamefont {Chen}, \citenamefont
  {Neilson}, \citenamefont {Kim},\ and\ \citenamefont
  {Carpenter}}]{Fontaine2019}%
  \BibitemOpen
  \bibfield  {author} {\bibinfo {author} {\bibfnamefont {N.~K.}\ \bibnamefont
  {Fontaine}}, \bibinfo {author} {\bibfnamefont {R.}~\bibnamefont {Ryf}},
  \bibinfo {author} {\bibfnamefont {H.}~\bibnamefont {Chen}}, \bibinfo {author}
  {\bibfnamefont {D.~T.}\ \bibnamefont {Neilson}}, \bibinfo {author}
  {\bibfnamefont {K.}~\bibnamefont {Kim}}, \ and\ \bibinfo {author}
  {\bibfnamefont {J.}~\bibnamefont {Carpenter}},\ }\href@noop {} {\bibfield
  {journal} {\bibinfo  {journal} {Nat. Commun.}\ }\textbf {\bibinfo {volume}
  {10}},\ \bibinfo {pages} {1865} (\bibinfo {year} {2019})}\BibitemShut
  {NoStop}%
\bibitem [{Note1()}]{Note1}%
  \BibitemOpen
  \bibinfo {note} {To describe sorting as well within this framework, we simply
  write more transmission functions, e.g. $T_j(\omega )$ for all the different
  input continua $j$, each leading to their own output continuum. Even more
  complicatedly, we could consider multiple outputs $i$ for a given input $j$
  and write $T_{i|j}(\omega )$. But even in this case we can focus on a
  particular $i$ and $j$ as we do this in this paper.}\BibitemShut {Stop}%
\bibitem [{\citenamefont {Tellinghuisen}(1975)}]{Tellinghuisen1975}%
  \BibitemOpen
  \bibfield  {author} {\bibinfo {author} {\bibfnamefont {J.}~\bibnamefont
  {Tellinghuisen}},\ }\href {\doibase 10.1103/physrevlett.34.1137} {\bibfield
  {journal} {\bibinfo  {journal} {Phys. Rev. Lett.}\ }\textbf {\bibinfo
  {volume} {34}},\ \bibinfo {pages} {1137} (\bibinfo {year}
  {1975})}\BibitemShut {NoStop}%
\bibitem [{\citenamefont {Odegard}(2002)}]{Odegard2002}%
  \BibitemOpen
  \bibfield  {author} {\bibinfo {author} {\bibfnamefont {G.}~\bibnamefont
  {Odegard}},\ }\href@noop {} {\bibfield  {journal} {\bibinfo  {journal}
  {Compos. Sci. Technol.}\ }\textbf {\bibinfo {volume} {62}},\ \bibinfo {pages}
  {1869} (\bibinfo {year} {2002})}\BibitemShut {NoStop}%
\bibitem [{\citenamefont {Nygaard}\ \emph {et~al.}(2008)\citenamefont
  {Nygaard}, \citenamefont {Piil},\ and\ \citenamefont
  {M{\o}lmer}}]{Nygaard2008}%
  \BibitemOpen
  \bibfield  {author} {\bibinfo {author} {\bibfnamefont {N.}~\bibnamefont
  {Nygaard}}, \bibinfo {author} {\bibfnamefont {R.}~\bibnamefont {Piil}}, \
  and\ \bibinfo {author} {\bibfnamefont {K.}~\bibnamefont {M{\o}lmer}},\
  }\href@noop {} {\ \textbf {\bibinfo {volume} {78}},\ \bibinfo {pages}
  {023617} (\bibinfo {year} {2008})}\BibitemShut {NoStop}%
\bibitem [{\citenamefont {Garraway}(1997{\natexlab{a}})}]{pseudomodes2}%
  \BibitemOpen
  \bibfield  {author} {\bibinfo {author} {\bibfnamefont {B.~M.}\ \bibnamefont
  {Garraway}},\ }\href
  {https://journals.aps.org/pra/abstract/10.1103/PhysRevA.55.2290} {\bibfield
  {journal} {\bibinfo  {journal} {Phys. Rev. A}\ }\textbf {\bibinfo {volume}
  {55}},\ \bibinfo {pages} {2290} (\bibinfo {year}
  {1997}{\natexlab{a}})}\BibitemShut {NoStop}%
\bibitem [{\citenamefont {Garraway}(1997{\natexlab{b}})}]{pseudomodes3}%
  \BibitemOpen
  \bibfield  {author} {\bibinfo {author} {\bibfnamefont {B.~M.}\ \bibnamefont
  {Garraway}},\ }\href@noop {} {\bibfield  {journal} {\bibinfo  {journal}
  {Phys. Rev. A}\ }\textbf {\bibinfo {volume} {55}},\ \bibinfo {pages} {4636}
  (\bibinfo {year} {1997}{\natexlab{b}})}\BibitemShut {NoStop}%
\bibitem [{\citenamefont {Dalton}\ \emph {et~al.}(2001)\citenamefont {Dalton},
  \citenamefont {Barnett},\ and\ \citenamefont {Garraway}}]{pseudomodes4}%
  \BibitemOpen
  \bibfield  {author} {\bibinfo {author} {\bibfnamefont {B.~J.}\ \bibnamefont
  {Dalton}}, \bibinfo {author} {\bibfnamefont {S.~M.}\ \bibnamefont {Barnett}},
  \ and\ \bibinfo {author} {\bibfnamefont {B.~M.}\ \bibnamefont {Garraway}},\
  }\href@noop {} {\bibfield  {journal} {\bibinfo  {journal} {Phys. Rev. A}\
  }\textbf {\bibinfo {volume} {64}},\ \bibinfo {pages} {053813} (\bibinfo
  {year} {2001})}\BibitemShut {NoStop}%
\bibitem [{\citenamefont {Mazzola}\ \emph {et~al.}(2008)\citenamefont
  {Mazzola}, \citenamefont {Maniscalco}, \citenamefont {Piilo}, \citenamefont
  {Suominen},\ and\ \citenamefont {Garraway}}]{pseudomodes1}%
  \BibitemOpen
  \bibfield  {author} {\bibinfo {author} {\bibfnamefont {L.}~\bibnamefont
  {Mazzola}}, \bibinfo {author} {\bibfnamefont {S.}~\bibnamefont {Maniscalco}},
  \bibinfo {author} {\bibfnamefont {J.}~\bibnamefont {Piilo}}, \bibinfo
  {author} {\bibfnamefont {K.~A.}\ \bibnamefont {Suominen}}, \ and\ \bibinfo
  {author} {\bibfnamefont {B.~M.}\ \bibnamefont {Garraway}},\ }\href@noop {}
  {\emph {\bibinfo {title} {Pseudomodes as an effective description of
  memory}}},\ \bibinfo {type} {Tech. Rep.}\ (\bibinfo {year}
  {2008})\BibitemShut {NoStop}%
\bibitem [{\citenamefont {Stinespring}(1955)}]{stinespring}%
  \BibitemOpen
  \bibfield  {author} {\bibinfo {author} {\bibfnamefont {W.~F.}\ \bibnamefont
  {Stinespring}},\ }\href {\doibase 10.2307/2032342} {\bibfield  {journal}
  {\bibinfo  {journal} {P. Am. Math. Soc.}\ }\textbf {\bibinfo {volume} {6}},\
  \bibinfo {pages} {211} (\bibinfo {year} {1955})}\BibitemShut {NoStop}%
\bibitem [{\citenamefont {Gardiner}\ and\ \citenamefont
  {Collett}(1985)}]{input1985}%
  \BibitemOpen
  \bibfield  {author} {\bibinfo {author} {\bibfnamefont {C.~W.}\ \bibnamefont
  {Gardiner}}\ and\ \bibinfo {author} {\bibfnamefont {M.~J.}\ \bibnamefont
  {Collett}},\ }\href@noop {} {\bibfield  {journal} {\bibinfo  {journal} {Phys.
  Rev. A}\ }\textbf {\bibinfo {volume} {31}},\ \bibinfo {pages} {3761}
  (\bibinfo {year} {1985})}\BibitemShut {NoStop}%
\bibitem [{Note2()}]{Note2}%
  \BibitemOpen
  \bibinfo {note} {Timing jitter for a photo detection has three contributing
  components \cite {vanenk2017}. The first comes from the temporal spread of
  the mode onto which the measurement projects \cite {spectralPOVM} and is
  intrinsic to the resonance structure of the photo detector. The second comes
  from integrating a continuously monitored continua to form discrete detection
  event, which is necessary for an accurate information-theoretic
  characterization of photo detection and influences photo detection efficiency
  (see Appendix A). It is this part of the jitter that $\protect \mathaccentV
  {tilde}07E{\Gamma }^{-1}$ is a lower bound for; it sets the timescale for
  integration where monochromatic states are detected with high efficiency. The
  third contribution comes from input signals with long temporal wave-packets
  and, since it is input-dependent, will be ignored in this analysis since here
  we assume no priors about the single photon input.}\BibitemShut {Stop}%
\bibitem [{\citenamefont {van Enk}(2017{\natexlab{a}})}]{spectralPOVM}%
  \BibitemOpen
  \bibfield  {author} {\bibinfo {author} {\bibfnamefont {S.~J.}\ \bibnamefont
  {van Enk}},\ }\href@noop {} {\bibfield  {journal} {\bibinfo  {journal} {Phys.
  Rev. A}\ }\textbf {\bibinfo {volume} {96}},\ \bibinfo {pages} {033834}
  (\bibinfo {year} {2017}{\natexlab{a}})}\BibitemShut {NoStop}%
\bibitem [{\citenamefont {Yariv}\ \emph {et~al.}(1999)\citenamefont {Yariv},
  \citenamefont {Xu}, \citenamefont {Lee},\ and\ \citenamefont
  {Scherer}}]{yariv1999}%
  \BibitemOpen
  \bibfield  {author} {\bibinfo {author} {\bibfnamefont {A.}~\bibnamefont
  {Yariv}}, \bibinfo {author} {\bibfnamefont {Y.}~\bibnamefont {Xu}}, \bibinfo
  {author} {\bibfnamefont {R.~K.}\ \bibnamefont {Lee}}, \ and\ \bibinfo
  {author} {\bibfnamefont {A.}~\bibnamefont {Scherer}},\ }\href@noop {}
  {\bibfield  {journal} {\bibinfo  {journal} {Opt. Lett.}\ }\textbf {\bibinfo
  {volume} {24}},\ \bibinfo {pages} {711} (\bibinfo {year} {1999})}\BibitemShut
  {NoStop}%
\bibitem [{\citenamefont {Poon}\ \emph {et~al.}(2004)\citenamefont {Poon},
  \citenamefont {Scheuer}, \citenamefont {Xu},\ and\ \citenamefont
  {Yariv}}]{poon2004}%
  \BibitemOpen
  \bibfield  {author} {\bibinfo {author} {\bibfnamefont {J.~K.}\ \bibnamefont
  {Poon}}, \bibinfo {author} {\bibfnamefont {J.}~\bibnamefont {Scheuer}},
  \bibinfo {author} {\bibfnamefont {Y.}~\bibnamefont {Xu}}, \ and\ \bibinfo
  {author} {\bibfnamefont {A.}~\bibnamefont {Yariv}},\ }\href@noop {}
  {\bibfield  {journal} {\bibinfo  {journal} {J. Opt. Soc. Am. B}\ }\textbf
  {\bibinfo {volume} {21}},\ \bibinfo {pages} {1665} (\bibinfo {year}
  {2004})}\BibitemShut {NoStop}%
\bibitem [{\citenamefont {Madsen}(2000)}]{madsen2000}%
  \BibitemOpen
  \bibfield  {author} {\bibinfo {author} {\bibfnamefont {C.~K.}\ \bibnamefont
  {Madsen}},\ }\href@noop {} {\bibfield  {journal} {\bibinfo  {journal} {J.
  Light. Technol.}\ }\textbf {\bibinfo {volume} {18}},\ \bibinfo {pages} {860}
  (\bibinfo {year} {2000})}\BibitemShut {NoStop}%
\bibitem [{\citenamefont {van Enk}(2017{\natexlab{b}})}]{vanenk2017}%
  \BibitemOpen
  \bibfield  {author} {\bibinfo {author} {\bibfnamefont {S.~J.}\ \bibnamefont
  {van Enk}},\ }\href@noop {} {\bibfield  {journal} {\bibinfo  {journal} {J.
  Phys. Comm.}\ }\textbf {\bibinfo {volume} {1}},\ \bibinfo {pages} {045001}
  (\bibinfo {year} {2017}{\natexlab{b}})}\BibitemShut {NoStop}%
\bibitem [{\citenamefont {Cirac}\ \emph {et~al.}(1997)\citenamefont {Cirac},
  \citenamefont {Zoller}, \citenamefont {Kimble},\ and\ \citenamefont
  {Mabuchi}}]{Cirac1997}%
  \BibitemOpen
  \bibfield  {author} {\bibinfo {author} {\bibfnamefont {J.~I.}\ \bibnamefont
  {Cirac}}, \bibinfo {author} {\bibfnamefont {P.}~\bibnamefont {Zoller}},
  \bibinfo {author} {\bibfnamefont {H.~J.}\ \bibnamefont {Kimble}}, \ and\
  \bibinfo {author} {\bibfnamefont {H.}~\bibnamefont {Mabuchi}},\ }\href@noop
  {} {\bibfield  {journal} {\bibinfo  {journal} {Phys. Rev. Lett.}\ }\textbf
  {\bibinfo {volume} {78}},\ \bibinfo {pages} {3221} (\bibinfo {year}
  {1997})}\BibitemShut {NoStop}%
\bibitem [{\citenamefont {van Enk}\ and\ \citenamefont
  {Kimble}(2000)}]{vanEnk2000}%
  \BibitemOpen
  \bibfield  {author} {\bibinfo {author} {\bibfnamefont {S.~J.}\ \bibnamefont
  {van Enk}}\ and\ \bibinfo {author} {\bibfnamefont {H.~J.}\ \bibnamefont
  {Kimble}},\ }\href@noop {} {\bibfield  {journal} {\bibinfo  {journal} {Phys.
  Rev. A}\ }\textbf {\bibinfo {volume} {61}},\ \bibinfo {pages} {051802(R)}
  (\bibinfo {year} {2000})}\BibitemShut {NoStop}%
\bibitem [{\citenamefont {Tey}\ \emph {et~al.}(2008)\citenamefont {Tey},
  \citenamefont {Chen}, \citenamefont {Aljunid}, \citenamefont {Chng},
  \citenamefont {Huber}, \citenamefont {Maslennikov},\ and\ \citenamefont
  {Kurtsiefer}}]{Tey2008}%
  \BibitemOpen
  \bibfield  {author} {\bibinfo {author} {\bibfnamefont {M.~K.}\ \bibnamefont
  {Tey}}, \bibinfo {author} {\bibfnamefont {Z.}~\bibnamefont {Chen}}, \bibinfo
  {author} {\bibfnamefont {S.~A.}\ \bibnamefont {Aljunid}}, \bibinfo {author}
  {\bibfnamefont {B.}~\bibnamefont {Chng}}, \bibinfo {author} {\bibfnamefont
  {F.}~\bibnamefont {Huber}}, \bibinfo {author} {\bibfnamefont
  {G.}~\bibnamefont {Maslennikov}}, \ and\ \bibinfo {author} {\bibfnamefont
  {C.}~\bibnamefont {Kurtsiefer}},\ }\href@noop {} {\bibfield  {journal}
  {\bibinfo  {journal} {Nat. Phys.}\ }\textbf {\bibinfo {volume} {4}},\
  \bibinfo {pages} {924} (\bibinfo {year} {2008})}\BibitemShut {NoStop}%
\bibitem [{\citenamefont {Pinotsi}\ and\ \citenamefont
  {Imamoglu}(2008)}]{Pinotsi2008}%
  \BibitemOpen
  \bibfield  {author} {\bibinfo {author} {\bibfnamefont {D.}~\bibnamefont
  {Pinotsi}}\ and\ \bibinfo {author} {\bibfnamefont {A.}~\bibnamefont
  {Imamoglu}},\ }\href@noop {} {\bibfield  {journal} {\bibinfo  {journal}
  {Phys. Rev. Lett.}\ }\textbf {\bibinfo {volume} {100}},\ \bibinfo {pages}
  {093603} (\bibinfo {year} {2008})}\BibitemShut {NoStop}%
\bibitem [{\citenamefont {Wang}\ \emph {et~al.}(2011)\citenamefont {Wang},
  \citenamefont {Min{\'{a}}{\v{r}}}, \citenamefont {Sheridan},\ and\
  \citenamefont {Scarani}}]{Wang2011}%
  \BibitemOpen
  \bibfield  {author} {\bibinfo {author} {\bibfnamefont {Y.}~\bibnamefont
  {Wang}}, \bibinfo {author} {\bibfnamefont {J.}~\bibnamefont
  {Min{\'{a}}{\v{r}}}}, \bibinfo {author} {\bibfnamefont {L.}~\bibnamefont
  {Sheridan}}, \ and\ \bibinfo {author} {\bibfnamefont {V.}~\bibnamefont
  {Scarani}},\ }\href@noop {} {\bibfield  {journal} {\bibinfo  {journal} {Phys.
  Rev. A}\ }\textbf {\bibinfo {volume} {83}},\ \bibinfo {pages} {063842}
  (\bibinfo {year} {2011})}\BibitemShut {NoStop}%
\bibitem [{\citenamefont {Wrigge}\ \emph {et~al.}(2007)\citenamefont {Wrigge},
  \citenamefont {Gerhardt}, \citenamefont {Hwang}, \citenamefont {Zumofen},\
  and\ \citenamefont {Sandoghdar}}]{Wrigge2007}%
  \BibitemOpen
  \bibfield  {author} {\bibinfo {author} {\bibfnamefont {G.}~\bibnamefont
  {Wrigge}}, \bibinfo {author} {\bibfnamefont {I.}~\bibnamefont {Gerhardt}},
  \bibinfo {author} {\bibfnamefont {J.}~\bibnamefont {Hwang}}, \bibinfo
  {author} {\bibfnamefont {G.}~\bibnamefont {Zumofen}}, \ and\ \bibinfo
  {author} {\bibfnamefont {V.}~\bibnamefont {Sandoghdar}},\ }\href@noop {}
  {\bibfield  {journal} {\bibinfo  {journal} {Nat. Phys.}\ }\textbf {\bibinfo
  {volume} {4}},\ \bibinfo {pages} {60} (\bibinfo {year} {2007})}\BibitemShut
  {NoStop}%
\bibitem [{\citenamefont {Harmon}\ and\ \citenamefont
  {Flatte}(2019)}]{iowaPOVM}%
  \BibitemOpen
  \bibfield  {author} {\bibinfo {author} {\bibfnamefont {N.~J.}\ \bibnamefont
  {Harmon}}\ and\ \bibinfo {author} {\bibfnamefont {M.~E.}\ \bibnamefont
  {Flatte}},\ }\href@noop {} {\bibfield  {journal} {\bibinfo  {journal} {arXiv
  preprint arXiv:1906.01800}\ } (\bibinfo {year} {2019})}\BibitemShut {NoStop}%
\bibitem [{\citenamefont {Zumofen}\ \emph {et~al.}(2008)\citenamefont
  {Zumofen}, \citenamefont {Mojarad}, \citenamefont {Sandoghdar},\ and\
  \citenamefont {Agio}}]{Zumofen2008}%
  \BibitemOpen
  \bibfield  {author} {\bibinfo {author} {\bibfnamefont {G.}~\bibnamefont
  {Zumofen}}, \bibinfo {author} {\bibfnamefont {N.~M.}\ \bibnamefont
  {Mojarad}}, \bibinfo {author} {\bibfnamefont {V.}~\bibnamefont {Sandoghdar}},
  \ and\ \bibinfo {author} {\bibfnamefont {M.}~\bibnamefont {Agio}},\
  }\href@noop {} {\bibfield  {journal} {\bibinfo  {journal} {Phys. Rev. Lett.}\
  }\textbf {\bibinfo {volume} {101}},\ \bibinfo {pages} {180404} (\bibinfo
  {year} {2008})}\BibitemShut {NoStop}%
\bibitem [{\citenamefont {Chen}\ \emph {et~al.}(2011)\citenamefont {Chen},
  \citenamefont {Wubs}, \citenamefont {M{\o}rk},\ and\ \citenamefont
  {Koenderink}}]{Chen2011}%
  \BibitemOpen
  \bibfield  {author} {\bibinfo {author} {\bibfnamefont {Y.}~\bibnamefont
  {Chen}}, \bibinfo {author} {\bibfnamefont {M.}~\bibnamefont {Wubs}}, \bibinfo
  {author} {\bibfnamefont {J.}~\bibnamefont {M{\o}rk}}, \ and\ \bibinfo
  {author} {\bibfnamefont {A.~F.}\ \bibnamefont {Koenderink}},\ }\href@noop {}
  {\bibfield  {journal} {\bibinfo  {journal} {New J. Phys.}\ }\textbf {\bibinfo
  {volume} {13}},\ \bibinfo {pages} {103010} (\bibinfo {year}
  {2011})}\BibitemShut {NoStop}%
\bibitem [{Note3()}]{Note3}%
  \BibitemOpen
  \bibinfo {note} {Since both operators and their expectation values (mode
  amplitudes) will satisfy the same systems of equations, we omit hats
  throughout this paper.}\BibitemShut {Stop}%
\bibitem [{Note4()}]{Note4}%
  \BibitemOpen
  \bibinfo {note} {In assuming these couplings to be frequency independent, we
  are invoking a modified version of the first Markov approximation. Formally,
  we define $\gamma _i=2\pi \protect \tmspace +\thinmuskip {.1667em}\kappa
  _i^2(\omega _i)$ where $\omega _i$ is the resonance frequency of the $i$th
  discrete state and $\kappa _i(\omega )$ is the coupling between the $i$th
  discrete state and the left continuum at the discrete state frequency. This
  implies that, in an experiment, the decays $\gamma _i$ and resonances $\omega
  _i$ cannot be varied independently, which is well known in the context of the
  Thomas-Reiche-Kuhn sum rule for electric dipole transitions \cite
  {kuhn1992}}\BibitemShut {NoStop}%
\bibitem [{\citenamefont {Vogel}\ and\ \citenamefont
  {Welsch}(2006)}]{vogelswelsch}%
  \BibitemOpen
  \bibfield  {author} {\bibinfo {author} {\bibfnamefont {W.}~\bibnamefont
  {Vogel}}\ and\ \bibinfo {author} {\bibfnamefont {D.-G.}\ \bibnamefont
  {Welsch}},\ }\href@noop {} {\emph {\bibinfo {title} {Quantum Optics}}},\
  \bibinfo {edition} {3rd}\ ed.\ (\bibinfo  {publisher} {Wiley-VCH},\ \bibinfo
  {year} {2006})\BibitemShut {NoStop}%
\bibitem [{\citenamefont {Siegman}(1986)}]{siegman86}%
  \BibitemOpen
  \bibfield  {author} {\bibinfo {author} {\bibfnamefont {A.~E.}\ \bibnamefont
  {Siegman}},\ }\href@noop {} {\emph {\bibinfo {title} {Lasers}}}\ (\bibinfo
  {publisher} {University Science Books},\ \bibinfo {year} {1986})\BibitemShut
  {NoStop}%
\bibitem [{Note5()}]{Note5}%
  \BibitemOpen
  \bibinfo {note} {Luckily, there is some redundancy. Once one has solved for
  one $c_i$ in terms of the input and output fields, one can permute the labels
  to generate the remaining solutions.}\BibitemShut {Stop}%
\bibitem [{Note6()}]{Note6}%
  \BibitemOpen
  \bibinfo {note} {One alternative approach for a general network is to take
  the weak-coupling limit ($2\protect \tmspace +\thinmuskip {.1667em}g_{ij}\ll
  \protect \sqrt {\gamma _i\protect \tmspace +\thinmuskip {.1667em}\gamma _j} +
  \protect \sqrt {\Gamma _i\protect \tmspace +\thinmuskip {.1667em}\Gamma
  _j}\protect \tmspace +\thinmuskip {.1667em}\forall i,j$) and truncate the
  solutions after some power in $g_{ij}$, making (\ref {RGen}) the $0$th order
  approximate solution with higher order corrections. However, this method
  fails if even a single discrete state decouples from both
  continua.}\BibitemShut {Stop}%
\bibitem [{\citenamefont {Datta}(2005)}]{datta2005}%
  \BibitemOpen
  \bibfield  {author} {\bibinfo {author} {\bibfnamefont {S.}~\bibnamefont
  {Datta}},\ }\href@noop {} {\emph {\bibinfo {title} {{Quantum Transport: Atom
  to Transistor}}}},\ \bibinfo {edition} {2nd}\ ed.\ (\bibinfo  {publisher}
  {Cambridge University Press},\ \bibinfo {year} {2005})\BibitemShut {NoStop}%
\bibitem [{\citenamefont {Higginbottom}\ \emph {et~al.}(2016)\citenamefont
  {Higginbottom}, \citenamefont {Slodi{\v{c}}ka}, \citenamefont {Araneda},
  \citenamefont {Lachman}, \citenamefont {Filip}, \citenamefont {Hennrich},\
  and\ \citenamefont {Blatt}}]{Higginbottom2016}%
  \BibitemOpen
  \bibfield  {author} {\bibinfo {author} {\bibfnamefont {D.~B.}\ \bibnamefont
  {Higginbottom}}, \bibinfo {author} {\bibfnamefont {L.}~\bibnamefont
  {Slodi{\v{c}}ka}}, \bibinfo {author} {\bibfnamefont {G.}~\bibnamefont
  {Araneda}}, \bibinfo {author} {\bibfnamefont {L.}~\bibnamefont {Lachman}},
  \bibinfo {author} {\bibfnamefont {R.}~\bibnamefont {Filip}}, \bibinfo
  {author} {\bibfnamefont {M.}~\bibnamefont {Hennrich}}, \ and\ \bibinfo
  {author} {\bibfnamefont {R.}~\bibnamefont {Blatt}},\ }\href@noop {}
  {\bibfield  {journal} {\bibinfo  {journal} {New J. Phys.}\ }\textbf {\bibinfo
  {volume} {18}},\ \bibinfo {pages} {093038} (\bibinfo {year}
  {2016})}\BibitemShut {NoStop}%
\bibitem [{\citenamefont {Livache}\ \emph {et~al.}(2019)\citenamefont
  {Livache}, \citenamefont {Martinez}, \citenamefont {Goubet}, \citenamefont
  {Gr{\'{e}}boval}, \citenamefont {Qu}, \citenamefont {Chu}, \citenamefont
  {Royer}, \citenamefont {Ithurria}, \citenamefont {Silly}, \citenamefont
  {Dubertret},\ and\ \citenamefont {Lhuillier}}]{Livache2019}%
  \BibitemOpen
  \bibfield  {author} {\bibinfo {author} {\bibfnamefont {C.}~\bibnamefont
  {Livache}}, \bibinfo {author} {\bibfnamefont {B.}~\bibnamefont {Martinez}},
  \bibinfo {author} {\bibfnamefont {N.}~\bibnamefont {Goubet}}, \bibinfo
  {author} {\bibfnamefont {C.}~\bibnamefont {Gr{\'{e}}boval}}, \bibinfo
  {author} {\bibfnamefont {J.}~\bibnamefont {Qu}}, \bibinfo {author}
  {\bibfnamefont {A.}~\bibnamefont {Chu}}, \bibinfo {author} {\bibfnamefont
  {S.}~\bibnamefont {Royer}}, \bibinfo {author} {\bibfnamefont
  {S.}~\bibnamefont {Ithurria}}, \bibinfo {author} {\bibfnamefont {M.~G.}\
  \bibnamefont {Silly}}, \bibinfo {author} {\bibfnamefont {B.}~\bibnamefont
  {Dubertret}}, \ and\ \bibinfo {author} {\bibfnamefont {E.}~\bibnamefont
  {Lhuillier}},\ }\href@noop {} {\bibfield  {journal} {\bibinfo  {journal}
  {Nat. Commun.}\ }\textbf {\bibinfo {volume} {10}},\ \bibinfo {pages} {2125}
  (\bibinfo {year} {2019})}\BibitemShut {NoStop}%
\bibitem [{\citenamefont {Hughes}\ \emph {et~al.}(2018)\citenamefont {Hughes},
  \citenamefont {Richter},\ and\ \citenamefont {Knorr}}]{Hughes2018}%
  \BibitemOpen
  \bibfield  {author} {\bibinfo {author} {\bibfnamefont {S.}~\bibnamefont
  {Hughes}}, \bibinfo {author} {\bibfnamefont {M.}~\bibnamefont {Richter}}, \
  and\ \bibinfo {author} {\bibfnamefont {A.}~\bibnamefont {Knorr}},\
  }\href@noop {} {\bibfield  {journal} {\bibinfo  {journal} {Opt. Lett.}\
  }\textbf {\bibinfo {volume} {43}},\ \bibinfo {pages} {1834} (\bibinfo {year}
  {2018})}\BibitemShut {NoStop}%
\bibitem [{\citenamefont {Rosfjord}\ \emph {et~al.}(2006)\citenamefont
  {Rosfjord}, \citenamefont {Yang}, \citenamefont {Dauler}, \citenamefont
  {Kerman}, \citenamefont {Anant}, \citenamefont {Voronov}, \citenamefont
  {Goltsman},\ and\ \citenamefont {Berggren}}]{Rosfjord2006}%
  \BibitemOpen
  \bibfield  {author} {\bibinfo {author} {\bibfnamefont {K.~M.}\ \bibnamefont
  {Rosfjord}}, \bibinfo {author} {\bibfnamefont {J.~K.~W.}\ \bibnamefont
  {Yang}}, \bibinfo {author} {\bibfnamefont {E.~A.}\ \bibnamefont {Dauler}},
  \bibinfo {author} {\bibfnamefont {A.~J.}\ \bibnamefont {Kerman}}, \bibinfo
  {author} {\bibfnamefont {V.}~\bibnamefont {Anant}}, \bibinfo {author}
  {\bibfnamefont {B.~M.}\ \bibnamefont {Voronov}}, \bibinfo {author}
  {\bibfnamefont {G.~N.}\ \bibnamefont {Goltsman}}, \ and\ \bibinfo {author}
  {\bibfnamefont {K.~K.}\ \bibnamefont {Berggren}},\ }\href@noop {} {\bibfield
  {journal} {\bibinfo  {journal} {Opt. Express}\ }\textbf {\bibinfo {volume}
  {14}},\ \bibinfo {pages} {527} (\bibinfo {year} {2006})}\BibitemShut
  {NoStop}%
\bibitem [{\citenamefont {Dilley}\ \emph {et~al.}(2012)\citenamefont {Dilley},
  \citenamefont {Nisbet-Jones}, \citenamefont {Shore},\ and\ \citenamefont
  {Kuhn}}]{Dilley2012}%
  \BibitemOpen
  \bibfield  {author} {\bibinfo {author} {\bibfnamefont {J.}~\bibnamefont
  {Dilley}}, \bibinfo {author} {\bibfnamefont {P.}~\bibnamefont
  {Nisbet-Jones}}, \bibinfo {author} {\bibfnamefont {B.~W.}\ \bibnamefont
  {Shore}}, \ and\ \bibinfo {author} {\bibfnamefont {A.}~\bibnamefont {Kuhn}},\
  }\href@noop {} {\bibfield  {journal} {\bibinfo  {journal} {Phys. Rev. A}\
  }\textbf {\bibinfo {volume} {85}},\ \bibinfo {pages} {023834} (\bibinfo
  {year} {2012})}\BibitemShut {NoStop}%
\bibitem [{\citenamefont {Song}\ \emph {et~al.}(2018)\citenamefont {Song},
  \citenamefont {Munro}, \citenamefont {Nie}, \citenamefont {Kwek},
  \citenamefont {Deng},\ and\ \citenamefont {Long}}]{Song2018}%
  \BibitemOpen
  \bibfield  {author} {\bibinfo {author} {\bibfnamefont {G.-Z.}\ \bibnamefont
  {Song}}, \bibinfo {author} {\bibfnamefont {E.}~\bibnamefont {Munro}},
  \bibinfo {author} {\bibfnamefont {W.}~\bibnamefont {Nie}}, \bibinfo {author}
  {\bibfnamefont {L.-C.}\ \bibnamefont {Kwek}}, \bibinfo {author}
  {\bibfnamefont {F.-G.}\ \bibnamefont {Deng}}, \ and\ \bibinfo {author}
  {\bibfnamefont {G.-L.}\ \bibnamefont {Long}},\ }\href@noop {} {\bibfield
  {journal} {\bibinfo  {journal} {Phys. Rev. A}\ }\textbf {\bibinfo {volume}
  {98}},\ \bibinfo {pages} {023814} (\bibinfo {year} {2018})}\BibitemShut
  {NoStop}%
\bibitem [{\citenamefont {Scholes}\ \emph {et~al.}(2017)\citenamefont
  {Scholes}, \citenamefont {Fleming}, \citenamefont {Chen}, \citenamefont
  {Aspuru-Guzik}, \citenamefont {Buchleitner}, \citenamefont {Coker},
  \citenamefont {Engel}, \citenamefont {van Grondelle}, \citenamefont
  {Ishizaki}, \citenamefont {Jonas}, \citenamefont {Lundeen}, \citenamefont
  {McCusker}, \citenamefont {Mukamel}, \citenamefont {Ogilvie}, \citenamefont
  {Olaya-Castro}, \citenamefont {Ratner}, \citenamefont {Spano}, \citenamefont
  {Whaley},\ and\ \citenamefont {Zhu}}]{Scholes2017}%
  \BibitemOpen
  \bibfield  {author} {\bibinfo {author} {\bibfnamefont {G.~D.}\ \bibnamefont
  {Scholes}}, \bibinfo {author} {\bibfnamefont {G.~R.}\ \bibnamefont
  {Fleming}}, \bibinfo {author} {\bibfnamefont {L.~X.}\ \bibnamefont {Chen}},
  \bibinfo {author} {\bibfnamefont {A.}~\bibnamefont {Aspuru-Guzik}}, \bibinfo
  {author} {\bibfnamefont {A.}~\bibnamefont {Buchleitner}}, \bibinfo {author}
  {\bibfnamefont {D.~F.}\ \bibnamefont {Coker}}, \bibinfo {author}
  {\bibfnamefont {G.~S.}\ \bibnamefont {Engel}}, \bibinfo {author}
  {\bibfnamefont {R.}~\bibnamefont {van Grondelle}}, \bibinfo {author}
  {\bibfnamefont {A.}~\bibnamefont {Ishizaki}}, \bibinfo {author}
  {\bibfnamefont {D.~M.}\ \bibnamefont {Jonas}}, \bibinfo {author}
  {\bibfnamefont {J.~S.}\ \bibnamefont {Lundeen}}, \bibinfo {author}
  {\bibfnamefont {J.~K.}\ \bibnamefont {McCusker}}, \bibinfo {author}
  {\bibfnamefont {S.}~\bibnamefont {Mukamel}}, \bibinfo {author} {\bibfnamefont
  {J.~P.}\ \bibnamefont {Ogilvie}}, \bibinfo {author} {\bibfnamefont
  {A.}~\bibnamefont {Olaya-Castro}}, \bibinfo {author} {\bibfnamefont {M.~A.}\
  \bibnamefont {Ratner}}, \bibinfo {author} {\bibfnamefont {F.~C.}\
  \bibnamefont {Spano}}, \bibinfo {author} {\bibfnamefont {K.~B.}\ \bibnamefont
  {Whaley}}, \ and\ \bibinfo {author} {\bibfnamefont {X.}~\bibnamefont {Zhu}},\
  }\href@noop {} {\bibfield  {journal} {\bibinfo  {journal} {Nature}\ }\textbf
  {\bibinfo {volume} {543}},\ \bibinfo {pages} {647} (\bibinfo {year}
  {2017})}\BibitemShut {NoStop}%
\bibitem [{\citenamefont {Valleau}\ \emph {et~al.}(2017)\citenamefont
  {Valleau}, \citenamefont {Studer}, \citenamefont {H\"{a}se}, \citenamefont
  {Kreisbeck}, \citenamefont {Saer}, \citenamefont {Blankenship}, \citenamefont
  {Shakhnovich},\ and\ \citenamefont {Aspuru-Guzik}}]{Valleau2017}%
  \BibitemOpen
  \bibfield  {author} {\bibinfo {author} {\bibfnamefont {S.}~\bibnamefont
  {Valleau}}, \bibinfo {author} {\bibfnamefont {R.~A.}\ \bibnamefont {Studer}},
  \bibinfo {author} {\bibfnamefont {F.}~\bibnamefont {H\"{a}se}}, \bibinfo
  {author} {\bibfnamefont {C.}~\bibnamefont {Kreisbeck}}, \bibinfo {author}
  {\bibfnamefont {R.~G.}\ \bibnamefont {Saer}}, \bibinfo {author}
  {\bibfnamefont {R.~E.}\ \bibnamefont {Blankenship}}, \bibinfo {author}
  {\bibfnamefont {E.~I.}\ \bibnamefont {Shakhnovich}}, \ and\ \bibinfo {author}
  {\bibfnamefont {A.}~\bibnamefont {Aspuru-Guzik}},\ }\href@noop {} {\bibfield
  {journal} {\bibinfo  {journal} {ACS Cent. Sci}\ }\textbf {\bibinfo {volume}
  {3}},\ \bibinfo {pages} {1086} (\bibinfo {year} {2017})}\BibitemShut
  {NoStop}%
\bibitem [{\citenamefont {Chan}\ \emph {et~al.}(2018)\citenamefont {Chan},
  \citenamefont {Gamel}, \citenamefont {Fleming},\ and\ \citenamefont
  {Whaley}}]{Chan2018}%
  \BibitemOpen
  \bibfield  {author} {\bibinfo {author} {\bibfnamefont {H.~C.~H.}\
  \bibnamefont {Chan}}, \bibinfo {author} {\bibfnamefont {O.~E.}\ \bibnamefont
  {Gamel}}, \bibinfo {author} {\bibfnamefont {G.~R.}\ \bibnamefont {Fleming}},
  \ and\ \bibinfo {author} {\bibfnamefont {K.~B.}\ \bibnamefont {Whaley}},\
  }\href@noop {} {\bibfield  {journal} {\bibinfo  {journal} {J. Phys. B}\
  }\textbf {\bibinfo {volume} {51}},\ \bibinfo {pages} {054002} (\bibinfo
  {year} {2018})}\BibitemShut {NoStop}%
\bibitem [{\citenamefont {Steck}(2018)}]{steckquoptics}%
  \BibitemOpen
  \bibfield  {author} {\bibinfo {author} {\bibfnamefont {D.~A.}\ \bibnamefont
  {Steck}},\ }\href@noop {} {\emph {\bibinfo {title} {Quantum and Atom
  Optics,}}}\ (\bibinfo  {publisher} {Available online at
  http://steck.us/teaching},\ \bibinfo {year} {revision 0.12.3, 25 October
  2018})\BibitemShut {NoStop}%
\bibitem [{Note7()}]{Note7}%
  \BibitemOpen
  \bibinfo {note} {We've included the assumption of a single input photon in
  our theory \protect \emph {ab initio}. Thus, discrete state expectation
  values must be correlated, manifesting as the purely virtual coupling in Eq.
  \ref {quantlangspectpar}. If we relax the assumption of a single input
  photon, we find that these correlations still manifest in the portion of the
  POVM that projects onto a single photon Hilbert space, that is, the part that
  is relevant for single photon detection! (See Ref.~\cite {spectralPOVM} for
  details on constructing POVMs that include general Fock states of
  photons.)}\BibitemShut {NoStop}%
\bibitem [{Note8()}]{Note8}%
  \BibitemOpen
  \bibinfo {note} {For $N=2$, the requirement for large spacing such that
  $R(\omega _i)=0$ is less stringent, we only need it much larger than
  difference in decay rates $|\omega _i-\omega _j|\gg (\protect \sqrt {\Gamma
  _j}-\protect \sqrt {\gamma _j})^2$.}\BibitemShut {Stop}%
\bibitem [{\citenamefont {Asenjo-Garcia}\ \emph {et~al.}(2017)\citenamefont
  {Asenjo-Garcia}, \citenamefont {Hood}, \citenamefont {Chang},\ and\
  \citenamefont {Kimble}}]{atomwire1}%
  \BibitemOpen
  \bibfield  {author} {\bibinfo {author} {\bibfnamefont {A.}~\bibnamefont
  {Asenjo-Garcia}}, \bibinfo {author} {\bibfnamefont {J.~D.}\ \bibnamefont
  {Hood}}, \bibinfo {author} {\bibfnamefont {D.~E.}\ \bibnamefont {Chang}}, \
  and\ \bibinfo {author} {\bibfnamefont {H.~J.}\ \bibnamefont {Kimble}},\
  }\href@noop {} {\bibfield  {journal} {\bibinfo  {journal} {Phys. Rev. A}\
  }\textbf {\bibinfo {volume} {95}},\ \bibinfo {pages} {033818} (\bibinfo
  {year} {2017})}\BibitemShut {NoStop}%
\bibitem [{\citenamefont {Qi}\ \emph {et~al.}(2016)\citenamefont {Qi},
  \citenamefont {Baragiola}, \citenamefont {Jessen},\ and\ \citenamefont
  {Deutsch}}]{atomwire2}%
  \BibitemOpen
  \bibfield  {author} {\bibinfo {author} {\bibfnamefont {X.}~\bibnamefont
  {Qi}}, \bibinfo {author} {\bibfnamefont {B.~Q.}\ \bibnamefont {Baragiola}},
  \bibinfo {author} {\bibfnamefont {P.~S.}\ \bibnamefont {Jessen}}, \ and\
  \bibinfo {author} {\bibfnamefont {I.~H.}\ \bibnamefont {Deutsch}},\
  }\href@noop {} {\bibfield  {journal} {\bibinfo  {journal} {Phys. Rev. A}\
  }\textbf {\bibinfo {volume} {93}},\ \bibinfo {pages} {023817} (\bibinfo
  {year} {2016})}\BibitemShut {NoStop}%
\bibitem [{\citenamefont {Lopez-Dieguez}\ \emph {et~al.}(2017)\citenamefont
  {Lopez-Dieguez}, \citenamefont {Estudillo-Ayala}, \citenamefont
  {Jauregui-Vazquez}, \citenamefont {Sierra-Hernandez}, \citenamefont
  {Herrera-Piad}, \citenamefont {Cruz-Duarte}, \citenamefont
  {Hernandez-Garcia},\ and\ \citenamefont {Rojas-Laguna}}]{multimodefabry}%
  \BibitemOpen
  \bibfield  {author} {\bibinfo {author} {\bibfnamefont {Y.}~\bibnamefont
  {Lopez-Dieguez}}, \bibinfo {author} {\bibfnamefont {J.}~\bibnamefont
  {Estudillo-Ayala}}, \bibinfo {author} {\bibfnamefont {D.}~\bibnamefont
  {Jauregui-Vazquez}}, \bibinfo {author} {\bibfnamefont {J.}~\bibnamefont
  {Sierra-Hernandez}}, \bibinfo {author} {\bibfnamefont {L.}~\bibnamefont
  {Herrera-Piad}}, \bibinfo {author} {\bibfnamefont {J.}~\bibnamefont
  {Cruz-Duarte}}, \bibinfo {author} {\bibfnamefont {J.}~\bibnamefont
  {Hernandez-Garcia}}, \ and\ \bibinfo {author} {\bibfnamefont
  {R.}~\bibnamefont {Rojas-Laguna}},\ }\href@noop {} {\bibfield  {journal}
  {\bibinfo  {journal} {Optik}\ }\textbf {\bibinfo {volume} {147}},\ \bibinfo
  {pages} {232} (\bibinfo {year} {2017})}\BibitemShut {NoStop}%
\bibitem [{Note9()}]{Note9}%
  \BibitemOpen
  \bibinfo {note} {We now can perform a small sanity check. Considering the two
  continuum fields together, we define total input and output flux operators
  $J_{in} = \protect \sqrt {\gamma } a_{in} + \protect \sqrt {\Gamma } b_{in}$
  and $J_{out} = \protect \sqrt {\gamma } a_{out} + \protect \sqrt {\Gamma }
  b_{out}$. Solving for the total outputs directly from (\ref {Nstatebound}),
  we find $J_{out} =\protect \frac {i+ \protect \frac {\gamma +\Gamma
  }{2}\protect \tmspace +\thinmuskip {.1667em}\DOTSB \sum@ \slimits@ _i
  \protect \frac {1}{\Delta _i}}{i - \protect \frac {\gamma +\Gamma
  }{2}\protect \tmspace +\thinmuskip {.1667em}\DOTSB \sum@ \slimits@ _i
  \protect \frac {1}{\Delta _i}}\protect \tmspace +\thinmuskip
  {.1667em}J_{in}$. We can immediately see that $|J_{out}|=|J_{in}|$; photon
  flux is preserved through the system at every frequency, as it must be since
  there are no side channels present.}\BibitemShut {Stop}%
\bibitem [{Note10()}]{Note10}%
  \BibitemOpen
  \bibinfo {note} {The uncertainty in frequency, as defined in \cite
  {vanenk2017} and calculated entropically from the spectral POVM \cite
  {spectralPOVM}, is in this case directly proportional to the bandwidth (and
  hence also independent of discrete state spacing).}\BibitemShut {Stop}%
\bibitem [{Note11()}]{Note11}%
  \BibitemOpen
  \bibinfo {note} {For $N=1$, the lower order equation is just the requirement
  that $\gamma =\Gamma $ and the higher order equation just requires $\Delta
  _1=0$ (on-resonance). For $N=2$, the lower order equation is $\gamma \Delta
  _2=\Gamma \Delta _1$ and the higher order equation is $g^2=\Delta _1 \Delta
  _2+\protect \frac {\gamma \Gamma }{4}$. The lower order equation is only
  frequency independent when $\gamma =\Gamma $ and $\Delta _2=\Delta _1$. When
  the discrete states are degenerate, we find that $g\geq \protect \frac
  {\protect \sqrt {\gamma \protect \tmspace +\thinmuskip {.1667em}\Gamma }}{2}$
  is the requirement for perfect transmission at one ($g= \protect \frac
  {\protect \sqrt {\gamma \protect \tmspace +\thinmuskip {.1667em}\Gamma
  }}{2}$) or two ($g> \protect \frac {\protect \sqrt {\gamma \protect \tmspace
  +\thinmuskip {.1667em}\Gamma }}{2}$) frequencies.}\BibitemShut {Stop}%
\bibitem [{Note12()}]{Note12}%
  \BibitemOpen
  \bibinfo {note} {The bound is somewhat stronger for critically coupled
  networks, with the number of perfectly transmitted frequencies bounded by
  $(M-\protect \frac {1}{2}(1+(-1)^M))\protect \tmspace +\thinmuskip
  {.1667em}Min\protect \{N_k\protect \}$ instead. This is due to the
  on-resonance broadening that occurs for critically coupled systems, which
  prevents the splitting of one peak to two when the number of manifolds $M$ is
  even.}\BibitemShut {Stop}%
\bibitem [{Note13()}]{Note13}%
  \BibitemOpen
  \bibinfo {note} {It's worth reiterating here that \protect \emph {all}
  networks reduce to (\ref {RGen}) in the very-weak coupling limit $2g_{ij}\ll
  \protect \sqrt {\gamma _i\gamma _j}+\protect \sqrt {\Gamma _i\Gamma _j}
  \protect \tmspace +\thinmuskip {.1667em}\forall i,j$}\BibitemShut {NoStop}%
\bibitem [{\citenamefont {Solli}\ \emph {et~al.}(2002)\citenamefont {Solli},
  \citenamefont {Chiao},\ and\ \citenamefont {Hickmann}}]{solli2002}%
  \BibitemOpen
  \bibfield  {author} {\bibinfo {author} {\bibfnamefont {D.}~\bibnamefont
  {Solli}}, \bibinfo {author} {\bibfnamefont {R.~Y.}\ \bibnamefont {Chiao}}, \
  and\ \bibinfo {author} {\bibfnamefont {J.~M.}\ \bibnamefont {Hickmann}},\
  }\href@noop {} {\bibfield  {journal} {\bibinfo  {journal} {Phys. Rev. E}\
  }\textbf {\bibinfo {volume} {66}},\ \bibinfo {pages} {056601} (\bibinfo
  {year} {2002})}\BibitemShut {NoStop}%
\bibitem [{Note14()}]{Note14}%
  \BibitemOpen
  \bibinfo {note} {Of course, this assumes a large spectral bandwidth is a
  desired. There is always an obvious tradeoff between the spectral bandwidth
  and the amount one learns about the incoming light; the more frequencies that
  can be detected efficiently, the less a single click tells you about the
  incoming light. Whether a high bandwidth or a low bandwidth is preferred will
  ultimately depend on the application of a SPD.}\BibitemShut {Stop}%
\bibitem [{\citenamefont {Propp}\ and\ \citenamefont {van Enk}()}]{Propp2}%
  \BibitemOpen
  \bibfield  {author} {\bibinfo {author} {\bibfnamefont {{\relax Tz}.~B.}\
  \bibnamefont {Propp}}\ and\ \bibinfo {author} {\bibfnamefont {S.~J.}\
  \bibnamefont {van Enk}},\ }\href@noop {} {\bibinfo  {journal} {in
  preparation}\ }\BibitemShut {NoStop}%
\bibitem [{\citenamefont {Baxter}(1992)}]{kuhn1992}%
  \BibitemOpen
\bibfield  {journal} {  }\bibfield  {author} {\bibinfo {author} {\bibfnamefont
  {C.}~\bibnamefont {Baxter}},\ }\href@noop {} {\bibfield  {journal} {\bibinfo
  {journal} {J. Phys. B}\ }\textbf {\bibinfo {volume} {25}},\ \bibinfo {pages}
  {L589} (\bibinfo {year} {1992})}\BibitemShut {NoStop}%
\bibitem [{\citenamefont {Molmer}\ and\ \citenamefont
  {Castin}(1996)}]{molmer1996}%
  \BibitemOpen
  \bibfield  {author} {\bibinfo {author} {\bibfnamefont {K.}~\bibnamefont
  {Molmer}}\ and\ \bibinfo {author} {\bibfnamefont {Y.}~\bibnamefont
  {Castin}},\ }\href@noop {} {\bibfield  {journal} {\bibinfo  {journal}
  {Quantum Semiclass. Opt.: J. Eur. Opt. Soc. B}\ }\textbf {\bibinfo {volume}
  {8}},\ \bibinfo {pages} {49} (\bibinfo {year} {1996})}\BibitemShut {NoStop}%
\end{thebibliography}%

\onecolumngrid
\vspace{\columnsep}
\vspace{3\columnsep}
\twocolumngrid

\appendix{
\section{Photo detection POVM}
So far in this paper, we have not discussed in detail what happens to the excitation after it ends up in the output mode $\bout$, except that it gives rise to a macroscopic photo detector click. Following Ref.~\cite{spectralPOVM}, we can construct the full POVM describing a photo detector from which all standard figures of merit can be obtained \cite{vanenk2017}. Assume that we finalize the photo detection process by simply ascertaining at some time $t$ whether the excitation is indeed in $\bout$ and also that the amplification is ideal and lossless so that, if a photon makes it to the monitored continuum, it will be detected (this assumption is revisited in more depth in \cite{proppamp}). We can then define normalized filtered photon states

\bea\label{state}
\ket{T\,\phi_t}=\frac{1}{\sqrt{\pi\,\tilde{\Gamma}}} \int_{0}^\infty\,d\omega \,T^*(\omega)\,e^{i\,\omega\,t}\,\hat{a}^{\dagger}(\omega)\,\ket{\textnormal{vac}}. 
\eea

From the quantum jump method \cite{molmer1996,pseudomodes1}, we know that a quantum jump from the last manifold of discrete states to the monitored continuum will occur in an infinitesimal time $dt$ with condition probability $\frac{\tilde{\Gamma}\,dt}{2}$. We can then infer the POVM element for detecting a photon at a particular time $t$ after the photo detector has been on for a time $dt$

\bea\label{povm}
\hat{ \Pi}_t=\frac{\tilde{\Gamma}\,dt}{2}\,\ket{T\,\phi_t}\bra{T\,\phi_t}.
\eea

(To reiterate, here $t$ refers to the time of detection, not the time evolution of the input state.) The probability of getting a click for a normalized input photon $\hat{\rho}$ is $\textnormal{Tr}\left(\hat{\Pi}_t\hat{\rho}\right)\leq1$. So the input state that will be detected with maximum probability is $\hat{\rho}=\ket{T\,\phi_t}\bra{T\,\phi_t}$, yielding an infinitesimal probability of detection of $\frac{\tilde{\Gamma}\,dt}{2}$. This is because this assumes that the detection events at times separated by a time $dt$ correspond to different measurement outcomes which is highly idealized; any realistic photo detection outcome corresponds to detection within an integrated time-window (see again footnote \footnotemark[1]).

To take a finite time-window into account, we consider a time-integrated POVM element $\hat{ \Pi}_\tau$, where a click corresponds to a detection event sometime between $t=0$ and $t=-\tau$. Then the time-integrated POVM element is
\bea\label{povm2}
\hat{ \Pi}_\tau=\int_{-\tau}^0 \,dt\,\frac{\tilde{\Gamma}}{2}\,\ket{T\,\phi_t}\bra{T\,\phi_t}\nonumber\\
 \approx \int_{-\infty}^{\infty}\,d\omega\,|T(\omega)|^2 \ket{\omega}\bra{\omega} \,\,\,\,\,\,(\tau\gg\tilde{\Gamma}^{-1})
\eea where, for $\tau\rightarrow\infty$, the projectors $\ket{\omega}\bra{\omega}$ are truly monochromatic because no timing information is obtained. With the time-integrated form in (\ref{povm2}), we can see that the conditions for perfect transmission discussed above correspond to perfect detection in the limit of $\tau\gg\tilde{\Gamma}^{-1}$ as we used previously in (\ref{povmlongtime}). Physically, a non-infinitesimal integration time $\tau$ means a photon incident on the photo detector at a time $t=-\tau$ has had sufficient time to propagate through the network before we end the integration at time $t=0$ and check whether there has been a click. This is in agreement with our observation that, for the simple model in (\ref{1state}), $\tau_g(\omega_0)=\tilde{\Gamma}^{-1} $: we must at least allow enough time for an on-resonance photon to travel through the network (interact with the device) to achieve perfect detection of a monochromatic on-resonance photon. (\ref{povm2}) is also illustrative of what the POVM element represents: the information a detector click reveals about what led up to it.

We can also use (\ref{povm2}) to construct the POVM element for not getting a click in the finite time $\tau$, which is simply $\hat{ \Pi}_{0}=\hat{ 1}-\hat{ \Pi}_\tau$, such that the full POVM $\{\hat{ \Pi}_{0},\,\hat{ \Pi}_{\tau}\}$ forms a partition of unity for the relevant Hilbert space; that is, the Hilbert space spanned by single photon states and the vacuum. (See Ref.~\cite{vanenk2017} for inclusion of general photon Fock states in the photo detection POVM.) This POVM corresponds to a photo detector that is reset after each integration time $\tau$. Similarly, one can divide up the the full detection window into $N$ time intervals $\tau_i$ without resetting the device in between so that the full POVM is $\{\hat{ \Pi}_{0},\,\hat{ \Pi}_{\tau_i}\}_{i=1...N}$. In this case, the measurement outcomes are not orthogonal \cite{spectralPOVM}; even if we know a photon is incident at a definite time $t'$ we cannot predict with certainty when it will be detected.

\end{document}